\documentclass[useAMS,usenatbib,twocolumn]{mnras}
\usepackage{newtxtext,newtxmath}
\usepackage[T1]{fontenc}
\usepackage{ae,aecompl}
\usepackage[dvipsnames]{xcolor}
\usepackage[normalem]{ulem}
\usepackage[bottom]{footmisc}
\usepackage{lineno}
\usepackage{amsfonts, dsfont}
\usepackage{savesym}
\usepackage[nointegrals]{wasysym}
\usepackage{color}
\usepackage[normalem]{ulem}
\usepackage[hang]{subfigure}
\usepackage{hyperref}
\usepackage{accents}
\usepackage{threeparttable, booktabs}
\savesymbol{tablenum}
\usepackage{physics}
\usepackage{siunitx}
\usepackage{mathtools}
\restoresymbol{SIX}{tablenum}
\usepackage{changes}
\usepackage{float}

\DeclareSIUnit{\erg}{erg}
\DeclareSIUnit{\jansky}{Jy}
\DeclareSIUnit{\parsec}{pc}
\DeclareSIUnit{\yr}{yr}
\DeclareSIUnit\msun{M\textsubscript{\astrosun}}
\DeclareSIUnit{\radian}{rad}
\DeclareSIUnit{\pixel}{px}
\DeclareSIUnit{\day}{d}
\DeclareSIUnit{\arcsectxt}{as}
\DeclareSIUnit{\arcsec}{arcsec}
\DeclareSIUnit{\correlation}{correlation}
\DeclareSIUnit{\strain}{strain}

\newcommand{\vct}[1]{\accentset{\rightharpoonup}{#1}}
\newcommand{\mtx}[1]{\mathbf{#1}}
\newcommand{\E}[1]{\langle #1 \rangle}
\newcommand{\cov}[2]{\mathrm{Cov}({#1,#2})}
\newcommand{\gw}{\mathrm{gw}}

\newcommand{\GWB}{\mathrm{GWB}}

\newcommand{\C}{(C^{-1})}
\newcommand{\trc}[1]{\mathrm{tr}[#1]}

\newcommand{\MPTAsigma}{3.4}

\newcommand{\radiometerpval}{0.015}
\newcommand{\cleanpval}{0.03}

\begin{document}

\title[MeerKAT Pulsar Timing Array: Search for Anisotropy]{The MeerKAT Pulsar Timing Array:\\
Maps of the gravitational-wave sky with the 4.5 year data release} 



\newcommand{\swinburne}{Centre for Astrophysics and Supercomputing, Swinburne University of Technology, P.O. Box 218, Hawthorn, Victoria 3122, Australia}
\newcommand{\SPA}{School of Physics and Astronomy, Monash University, Clayton VIC 3800, Australia}
\newcommand{\OzGravMonash}{OzGrav: The ARC Centre of Excellence for Gravitational Wave Discovery, Clayton VIC 3800, Australia}
\newcommand{\OzGravSwinburne}{OzGrav: The ARC Centre of Excellence for Gravitational Wave Discovery, Hawthorn, Victoria 3122, Australia}
\newcommand{\ozgrav}{Australia Research Council Centre for Excellence for Gravitational Wave Discovery (OzGrav)}
\newcommand{\CSIRO}{Australia Telescope National Facility, CSIRO, Space and Astronomy, PO Box 76, Epping, NSW 1710, Australia}

\newcommand{\mpifr}{Max-Planck-Institut f{\"u}r Radioastronomie, Auf dem H{\"u}gel 69, 53121 Bonn}
\newcommand{\aei}{Max Planck Institute for Gravitational Physics (Albert Einstein Institute), Am M{\"u}hlenberg 1, 14476 Potsdam, Germany}
\newcommand{\jbca}{Jodrell Bank Centre for Astrophysics, University of Manchester,\\ Department of Physics and Astronomy, Alan-Turing Building, Oxford Street, Manchester M13 9PL, UK}
\newcommand{\astron}{ASTRON, Netherlands Institute for Radio Astronomy, Oude Hoogeveensedijk 4, 7991 PD, Dwingeloo, The Netherlands}
\newcommand{\apc}{Universit{\'e} Paris Cit{\'e}, CNRS, Astroparticule et Cosmologie, 75013 Paris, France}
\newcommand{\nancay}{Observatoire Radioastronomique de Nan\c{c}ay, Observatoire de Paris, Universit\'e PSL, Université d'Orl\'eans, CNRS, 18330 Nan\c{c}ay, France}
\newcommand{\unimib}{Dipartimento di Fisica ``G. Occhialini", Universit{\'a} degli Studi di Milano-Bicocca, Piazza della Scienza 3, I-20126 Milano, Italy}
\newcommand{\infnunimib}{INFN, Sezione di Milano-Bicocca, Piazza della Scienza 3, I-20126 Milano, Italy}
\newcommand{\inafbrera}{INAF - Osservatorio Astronomico di Brera, via Brera 20, I-20121 Milano, Italy}
\newcommand{\inafoac}{INAF - Osservatorio Astronomico di Cagliari, via della Scienza 5, 09047 Selargius (CA), Italy}
\newcommand{\amst}{Anton Pannekoek Institute for Astronomy, University of Amsterdam, Science Park 904, 1098 XH Amsterdam, The Netherlands}

\newcommand{\uct}{High Energy Physics, Cosmology \& Astrophysics Theory (HEPCAT) Group,\\ Department of Mathematics and Applied Mathematics, University of Cape Town, Cape Town 7700, South Africa}
\newcommand{\sarao}{South African Radio Astronomy Observatory, 2 Fir Street, Observatory 7925, South Africa}


\author[Grunthal \& Nathan et al.]
{Kathrin Grunthal$^{1}$\thanks{kgrunthal@mpifr-bonn.mpg.de rowina.nathan@monash.edu}, Rowina S.~Nathan$^{2,3}$\footnotemark[1], Eric Thrane$^{2,3}$, David J.~Champion$^{1}$, Matthew T.~Miles$^{4,5}$,\newauthor
Ryan M.~Shannon$^{4,5}$, Atharva D.~Kulkarni$^{4,5}$, Federico Abbate$^{6,1}$, Sarah Buchner$^{7}$, Andrew D.~Cameron$^{5}$,\newauthor
Marisa Geyer$^{8}$, Pratyasha Gitika$^{4,5}$, Michael J.~Keith$^{9}$, Michael Kramer$^{1,9}$, Paul D.~Lasky$^{2,3}$,\newauthor
Aditya Parthasarathy$^{10,11,1}$, Daniel J.~Reardon$^{4,5}$, Jaikhomba Singha$^{8}$, Vivek Venkatraman Krishnan$^{1}$
\\
$^{1}$\mpifr\\
$^{2}$\SPA\\
$^{3}$\OzGravMonash\\
$^{4}$\swinburne\\
$^{5}$\OzGravSwinburne\\
$^{6}$\inafoac\\
$^{7}$\sarao\\
$^{8}$\uct \\
$^{9}$\jbca\\
$^{10}$\astron \\
$^{11}$\amst
}












\maketitle

\begin{abstract}
In an accompanying publication, the MeerKAT Pulsar Timing Array (MPTA) collaboration reports tentative evidence for the presence of a stochastic gravitational-wave background, following observations of similar signals from the European and Indian Pulsar Timing Arrays, NANOGrav, the Parkes Pulsar Timing Array and the Chinese Pulsar Timing Array. 
If such a gravitational-wave background signal originates from a population of inspiraling supermassive black-hole binaries, the signal may be anisotropically distributed on the sky. 
In this \textit{Letter} we evaluate the anisotropy of the MPTA signal using a spherical harmonic decomposition.
We discuss complications arising from the covariance between pulsar pairs and regularisation of the Fisher matrix. 
Applying our method to the $\SI{4.5}{\yr}$ dataset, we obtain two forms of sky maps for the three most sensitive MPTA frequency bins between $\SIrange{7}{21}{\nano\hertz}$.
Our ``clean maps'' estimate the distribution of gravitational-wave strain power with minimal assumptions.
Our radiometer maps answer the question: is there a statistically significant point source?
We find a noteworthy hotspot in the $\SI{7}{\nano\hertz}$ clean map with a $p$-factor of $p=\radiometerpval$ (not including trial factors).
Future observations are required to determine if this hotspot is of astrophysical origin.
\end{abstract}

\begin{keywords}
gravitational waves - methods: data analysis - methods: statistical - pulsars: general - stars: black holes
\end{keywords}

\section{Introduction} \label{sec:intro}
Acting as cosmic clocks, millisecond pulsars emit extraordinarily regular pulses at radio frequencies that are observed with ground-based radio telescopes. The stability of millisecond pulsar rotation allows us to build accurate timing models that account for the physics both intrinsic and extrinsic to the pulsar, with the most accurate models able to predict pulse times of arrival (ToAs) to a precision of tens of nanoseconds \citep[e.g.][]{NANOGrav15yObs, Reardon_2024, Wang2024}. 
Distortions in spacetime caused by gravitational waves induce a temporally correlated signal in the arrival times and hence residuals of each pulsar.
In a collection of pulsars, a so-called pulsar timing array (PTA), this gravitational-wave signal appears as a correlation between the residuals of different pulsars depending on the separation angle between these pulsars \citep{HellingsDowns}. 

Pulsar timing arrays are sensitive to nanohertz gravitational waves with periods of years to decades.
In this frequency range, gravitational waves are likely to create a stochastic gravitational-wave background from the superposition of incoherent gravitational waves radiated from many sources \citep{AllenRomano_1999,Rosado_2015}.
The signal is expected to first appear as a time-correlated red noise process \citep{Siemens_2013}, characterized by a power spectral density with a steep power law index. 
However, this red process may be mimicked by pulsar noise processes \citep{Hazboun_2020,ppta_dr2_noise,common_red_noise,Zic2022}, and so angular correlations are crucial for establishing a confident detection.

An isotropic gravitational-wave background induces an angular correlation function known as the Hellings-Downs curve, which depends only on the angular separation of pairs of pulsars \citep{HellingsDowns}. 
The unique (nearly quadrupolar) correlations of the Hellings-Downs curve are distinct from other correlations that may be induced from systematic errors such as clock errors (which induce monopole moments) and Solar System ephemeris errors (which induce dipole moments) \citep{Tiburzi_2016}.
A statistically significant Hellings-Downs correlation is the evidence required for a gravitational-wave background detection.

In June 2023 there was a coordinated release of pulsar timing papers by members of the International Pulsar Timing Array (IPTA), with each presenting evidence for a gravitational-wave background signal. The North American Nanohertz Observatory for Gravitational Waves (NANOGrav) reported evidence for a gravitational-wave background with a significance of $\sim3-4 \sigma$ \citep[Bayesian vs frequentist methods provide slightly different levels of evidence; see][]{NG15_GWB},
while the joint publication from the European and Indian Pulsar Timing Arrays (EPTA and InPTA respectively) reported $3-4 \sigma$ \citep[using a subset of the full data; see][]{EPTA_DR2_GWB}. 
The Parkes Pulsar Timing Array (PPTA) simultaneously found $\sim2 \sigma$ evidence for a gravitational-wave background signal \citep{PPTA_GWB}. In addition to these IPTA publications, the Chinese Pulsar Timing Array (CPTA) reported a $4.6 \sigma$ evidence for the presence of a Hellings-Downs correlation in the data at the frequency slice of \SI{14}{\nano\hertz}\citep{CPTA_GWB}, although due to the frequency-dependent analysis, the CPTA result is not directly comparable to the results from the other PTAs. 

Once a gravitational-wave background has been established, one of the next key steps is to characterize how the signal is distributed on the sky. In this \textit{Letter} we assess the anisotropy of the nanohertz gravitational-wave background in the hope of determining its source.

The most likely source of the gravitational-wave background is a population of inspiraling supermassive black-hole binaries \citep{Rajagopal_1995}. Supermassive black holes are thought to be found in the centre of most massive galaxies \citep{Kormendy_1995}, and mergers between these galaxies gravitationally bind their central black holes, leading to the formation of supermassive black-hole binaries \citep{Begelman_1980}. 
Circular, gravitationally-driven supermassive binary black holes are expected to stall in a phenomenon known as the final parsec problem~\citep{Milosavljevic_2003}, though various solutions have been proposed including multiple-body interactions \citep{Ryu_2018, Bonetti_2019}, hardening from interactions with stars \citep{Gualandris_2017} and galaxy rotation \citep{HolleyBockelmann_2015}. 

A gravitational-wave background arising from supermassive black-hole binaries is likely to exhibit anisotropy from the finite number of merging binaries in the PTA observing band. While \cite{Mingarelli_2017} estimated that the level of anisotropy due to undetected continuous wave sources is about 20\%, other studies \cite[e.g.,][]{Sesana_2004,Simon_2023} argue that there is a large uncertainty involved, due to our relatively poor knowledge of the supermassive binary black hole population .

Alternatively, a gravitational-wave background may arise from processes in the early Universe \citep{Maggiore_2000}. 
Primordial gravitational waves from quantum fluctuations red-shifted by inflation \citep{Guzzetti_2016, Lasky_2016, Yuan_2021, Domenech_2021}, phase transitions \citep{Caprini_2020, Hindmarsh_2021} or cosmic strings \citep{Hindmarsh_1995, Saikawa_2017} are likely to give rise to an isotropic signal. 
Understanding the degree of anisotropy in the gravitational-wave background therefore provides clues about its source.
Recent analyses from NANOGrav \citep{Agazie_2023} and the EPTA \citep{Taylor_2015} have yielded no evidence for anisotropy in the nanohertz gravitational-wave background.

The remainder of this \textit{Letter} is organised as follows.
We start by giving a brief overview of the MPTA and the dataset used in this analysis in Section~\ref{sec:MPTA}. This is then followed by an in-depth description of the methods we use for evaluating the anisotropy of the nanohertz gravitational-wave background in Section~\ref{sec:methods}. 
Our method follows techniques developed in the context of LIGO and subsequently PTAs.
However, we extend this framework to take into account the contribution of noise from the gravitational-wave background itself. In Section~\ref{sec:results} we present the results obtained from our analysis of the MeerKAT Pulsar Timing Array (MPTA) 4.5-year dataset. 
We conclude in Section~\ref{sec:discussion} with a discussion of these findings and the future outlook. Lastly, the Appendix of this paper contains a series of auxiliary calculations and tests documenting the performance of our data analysis pipeline.

\section{The MeerKAT Pulsar Timing Array}
\label{sec:MPTA}
The MPTA uses the MeerKAT radio telescope, a 64-antenna array, located in the Northern Cape, South Africa \citep{Jonas_2018}. Together these antennas provide a gain of \SI{2.8}{\kelvin\per\jansky} making MeerKAT the most sensitive radio telescope in the Southern Hemisphere.
The MPTA is one sub-project of the MeerTime large survey project \citep{Bailes_2020}.
For pulsar timing work the beam formed data are coherently de-dispersed (to remove the dispersive effect of the ionised interstellar medium) and folded at the topocentric period of the pulsar. They are processed to remove interference signals and calibrate polarisation resulting Stokes~$I$ pulse profiles with 32 frequency channels.  In most cases the observations were fully  time averaged; however some long observations (usually observed as part of other sub-projects) were subdivided to better match the integration time of standard MPTA observations \citep{MPTA2024_data+noise}.

A frequency-resolved template (describing the pulse profile) is produced for each pulsar using the \texttt{PulsePortraiture} software \citep{Pennucci2019}. The ToAs (of the pulses closest to the middle of the observations) are determined using the Fourier domain Monte-Carlo method in the \texttt{psrchive} software suite \citep{Hotan2004} for each channel.
The data used in this work come from a 4.5 year period from February 2019 to August 2024. The MPTA observes 83 pulsars with an approximately 14 day cadence. Observations were done using the $L$-band receiver covering a bandwidth of 856 MHz centred at $\SI{1284}{\mega\hertz}$; see \cite{MPTA2024_data+noise} for more detail.

The frequency-resolved information is used to constrain models of chromatic effects, such as variations in the ionised interstellar medium, and the effect of the solar wind. Achromatic noise is modelled for both red and white noise processes. Details of this noise modelling can be found in \cite{MPTA2024_data+noise}.
The MPTA reports a \MPTAsigma $\sigma$ evidence for a stochastic isotropic gravitational-wave background in the $\sim \SIrange{7}{21}{\nano\hertz}$ frequency range \citep{MPTA2024_GWB}. This analysis uses the results from the ALT analysis in \citep{MPTA2024_GWB}, a conservative set of noise models where additional achromatic noise process have been added (for all pulsars barring PSR J2129$-$5721).
The excellent sensitivity of MeerKAT enables us to study the gravitational-wave background at high frequencies and high observational cadence. 
High frequency datasets are particularly useful in searches for anisotropy because the number density of binary black holes is expected to fall sharply with frequency $p(f) \propto f^{-11/3}$ \citep[assuming gravitationally-driven, circular binaries][]{Peters_1964,Rajagopal_1995,Smith_2018,Gardiner_2024}.
Thus, at higher frequencies, we expect the background to be created by a relatively smaller number of binaries.

\section{Methods} \label{sec:methods}
\subsection{Overview}\label{ssec:Overview}
Gravitational-wave cartography was pioneered in the context of ground-based detector community \citep[e.g.][]{Allen_1997,Ballmer_2006,Mitra_2008,Thrane_2009,Abadie_2011}. 
In the PTA community, \cite{Mingarelli_2013} derived a generalised overlap reduction function for PTA analyses, leading to the first Bayesian PTA analysis strategy towards estimating the anisotropy of a gravitational wave background by \cite{Taylor_2013}, which was later generalised by \cite{Gair_2014}.
A frequentist analysis pipeline using optimal statistics \citep[e.g.][]{Vigeland_2018} and the likelihood maximisation developed by \cite{Thrane_2009,Romano_2017} was recently put forward by \cite{Pol_2022}.

While there are differences, these methods employ many common features:
They define a basis consisting of either pixels or spherical harmonic functions \citep{Cornish2014,Ali-Haimoud_2020,Ali-Haimoud_2021,Taylor_2020,Banagiri_2021,Pol_2022,Agazie_2023} and estimate the gravitational-wave power (proportional to strain squared) associated with each basis element by maximizing the likelihood.
These are the most common choices. Other bases such as PTA eigenmaps \cite{Cornish2014,Ali-Haimoud_2020,Ali-Haimoud_2021} are also considered.
Here, we follow the method outlined in \cite{Thrane_2009,Abadie_2011}---and adapted for PTAs in \cite{Pol_2022} and references therein---which employs a spherical harmonic basis. In this framework, the reconstructed sky map is allowed to have regions with negative gravitational-wave power. 

Since the stochastic background can only induce positive power, these patches of negative power represent noise fluctuations.
While some authors chose to enforce positive power (e.g., using a ``square root basis'' \citep{Payne_2020,Taylor_2020,Pol_2022}, or through the use of priors \citep{Taylor_2013,Gair_2014},) we prefer to allow negative power as a way of visualizing the noise fluctuations.
Indeed, we use the behavior of the sky map noise fluctuations as a diagnostic to check that our pipeline produces reasonable results.

We highlight two dedicated PTA searches for anisotropy.
The first one was carried out by \cite{Taylor_2015} as part of the European Pulsar Timing Array (EPTA) collaboration \citep{Kramer_2013} using the timing data from the six best-quality pulsars. 
The second and most recent anisotropy analysis by \cite{Agazie_2023} was presented in the papers series following the 15-year NANOGrav data release. 
Neither found evidence for anisotropy.

In the following subsections, we introduce the formalism for our approach. Speaking broadly to give a high-level overview, we follow the spherical harmonic formalism from \cite{Mitra_2008} to derive ``clean maps,'' through the regularisation scheme described in \cite{Thrane_2009} using the generalised overlap reduction functions from \cite{Mingarelli_2013}.
We make some modifications to relax the assumption of the weak-signal limit.

\subsection{The overlap reduction function}
A PTA consisting of $N_\mathrm{psr}$ pulsars allows for the calculation of $N_\text{pairs} = (N_\mathrm{psr}^2 - N_\mathrm{psr})/2$ correlations between distinct pulsar pairs. 
These correlations encode the distribution of gravitational-wave power on the sky via the so-called overlap reduction function \citep{Christensen_1992,Allen_1997,Finn_2009,Mingarelli_2013}:
\begin{equation}\label{eq:Gamma}
    \Gamma_{ab} \propto \int_{S^2} \! \mathrm{d}\Hat{\Omega} \; \mathcal{P}(\Hat{\Omega}) \left[ \mathcal{F}_a^+(\Hat{\Omega})\mathcal{F}_b^+(\Hat{\Omega})
    +  \mathcal{F}_a^\times(\Hat{\Omega})\mathcal{F}_b^\times(\Hat{\Omega})\right] .
\end{equation}
Here, $\mathcal{P}(\Hat{\Omega})$ is the probability density function for gravitational-wave power at different locations on the sky $\hat{\Omega}$ as seen from the Solar System Barycenter---this is the distribution that we seek to measure.
The $a,b$ indices label different pulsars.
The quantity
\begin{equation}
    \mathcal{F}_a^A(\hat{p}, \Hat{\Omega}) = \frac{1}{2}\frac{\Hat{p}_a^i\Hat{p}_a^j}{1-\Hat{\Omega}\cdot\hat{p}_a}\, e_{ij}^A(\Hat{\Omega}) = \frac{1}{2}\frac{\Hat{p}_a^i\Hat{p}_a^j}{1+\hat{k}\cdot\hat{p}_a}\, e_{ij}^A(\Hat{\Omega})
\end{equation}
is the antenna factor for a gravitational wave propagating along the vector $\hat{k}$, i.e.\ the source is located at position $\hat{\Omega}$ with $\hat{k}$ = -$\hat{\Omega}$, with polarisation state $A$ as measured by a pulsar located in the direction $\hat p_a$.
The term $e^A_{ij}(\hat\Omega)$ is the polarisation tensor for a gravitational wave with polarisation $A$; $i,j$ are indices for spatial coordinates. In our analysis we adopt the definition of the polarisation basis from \cite{Taylor_2020}.

\subsection{Spherical harmonic decomposition}
We expand the distribution of gravitational-wave power in the complex spherical harmonics basis
\begin{equation}\label{eq:cSpH}
    \mathcal{P}(\Hat{\Omega}) = \sum_{l=0}^{\ell_\text{max}}\sum_{m=-\ell}^{\ell} P_{\ell m} Y_{\ell m}(\Hat{\Omega}) ,
\end{equation}
where $Y_{\ell m}(\Hat{\Omega})$ are the complex-valued spherical harmonics and $P_{\ell m}$ the respective coefficients \citep{Cornish_2001, Kudoh_2005, Cornish2014}.
The sum truncates at $\ell_\text{max}$, which we choose in order to match with the intrinsic resolution of our PTA.
Since we have 83 pulsars in our array, a single-frequency-bin map is characterized by at most $\lesssim 83$ independent parameters.
As we discuss below, the number is actually much smaller in practice owing to degeneracies between measurements.
Thus we choose $\ell_\text{max}=8$, which yields 81 degrees of freedom.
It follows that our angular resolution is $180^\circ/\ell_\text{max} \approx 23^\circ$.

\subsection{Response matrix}
Some additional book-keeping is necessary before we proceed.
We introduce the Greek indices $\alpha$ and $\beta$ to denote pulsar pairs $(ab)$ so, for example, we can write $\Gamma_\alpha$ as shorthand for $\Gamma_{ab}$.
Meanwhile, we use the indices $\mu$ and $\nu$ to denote the distinct spherical harmonics, i.e. for example we write the spherical harmonic basis functions as $Y_\mu$.
A summary of indices is provided in Table~\ref{tab:indices}.

\begin{table}
    \centering
    \caption{A summary of different indices used in this paper.}
    \begin{tabular}{c|c|c}
        index & represents & maximum range \\ \hline
        $a,b$  & pulsar & 83 \\
        $\alpha,\beta$ & pairs & 3403 \\
        $\mu,\nu$ & spherical harmonic & 81 
    \end{tabular}
    \label{tab:indices}
\end{table}

With this notation we can relate the overlap reduction function for some pair $\alpha$ to the spherical harmonic coefficients $P_\mu$ with a matrix equation:
\begin{align}
    \Gamma_\alpha = R_{\alpha \mu} P_\mu ,
\end{align}
where 
\begin{equation}\label{eq:R_matrix_sph}
    R_{\alpha\mu} = \int d\Omega \left[ \mathcal{F}^+_a(\hat\Omega) \mathcal{F}^+_b(\hat\Omega) + \mathcal{F}^\times_a(\hat\Omega) \mathcal{F}^\times_b(\hat\Omega)\right] Y_\mu(\hat\Omega) , 
\end{equation}
is referred to as the ``response matrix'' and we adopt the Einstein convention, so that repeated indices imply summation\footnote{Numerically evaluating the integral in Equation~\eqref{eq:Gamma} requires a suitable choice of the discretisation resolution. In this analysis we use \texttt{healpy} \citep{Gorski_2005,Zonca_2019} to create equal-area pixels, governed by the parameter \texttt{NSIDE}, which takes values of $2^n$, where $n \in \mathbb{N}^+$. If the sky patch size is too similar to the spherical harmonics resolution, this under-resolving strongly impairs the sky maps, i.e.\ they vary as a function of \texttt{NSIDE} until the resolution is adequate. For the MPTA dataset, we find it converges with \texttt{NSIDE} $\geq16$. }.

\subsection{Covariance matrices}
Pulsar timing array data consist of $N_\mathrm{psr}$ vectors of pulse times-of-arrival (ToAs) $\vct{\delta t}$. 
By taking the expectation values of the product of different residuals, we construct covariance matrices:
\begin{align}
    \mtx{C}_a &= \langle\vct{\delta t}_a\vct{\delta t}_a^T\rangle \\
    \mtx{S}_{ab} &= \langle\vct{\delta t}_a\vct{\delta t}_b^T\rangle_{a\neq b} .
\end{align}
Here $C_a$ is the auto-correlation variance and $S_{ab}$ is the cross-correlation variance matrix.\footnote{In the literature, the auto-covariance matrix is often also denoted as $P_a$, e.g. \cite[e.g.][]{Chamberlin_2015, Pol_2022}, but we adopt the notation of \cite{Vigeland_2018} as it is the most commonly used reference, and also matches the notation of the analysis script that we use.} 
The auto-correlation matrix is described by the sum of all individual noise processes determined in the single-pulsar noise analysis.

The cross-power matrix, on the other hand, depends only on the gravitational-wave signal that is common to all the pulsars in the array:
\begin{equation}\label{eq:S2}
    S_{ab} = \vct{F}_a \; \Gamma_{ab} P_\GWB(f) \; \vct{F}_b^\dag .
\end{equation} 
Here $P_\GWB(f)$ is the power spectral density of the gravitational-wave signal
\begin{equation}
    P_\GWB(f) = \frac{A_\GWB^2}{12\pi^2}\left(\frac{f}{f_\mathrm{yr}}\right)^{-\gamma_\GWB} ,
\end{equation}
which we parameterize in terms of an amplitude $A_\GWB$ and a spectral index $\gamma_\GWB$. For gravitationally driven binary inspirals, a value of $\gamma_\GWB=13/3$ is expected.
The Fourier basis vectors $\vct{F}_{a,b}$ convert the power spectral density from the frequency domain to the time domain.

\subsection{Likelihood}
Following  \cite{Demorest_2013, Chamberlin_2015,Pol_2022}, we define pulsar cross correlations $\rho_\alpha$ and their respective uncertainties $\sigma_\alpha$:
\begin{align}\label{eq:rho}
    \rho_\alpha \equiv \rho_{ab} &= \frac{\vct{\delta t}^\dag_a \mtx{C}_a^{-1}\mtx{\Hat{S}}_{ab}\mtx{C}_b^{-1}\vct{\delta t}_b}{\mathrm{tr}\left[\mtx{C}_a^{-1}\mtx{\Hat{S}}_{ab}\mtx{C}_b^{-1}\mtx{\Hat{S}}_{ba}\right]} \\
    \label{eq:delta_rho}
    \sigma_\alpha \equiv \sigma_{ab} &= \left(\mathrm{tr}\left[\mtx{C}_a^{-1}\mtx{\Hat{S}}_{ab}\mtx{C}_b^{-1}\mtx{\Hat{S}}_{ba}\right]\right)^{-1/2}.
\end{align}
where
\begin{align}
    \Hat{\mtx{S}}_{ab} = \mtx{S}_{ab} / A_\mathrm{GW}^2 \Gamma_{ab} .
\end{align}
The traces in Equations~\eqref{eq:rho},\eqref{eq:delta_rho} are taken over the ToA indices.
The cross correlations are a convenient reduced data product, which enables us to work with a simple likelihood function.

Assuming stationary Gaussian noise (and a stationary Gaussian stochastic background), the likelihood for the cross correlations can be written as \citep{Mitra_2008,Thrane_2009, Pol_2022}
\begin{equation}\label{eq:likelihood}
    {\cal L}(\rho |\mathcal{P}) = \frac{\exp\left[-\frac{1}{2}\left(\rho_\alpha - R_{\alpha\mu}\mathcal{P}_\mu\right)^\dag \Sigma_{\alpha \beta}^{-1}\left(\rho_\beta - R_{\beta \nu} \mathcal{P}_\nu\right)\right]}{\sqrt{\det(2\pi\mtx{\Sigma})}} .
\end{equation}
Here, $\mtx{\Sigma}$ is the \textit{second moment}\footnote{We avoid calling $\Sigma$ a covariance matrix---which it is---to avoid confusion with $\mtx{C}$ and $\mtx{S}$, which are also covariance matrices. The former is a covariance matrix for the correlations while the latter are covariance matrices for the residuals.} of the cross-correlations \citep{Allen_2023_HD71}:
\begin{equation}\label{eq:Sigma_effective}
    \Sigma_{\alpha\beta} = \mathrm{diag}(\sigma_\alpha^2)\, \delta_{\alpha\beta} + \varsigma_{\alpha\beta}({\cal P}, A_\text{GWB}) .
\end{equation}
The diagonal of $\Sigma$ is just the variance associated with each cross correlation.
However, when a gravitational-wave signal is present the term $\varsigma_{\alpha\beta}$ (which includes off-diagonal components) becomes important \citep{Allen_2023_HD71}.
The rather complicated expressions for $\varsigma$ are provided in Appendix~\ref{app:ssec:Sigma_cc}.

The likelihood is maximized with the following solution \citep{Mitra_2008,Thrane_2009,Pol_2022}:
\begin{equation}\label{eq:ml_estimator}
    \mathcal{P}_\mu' = M_{\mu\nu}^{-1} X_\nu .
\end{equation}
where 
\begin{align}
X_\nu &= R_{\mu \alpha}^\dag \Sigma_{\alpha \beta}^{-1} \, \rho_\beta,  
\label{eq:dirty_map}
\end{align}
is known as the \textit{dirty map} and 
\begin{align}
M_{\mu\nu} &= R_{\mu \alpha}^\dag\Sigma_{\alpha \beta}^{-1}R_{\beta \nu} ,
\label{eq:Fisher_matrix}
\end{align}
is the Fisher matrix of the maximum likelihood estimators\footnote{The apostrophe indicates the maximum likelihood estimate of a quantity} $\mathcal{P}_\mu'$, as shown in \cite{Thrane_2009} and derived in Appendix~\ref{app:ssec:fisher} in the notation used in this work, also accounting for the presence of a gravitational-wave signal. 
The dirty map is the inverse-noise weighted representation of the sky-distributed gravitational-wave power as seen through the response of the pulsars \citep{Thrane_2009,Pol_2022,Agazie_2023,Ali-Haimoud_2021}, i.e., it is a ``blurred'' picture of the true gravitational-wave power distribution. 
The maximum-likelihood estimator in Equation~\eqref{eq:ml_estimator} is referred to as the \textit{clean map} $\vct{\mathcal{P}}'$.\footnote{Our clean maps have units of $\si{\square\correlation\per\steradian}$. The analogous maps produced by LIGO--Virgo \citep{Abadie_2011} have units of $\si{\square\strain\per\hertz\per\steradian}$. } 
Equation~\eqref{eq:ml_estimator} was derived in e.g.\ \cite{Thrane_2009} for the weak-signal limit (i.e.\ where $\zeta$ is small), but it holds also as $\zeta$ increases, given that the corrections to $\mtx{\Sigma}$ are correctly applied.
As $\zeta$ is increased, the maximum-likelihood solution changes from Equation~\eqref{eq:ml_estimator}. 
We return to this detail later.

\subsection{Regularisation}\label{sssec:regularisation}
The inversion of the Fisher matrix in Equation~\eqref{eq:ml_estimator} involves an interesting subtlety.
The Fisher matrix can be diagonalized into eigenmodes, each of which represents some patch of sky \citep{Mitra_2008,Thrane_2009,Ali-Haimoud_2021}; see Fig.~\ref{fig:singular_value_spectrum} in the Appendix.
However, due to the irregular sky distribution and sensitivity of the pulsars, some eigenmodes can be measured far more precisely than others.
As a result, naive inversion of the Fisher matrix produces results that are dominated by the uncertainty associated with the least well-measured modes.

We therefore regularize the Fisher matrix with a singular value decomposition scheme outlined in \cite{Thrane_2009,Abadie_2011}.
We project out the least well-measured modes in and keep only $N_\text{cutoff}$ eigenmodes, i.e.\ formally, we set the smallest eigenvalues of $M$ to infinity.
With these poorly-measured modes removed, we calculate the inverse of the regularized Fisher matrix denoted $\Tilde{M}_{\mu\nu}^{-1}$.

Choosing the number of modes to keep $N_\text{cutoff}$ is a trade-off.
By throwing out some modes, the sky map becomes biased because these rejected modes are no longer included in our sky maps.
However, we obtain smaller uncertainties on the modes that we choose to keep.
We experimented with different cutoffs and chose $N_\text{cutoff}=32$ based on our ability to produce reliable reconstructions with simulated data.
We discuss below how our results change for different values of $N_\text{cutoff}$ ; see Appendix \ref{app:sec:regularisation}.

\subsection{Clean map}
Using the regularized Fisher matrix we obtain a regularized clean map \citep{Thrane_2009}
\begin{equation}
    \Tilde{\mathcal{P}}'_\mu = \Tilde{M}^{-1}_{\mu\nu} X_\nu ,
    \label{eq:power_sph}
\end{equation}
with associated uncertainty
\begin{equation}
    \Tilde{\sigma}^{\mathcal{P}'}_\mu = \sqrt{\left( \mtx{\Tilde{M}}^{-1} \mtx{M} \mtx{\Tilde{M}}^{-1} \right)_{\mu\mu} } = \sqrt{ \mtx{\Tilde{M}}^{-1}_{\mu\mu}},\label{eq:var_power_sph}
\end{equation}
where the tildes denote regularized quantities.
The clean map signal-to-noise ratio is given by 
\begin{align}\label{eq:clean_snr}
    S/N = \frac{\tilde{{\cal P}}'_{\hat\Omega}}{\tilde\sigma^{\cal P'}_{\hat\Omega}} .
\end{align}
Here the subscript $\hat\Omega$ indicate that we have transformed the numerator and denominator from the spherical harmonic basis to the pixel basis. 

Furthermore, throughout the paper a division of non-scalar quantities (mostly in calculation of signal-to-noise ratios) denotes an element-wise division.

Well regularized, clean, signal-to-noise ratio maps exhibit fluctuations of $\approx\pm 3$ about a mean value, which is determined by the magnitude of the isotropic gravitational-wave signal.
Large positive excursions from the mean can be indications of anisotropy, whereas large negative excursions from the mean are typically evidence of a problem in the upstream noise analysis.
We use Equation~\eqref{eq:clean_snr} to derive the main results of this paper.

\subsection{Radiometer map}
\label{ssec:radiometer_map}

The clean map $\tilde{\cal P}$ represents our best guess for how gravitational-wave power is distributed on the sky with minimal assumptions.
However, we can instead assume that there is precisely one point source somewhere on the sky and ask the question: what is the inferred strain power that would be coming from this one point source as a function of sky location?
The answer to this question is the radiometer map \citep{Mitra_2008,Thrane_2009}.
While it does not produce a reliable estimate for broad-scale anisotropy, it is optimized for point source detection.
The radiometer map is given by
\begin{align}
    \eta_{\hat\Omega} =  X_{\hat\Omega} / M_{\hat\Omega\hat\Omega} ,
\end{align}
with associated uncertainty
\begin{align}\label{eq:radiometer_snr}
    \sigma^\eta_{\hat\Omega} = (M_{\hat\Omega\hat\Omega})^{-1/2} .
\end{align}
As above, the $\hat\Omega$ subscripts denote that we are working in the pixel basis.
We emphasize that the radiometer map does not require regularisation because we invert elements of the Fisher matrix rather than the entire Fisher matrix.
This is because the radiometer map does not take into account the covariance between different patches of sky.
The radiometer signal-to-noise ratio is
\begin{align}
    S/N = \frac{\eta_{\hat\Omega}}{\sigma_{\hat\Omega}^\eta}.
\end{align}

\subsection{Significance}\label{sec:pval}
The final step is to quantify the statistical significance of deviations from isotropy. 
There are numerous ways to estimate the significance of anisotropy in sky maps, each of which answers a different question.
We opt to answer the question: is the maximum value of $S/N$ on our sky maps consistent with what one would expect from an isotropic background?
This question reflects our prior belief that the first deviation from isotropy is likely to appear as a hotspot ---a patch of sky with an anomalously high signal-to-noise ratio.

We calculate a frequentist $p$-value.
First, following \cite{Abadie_2011}, we generate a distribution of dirty maps; details are provided in  Appendix~\ref{app:ssec:fisher}.
We then construct the clean map from each of these realisations following Equation~\eqref{eq:ml_estimator}.
In order to determine the significance of the fluctuations about an isotropic background, we subtract the monopole from these maps by setting the monopole component to zero (i.e., for the clean map $\mathcal{P}_{\ell=0,m=0}=0$, and for the radiometer $\eta_{\ell=0,m=0}=0$) before converting back to the pixel basis.
For each clean map realisation we determine the minimum and maximum $S/N$ across the whole map and combine these values into two histograms, in order to obtain the distributions of the global minimum and maximum $S/N$. 
We compare the measured maximum $S/N$ to the expected distribution in order to obtain a $p$-value. 
A small $p$-value $\ll 0=1\%$ can indicate statistically significant anisotropy.

\section{Results} \label{sec:results}
We create both regularized clean maps and radiometer maps for three frequency bins: \SI{7}{\nano\hertz}, \SI{14}{\nano\hertz}, \SI{21}{\nano\hertz} (all multiples of the $1/t_\text{obs}$).
We focus on narrow-band maps since supermassive black hole binaries evolve slowly compared to our observation time, it is unlikely that the same source appears in more than one frequency bin\footnote{Assuming a circular orbit. Gravitational wave emission is no longer monochromatic over the orbital period in the presence of eccentricity \citep{Peters_1963}.}.
Each map indicates the sky position of all MPTA pulsars with white stars.
The size of each star is scaled by the inverse of the root mean square of the pulsar's residuals so that bigger stars provide more information about the stochastic background than small stars.

\begin{figure}
\centering
    \subfigure[$\SI{7}{\nano\hertz}$]{\includegraphics[width=\linewidth]{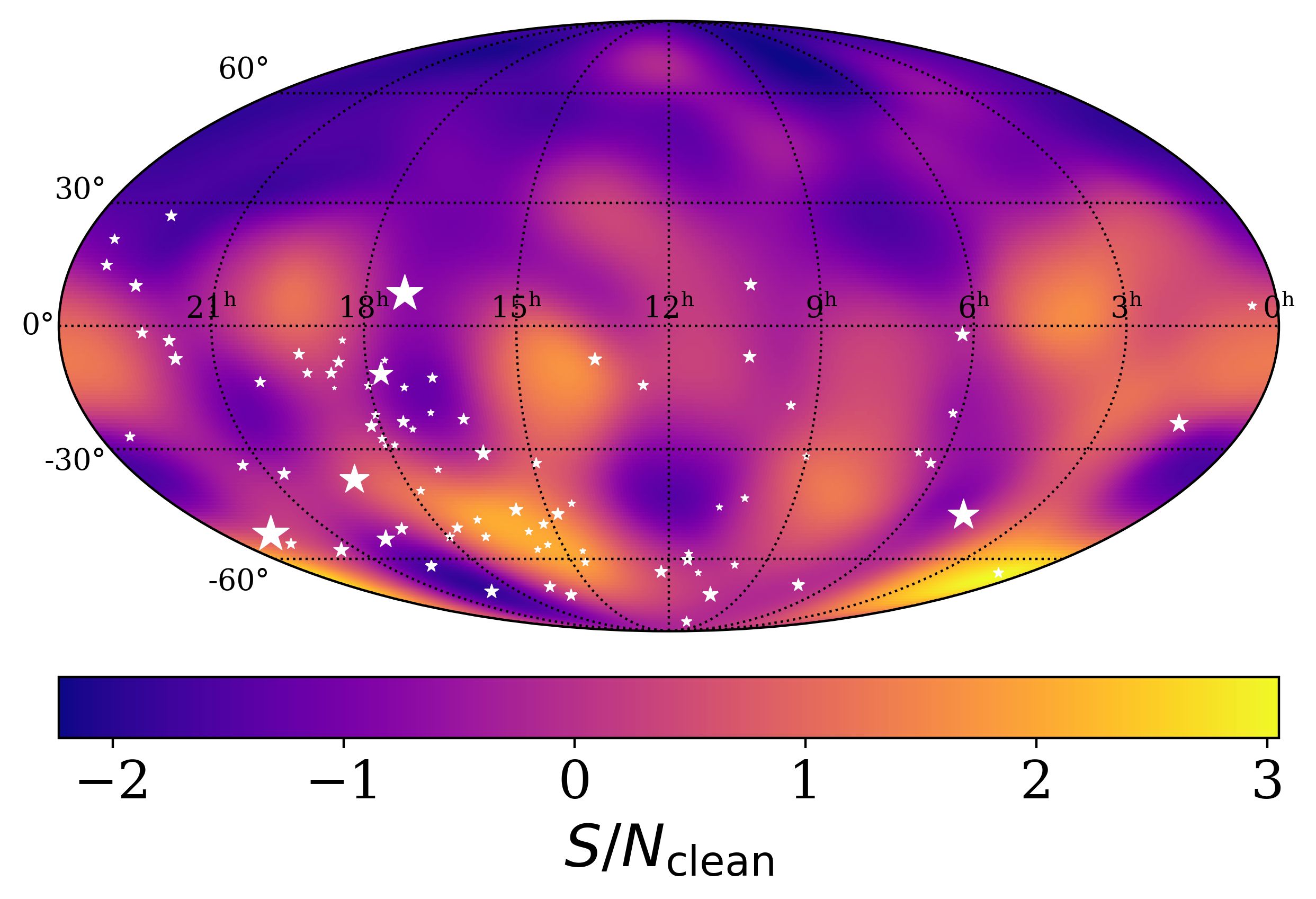}
    \label{fig:snr_data1}}
    \subfigure[$\SI{14}{\nano\hertz}$]{\includegraphics[width=\linewidth]{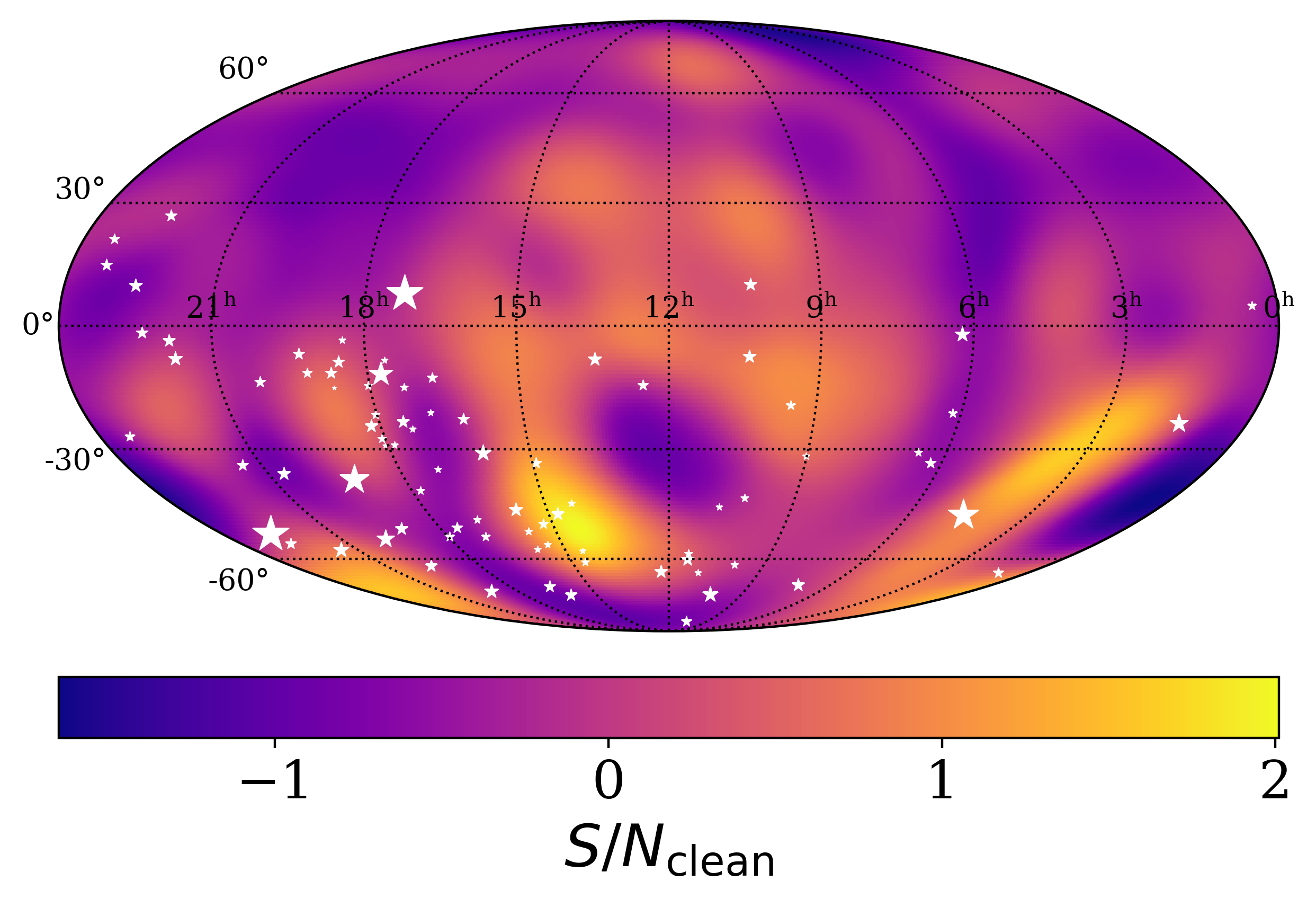}
    \label{fig:snr_data2}}
    \subfigure[$\SI{21}{\nano\hertz}$]{\includegraphics[width=\linewidth]{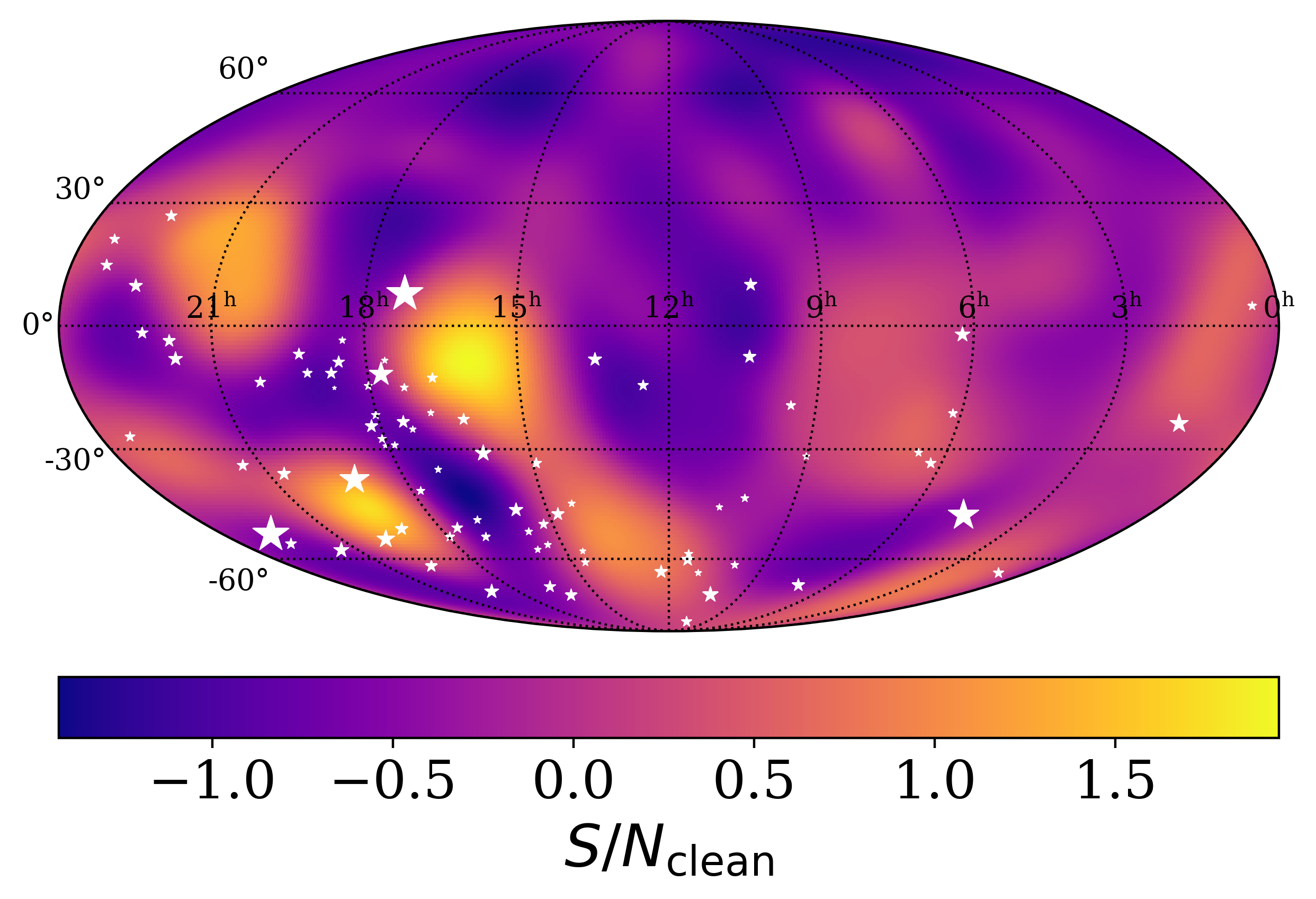}
    \label{fig:snr_data3}}
    \caption{Regularized clean signal-to-noise ratio maps of the gravitational wave power across the three frequency bins. The white stars are the sky locations of the pulsars in the MPTA, where the size corresponds to the inverse of the residuals' root-mean-square of each pulsar. 
    All three maps are consistent with an isotropic background signal at the $\lesssim 2.3\sigma$ level. There is a hotspot in the $\SI{7}{\nano\hertz}$ frequency bin with a modest statistical significance of $p=\cleanpval$ located at RA~1h DEC~\SI{-70}{\degree}.}
    \label{fig:full-snr}
\end{figure}

In Fig.~\ref{fig:full-snr}, we plot the regularized signal-to-noise ratio clean maps, which provide minimum-assumption reconstructions of the gravitational-wave sky.
The $p$-values for maximum and minimum $S/N$ are shown in the ``clean map''-column of Tab.~\ref{tab:p-values}.
The $p$-values for the minimum $S/N$ are between 0.15-0.96, which is encouraging because small $p$-values for minimum $S/N$ typically indicate a problem in the data analysis pipeline\footnote{As outlined in Section \ref{ssec:Overview}, negative power in the map is due to noise fluctuations and therefore should not be statistically significant.}.
While two of the maximum $S/N$ values are typical for isotropic backgrounds, the \SI{7}{\nano\hertz} maximum $S/N$ is somewhat unusual with $p=\cleanpval$.

\begin{figure}
\centering
    \subfigure[$\SI{7}{\nano\hertz}$]{\includegraphics[width=\linewidth]{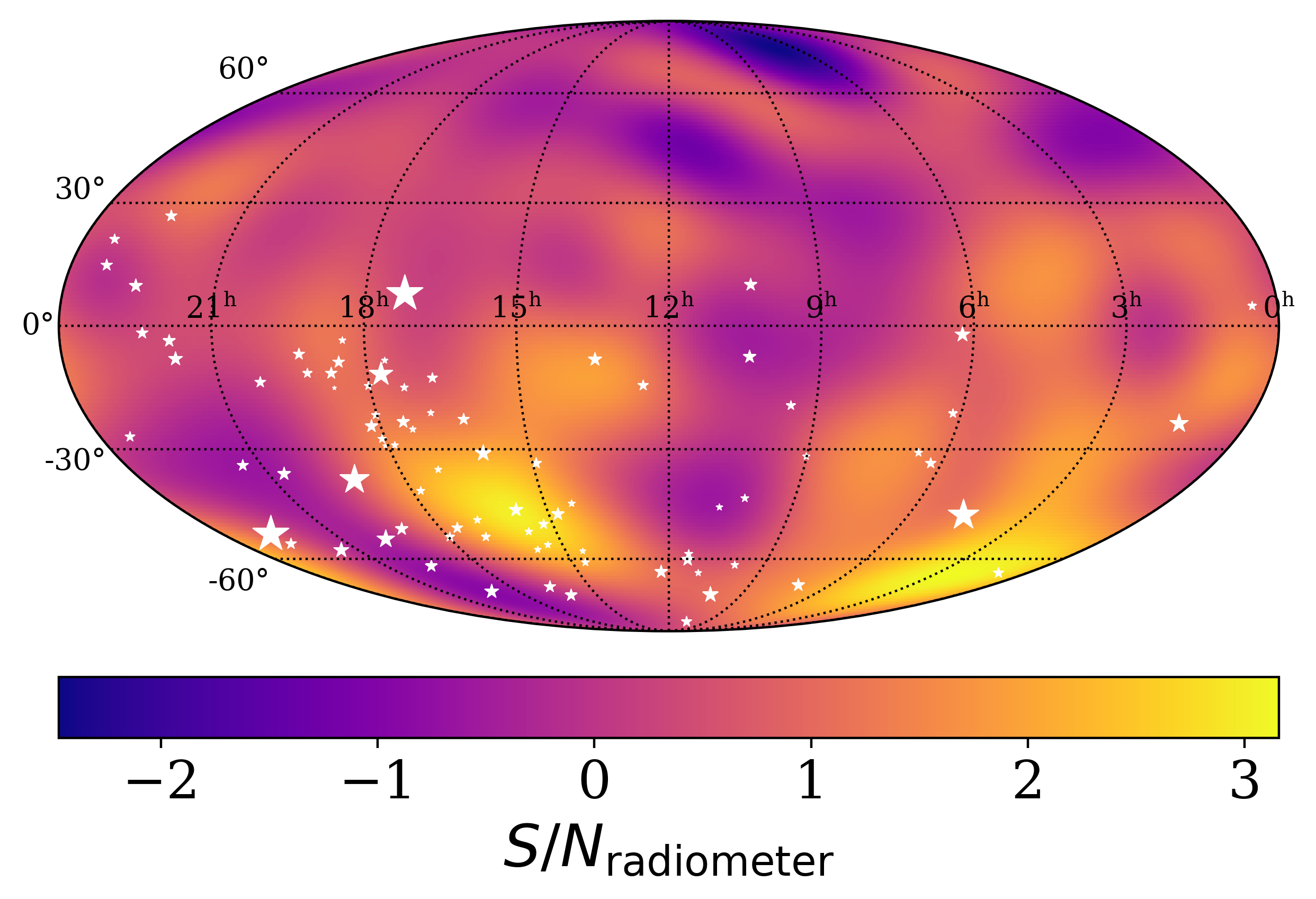}}
    \subfigure[$\SI{14}{\nano\hertz}$]{\includegraphics[width=\linewidth]{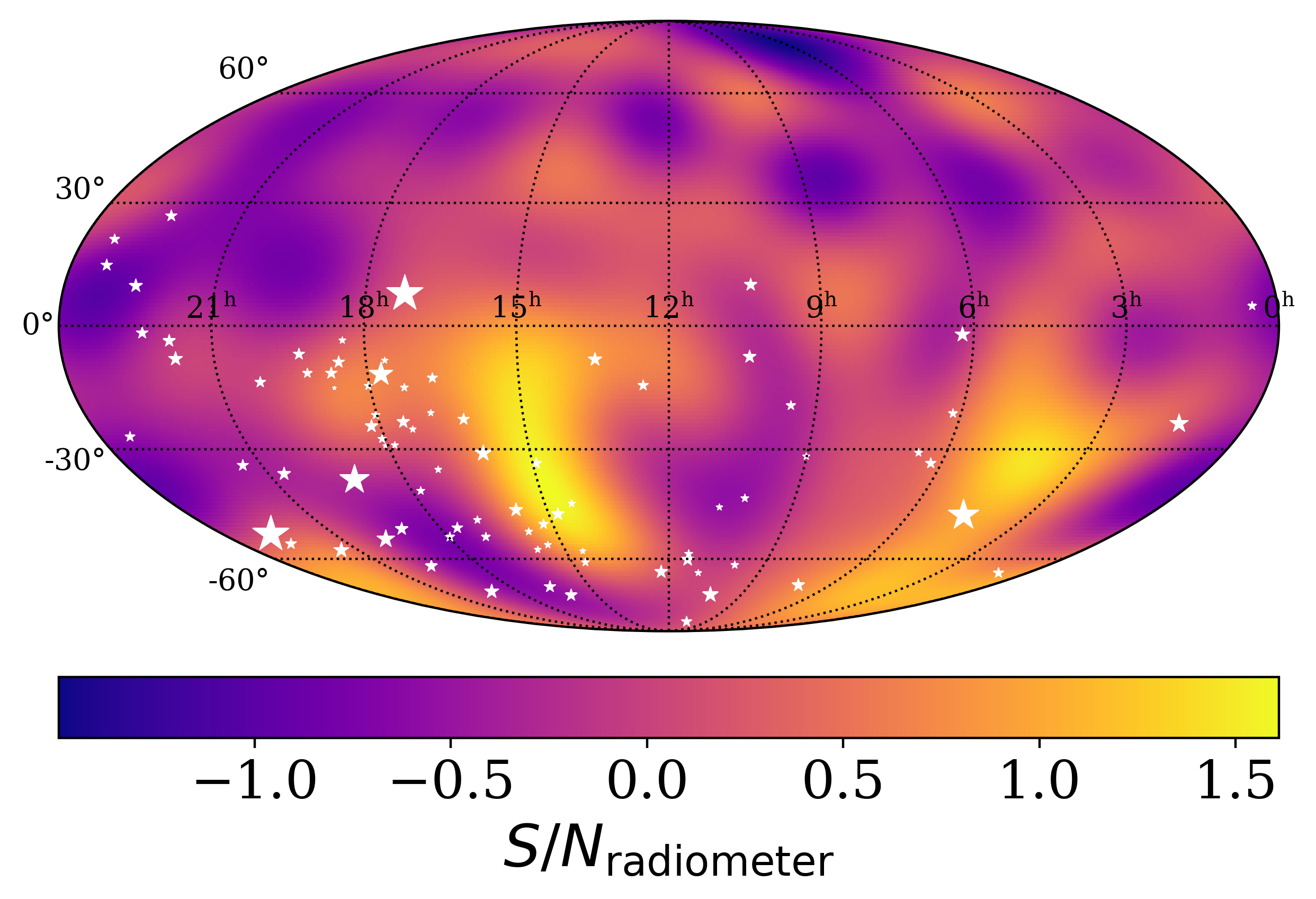}}
    \subfigure[$\SI{21}{\nano\hertz}$]{\includegraphics[width=\linewidth]{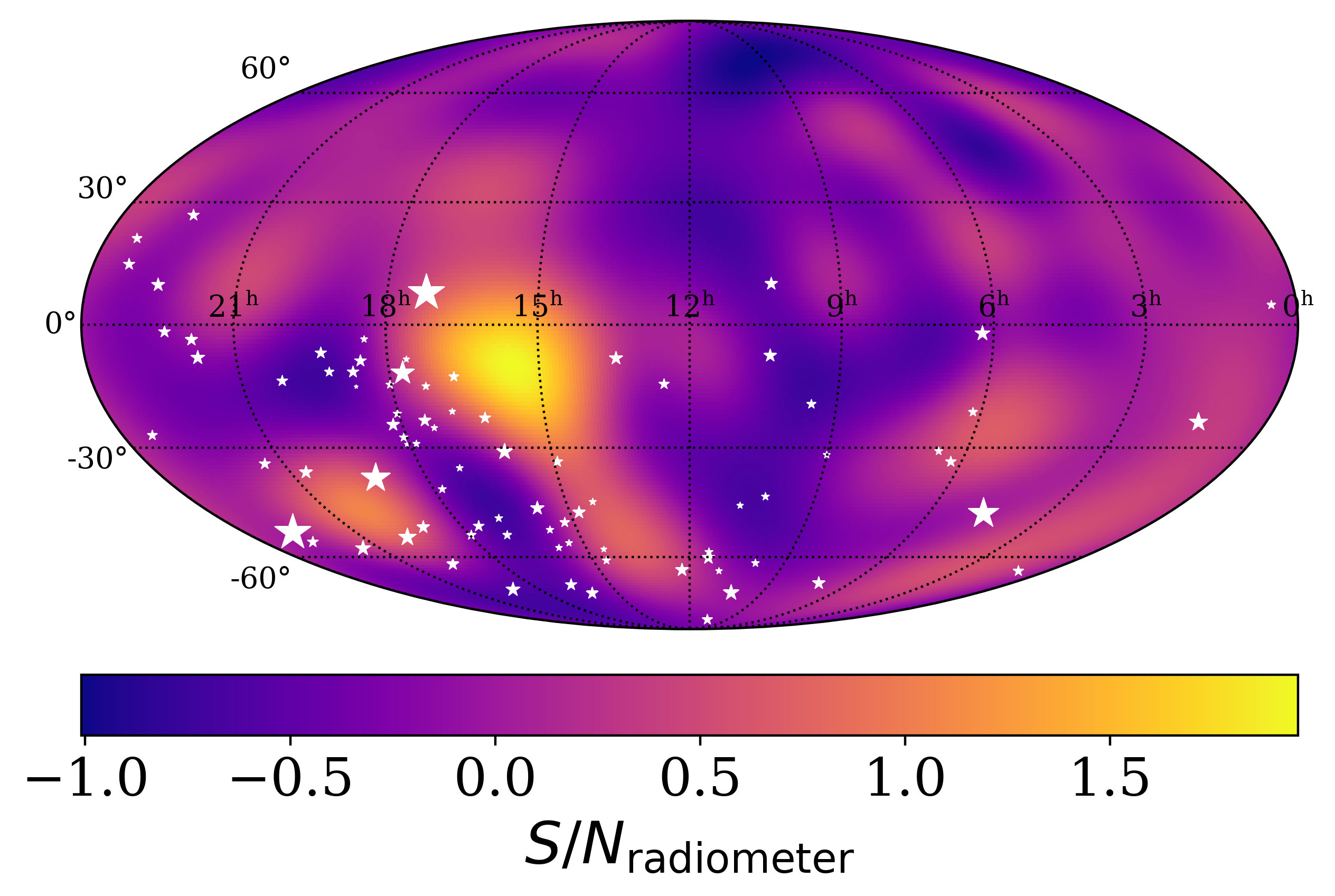}}
    \caption{Radiometer signal-to-noise ratio of the gravitational-wave power across the three frequency bins. The white stars are the sky locations of the pulsars in the MPTA, where the size corresponds to the inverse of the residuals' root-mean-square of each pulsar. There is a hotspot with a $p$-value of \radiometerpval  in the first map. The results are consistent with isotropy at the $\lesssim 2.3\sigma$ level.
    }
    \label{fig:full-radiometer}
\end{figure}

In Fig.~\ref{fig:full-radiometer}, we plot the radiometer maps, which are optimized for the detection of point sources.
Comparing the radiometer to the clean maps in Fig.~\ref{fig:full-snr}, we can see the impact of the cleaning. While there is a qualitative correspondence between hotspots on each map, the clean maps represent better estimates for the distribution of gravitational waves on the sky.
The $p$-values for the maximum and minimum $S/N$ are shown in the ``radiometer map''-column of  in Tab.~\ref{tab:p-values}.
As we saw with the clean maps, most of the $p$-values are ${\cal O}(1)$, suggesting that the data are consistent with an isotropic background.
However, the \SI{7}{\nano\hertz} maximum $S/N$ is again somewhat unusual with $p=\radiometerpval$.

\begin{table}
    \centering
    \caption{
    Quantifying \textit{radiometer map} and \textit{clean map} anisotropy.
    Frequentist $p$-values for the minimum and maximum $S/N$ in each frequency bins.}
    \begin{threeparttable}
    \begin{tabular}{c||c|c||c|c}
        & \multicolumn{2}{c||}{radiometer map} & \multicolumn{2}{c}{clean map} \\
        Frequency & $p(\mathrm{max})$\tnote{a} & $p(\mathrm{min})$\tnote{b} & $p(\mathrm{max})$\tnote{a} & $p(\mathrm{min})$\tnote{b} \\ 
        &&&& \\ [-1.5em]\midrule \midrule
        \SI{7}{\nano\hertz}  & 0.015 & 0.15 & 0.03 & 0.37\\
        \SI{14}{\nano\hertz} & 0.72 & 0.78 & 0.62 & 0.89 \\
        \SI{21}{\nano\hertz} & 0.42 & 0.94 & 0.67 & 0.96
    \end{tabular}
    \begin{tablenotes}\footnotesize
        \item[a] shorthand notation for $p(S/N_\mathrm{max}^\mathrm{means})$
        \item[b] shorthand notation for $p(S/N_\mathrm{min}^\mathrm{means})$
    \end{tablenotes}
    \end{threeparttable}
    \label{tab:p-values}
\end{table}

Further auxiliary data products characterising the information content of the MPTA data set include the cleaned gravitational wave power sky map and a map of the PTA's sensitivity towards gravitational waves from different areas on the sky. The respective maps are shown in Figures~\ref{fig:full_clean} and \ref{fig:full_sensitivity} in Appendix~\ref{app:sec:additional_results}. For the interested expert reader we also explain the mathematical details of the sensitivity estimation in subsections \ref{app:ssec:sensitivity_clean} and \ref{app:ssec:radiometer_sensitivity}

\section{Discussion} \label{sec:discussion}
The presence of a hotspot in our $\SI{7}{\nano\hertz}$ maps at approximately RA~1h DEC~\SI{-70}{\degree} is intriguing; however, the modest $p=\radiometerpval$ statistical significance suggests that it could be due to a noise fluctuation.
While the hotspot we have recovered may be due to noise, we show in Appendix~\ref{app:sec:simulation} that we are able to recover injected point sources accurately, which illustrates our ability to detect point sources should they exist in the data.

The statistical significance of the hotspot is strongly affected by the PSR~J2129$-$5721, less so by our most sensitive pulsars; see Appendix~\ref{app:ssec:dropout} for details. We note that this is the pulsar that lead to the creation of the ALT model in the accompanying isotropic gravitational wave background analysis \citep[see][]{MPTA2024_GWB}.

The hotspot is also, of course, affected by our choice of regularisation cutoff; see Appendix~\ref{app:sec:regularisation} for details.
Reducing the number of eigenmodes in our analysis, the hotspot is still visible (and a second hotspot emerges), but the significance of these hotspots is diminished compared to our fiducial analysis.
This is not unexpected: reducing $N_\text{SpH}$ throws out some of our signal, but provides a better measurement of the remaining modes.
Increasing the number of eigenmodes, the hotspot becomes indistinguishable from other fluctuations in the sky map.
We interpret this to mean that the map is dominated by uncertainty in the poorly resolved eigenmodes.
While we believe our choice of $N_\text{SpH}$ is reasonable, we recommend additional work to develop a more objective method for determining $N_\text{SpH}$, perhaps by optimising the ability to resolve specific anisotropic signals.

If the hotspot is subsequently shown to be of astrophysical in origin, it could indicate that the gravitational-wave background arises from supermassive black hole binaries.
Our current angular resolution of $\gtrsim 23^\circ$ makes it difficult to speculate as about potential host galaxies for a nearby supermassive black hole binary.
Additional pulsars must be added to the MPTA network in order to improve the angular resolution of our sky maps.
This would also serve to increase the overall sensitivity of network.
Since supermassive black hole binaries evolve slowly, the hotspot can be confirmed or ruled out with subsequent observations.

We have not attempted to account for cosmic variance, which means that our $p$-values are probably overestimating the presence of anisotropy.
Fluctuations in the gravitational-wave background from a statistically gravitational-wave background (one that is isotropic on average) yield individual realisations with anisotropy \citep{Allen_2023_VarianceHD,Romano_2024_FAQ}. 
Future work is required to take into cosmic variance in our $p$-value calculations.

Another topic for future work is the inclusion of gravitational-wave noise in our analysis, which becomes increasingly important as we depart from the low signal-to-noise ratio regime.
While our Fisher matrix includes contributions from gravitational-wave noise, we do not take this into account when maximizing the likelihood function.
As a result, our maximum-likelihood estimators are only approximate solutions.
Since our gravitational-wave signal is not so large in each frequency bin, we expect the exact solutions to be qualitatively similar to our approximate solution.
However, as the sensitivity of MPTA improves, it will become increasingly important to include a self-consistent treatment of gravitational-wave noise.

One possible solution is a recursive approach.
The initial maximum likelihood estimators for the spherical harmonics coefficients are used in the next iteration to calculate the overlap reduction function, leading to a new pulsar pair correlation covariance matrix and an updated maximum likelihood estimate for $\Tilde{\mathcal{P}}'$.
This method is currently under development.

\section*{Acknowledgements}
We thank Matthew Bailes for his ongoing valuable support as chair of the MeerTime collaboration. Furthermore, we appreciate the helpful comments on this work by Rutger van Haasteren, Huanchen Hu, Steve Taylor, Gosiya Curylo and Levi Schult.

The MeerKAT telescope is operated by the South African Radio Astronomy Observatory (SARAO), which is a facility of the National Research Foundation, an agency of the Department of Science and Innovation. SARAO acknowledges the ongoing advice and calibration of GPS systems by the National Metrology Institute of South Africa (NMISA) and the time space reference systems department of the Paris Observatory.
MeerTime observations used the PTUSE computing cluster. This cluster was funded in part by the Max-Planck-Institut für Radioastronomie (MPIfR) and the MaxPlanck-Gesellschaft.
MeerTime data is stored and processed on the OzStar and Ngarrgu Tindebeek supercomputers, operated by the Swinburne University of Technology. This project utilises the MeerTime data portal, which is supported by Nick Swainston and the ADACS team. We acknowledge and pay respects to the Elders and Traditional Owners of the land on which the Australian institutions stand, the Bunurong and Wurundjeri Peoples of the Kulin Nation.

We acknowledge support from the Australian Research Council (ARC) Centres of Excellence for Gravitational Wave Discovery (OzGrav) CE170100004 and CE230100016. KG, DC, FA, MK, VVK acknowledge continuing valuable support from the Max-Planck society. KG acknowledges support from the International Max Planck Research School (IMPRS) for Astronomy and Astrophysics at the Universities of Bonn and Cologne. 
RSN acknowledges support from the Astronomical Society of Australia.
RMS acknowledges support through ARC Future Fellowship FT190100155.
FA acknowledges that part of the research activities described in this paper were carried out with the contribution of the NextGenerationEU funds within the National Recovery and Resilience Plan (PNRR), Mission 4 - Education and Research, Component 2 - From Research to Business (M4C2), Investment Line 3.1 - Strengthening and creation of Research Infrastructures, Project IR0000034 – “STILES -Strengthening the Italian Leadership in ELT and SKA”.
Pulsar research at Jodrell Bank Centre for Astrophysics is supported by an STFC Consolidated Grant (ST/T000414/1; ST/X001229/1).
MKr acknowlegdes support by the CAS-MPG legacy Programme.
AP acknowledges financial support from the European Research Council (ERC) starting grant 'GIGA' (grant agreement number: 101116134) and through the NWO-I Veni fellowship.
JS acknowledges funding from the South African Research Chairs Initiative of the Depart of Science and Technology and the National Research Foundation of South Africa. 
VVK acknowledges financial support from the European Research Council (ERC) starting grant "COMPACT" (Grant agreement number 101078094). 
PG acknowledges support through SUT stipend SUPRA.
AP is funded/Co-funded by the European Union (ERC Starting Grant, GIGA, 101116134). Views and opinions expressed are however those of the author(s) only and do not necessarily reflect those of the European Union or the European Research Council. Neither the European Union nor the granting authority can be held responsible for them.

This publication made use of open source python libraries including \textsc{numpy} \citep{numpy} and \textsc{matplotlib} \citep{matplotlib}, as well as the python-based pulsar analysis packages, \textsc{enterprise} \citep{enterprise} and \textsc{enterprise\_extensions} \citep{enterprise_extensions}.

\section*{Data Availability}
Data used in this analysis will be available on the MeerTime data portal \url{https://pulsars.org.au/}.

\bibliographystyle{mnras}
\bibliography{refs}
\vspace{1em}
\noindent

\appendix
\onecolumn

\section{Additional Methodology Details: Beyond the weak signal approximation}

\subsection{Fisher matrix}\label{app:ssec:fisher}

\cite{Thrane_2009} showed that in the weak signal limit, $\mtx{M} = \mtx{R}^T\mtx{\Sigma}^{-1}\mtx{R}$ is the Fisher matrix of the maximum likelihood estimators $\vct{\mathcal{P}}'$. 
This is appropriate in the context of audio-band detectors like LIGO because the stochastic background is extremely small compared to the instrumental noise in any given measurement.
The audio-band stochastic background is only detectable by stacking a large number of measurements to dig beneath the noise \citep{Romano_2017}.

However, the situation is different in pulsar timing where the gravitational-wave signal can be comparable to the pulsar noise.
In this work, we include some corrections that arise when we relax the small-signal assumption.
We here show that $\cov{\vct{\mathcal{P}}'}{\vct{\mathcal{P}}'} = \left[\mtx{R}^T\mtx{\Sigma}^{-1}\mtx{R}\right] = \mtx{M}^{-1}$ still holds irrespective of the signal-to-noise ratio. 

We begin by simplifying the expression for the covariance as
\begin{equation}
    \label{eq:B1}
    \cov{\mathcal{P}'_\mu}{\mathcal{P}'_\nu} = M^{-1}_{\mu\kappa}M^{-1}_{\nu\lambda} \cov{X_\kappa}{X_\lambda} = M^{-1}_{\mu\kappa}M^{-1}_{\nu\lambda}  \left[\E{X_\kappa X_\lambda} - \E{X_\kappa}\E{X_\lambda} \right],
\end{equation}
where we used the definition Equation~\eqref{eq:ml_estimator} to substitute $\vct{\mathcal{P}}'$. In the next step we further break down the covariance of the dirty map $X$ by substituting Equation~\eqref{eq:dirty_map}. This yields for the expectation values in Equation~\eqref{eq:B1}: 
\begin{align}
    \E{X_\kappa} &= R^T_{\kappa\alpha}\Sigma^{-1}_{\alpha\beta}\E{\rho_\beta}, \\
    \E{X_\kappa X_\lambda^T} &= R^T_{\kappa\alpha}\Sigma^{-1}_{\alpha\beta}\; R_{\lambda\delta}\Sigma^{-1}_{\delta\epsilon}\E{ \rho_\beta \rho_\epsilon^T},  \nonumber\\
    &=  R^T_{\kappa\alpha}\Sigma^{-1}_{\alpha\beta}\; R^T_{\lambda\delta}\Sigma^{-1}_{\delta\epsilon} \left[\Sigma_{\beta\epsilon} + \E{\rho_\beta}\E{\rho_\epsilon^T}\right].
\end{align}
Putting everything together, the components of the covariance matrix of the dirty map read
\begin{align}
    \cov{X_\kappa}{X_\lambda} &= R^T_{\kappa\alpha}\Sigma^{-1}_{\alpha\beta} R_{\lambda\delta}\Sigma^{-1}_{\delta\epsilon} \left[ \Sigma_{\beta\epsilon} + \E{\rho_\beta}\E{\rho_\epsilon^T} - \E{\rho_\beta}\E{\rho_\epsilon^T} \right] \nonumber\\
    &= R^T_{\kappa\alpha}\Sigma^{-1}_{\alpha\beta} R_{\lambda\delta}\Sigma^{-1}_{\delta\epsilon}\Sigma_{\beta\epsilon} \nonumber\\
    &= R^T_{\kappa\alpha}\Sigma^{-1}_{\alpha\beta}R_{\delta\lambda} \delta_{\delta\beta} \nonumber\\
    &= R^T_{\kappa\alpha}\Sigma^{-1}_{\alpha\beta}R_{\beta\lambda} \nonumber\\
    &= M_{\kappa\lambda}.
\end{align}
Inserting back into Equation~\eqref{eq:B1}, we find
\begin{equation}
    \cov{\mathcal{P}'_\mu}{\mathcal{P}'_\nu} =  M^{-1}_{\mu\kappa}M^{-1}_{\nu\lambda} \cov{X_\kappa}{X_\lambda} = M^{-1}_{\mu\kappa}\underbrace{M^{-1}_{\nu\lambda} M_{\kappa\lambda}}_{=\delta_{\kappa\nu}} = M^{-1}_{\mu\nu}.
\end{equation}

Thus, the expression for $\mtx{M}$ as defined already in \cite{Thrane_2009,Pol_2022}; $\mtx{M} = \mtx{R}^T\mtx{\Sigma}^{-1}\mtx{R}$, holds as the Fisher matrix of the clean map, regardless of the signal strength. The correction with respect to the weak signal limit is completely absorbed in the correct calculation of the correlation covariance matrix $\mtx{\Sigma}$, i.e.\ the only change to the formalism lies in the calculation of $\mtx{\Sigma}$.

\subsection{Corrections to the Sigma matrix}\label{app:ssec:Sigma_cc}

When we relax the weak-signal approximation, the off-diagonal entries of the cross-correlation covariance matrix are not negligible anymore and need to be included into the analysis. A general mathematical expression for the full matrix has been derived in \cite{Allen_2023_HD71}. By distinguishing between different cases of pulsar pair constituents, this general expression can be simplified and written in terms of known quantities such as the auto- and cross-correlation matrices of the pulsars. This is also the key step towards any implementation of the full matrix. In the following, we therefore derive these simplified expressions in order to bridge the gap between the work done by \cite{Allen_2023_HD71} and the code provided for the analysis done in this work.\footnote{These equations were simultaneously and independently derived in \citep{Gersbach_2024}, and were published while this paper was in preparation.}

We aim to calculate the covariance of the cross-correlations, i.e.,
\begin{align}
    \Sigma_{\alpha\beta} = \cov{\rho_\alpha}{\rho_\beta} = \E{\rho_\alpha\rho_\beta}-\E{\rho_\alpha}\E{\rho_\beta}.
\end{align}
First, we adopt index notation, explicitly writing out the vector and matrix products in Equation~\eqref{eq:rho}. In order to maintain a human level of readability, all pulsar indices $a,b,c,d$ are upper indices, while the lower indices $i,j,k,l,m,n,p,q$ run over the number of ToAs in each pulsar. We furthermore adopt the Einstein summation convention for lower indices. 

In this notation we can rewrite the pulsar cross-correlation calculation from Equation~\eqref{eq:rho} as
\begin{equation}
    \rho_{ab} = \sigma_{ab}^2 \delta t^a_i (C^{-1})^a_{ij} \Hat{S}^{ab}_jk (C^{-1})^b_{kl} \delta t^b_l,
\end{equation}
where we used Equation~\eqref{eq:delta_rho} to replace $\left(\tr[(C_a^{-1}\Hat{S}_{ab}C_b^{-1}\Hat{S}_{ab})]\right)^{-1} = \sigma_{ab}^2$.

\begin{align}
    \cov{\rho_{ab}}{\rho_{cd}} &= \sigma_{ab}^2\sigma_{cd}^2 \left[\C^a_{ij} \Hat{S}_{jk}^{ab}\C^b_{kl}\right]\;\left[\C_{mn}^c \Hat{S}_{np}^{cd}\C^d_{pq}\right] \underbrace{\left\{\E{\delta t^a_i \delta t^b_l \delta t^c_m \delta t^d_q} - \E{\delta t^a_i \delta t^b_l} \E{\delta t^c_m \delta t^d_q} \right\}}_{\mathclap{\substack{\sim\; \E{ab}\E{cd} + \E{ac}\E{bd} + \E{ad}\E{bc}\,\,-\E{ab}\E{cd} \\ = \E{ac}\E{bd} + \E{ad}\E{bc}}}} \nonumber \\
    &= \sigma_{ab}^2\sigma_{cd}^2 \left[\C^a_{ij} \Hat{S}_{jk}^{ab}\C^b_{kl}\right]\;\left[\C_{mn}^c \Hat{S}_{np}^{cd}\C^d_{pq}\right] \underbrace{\left\{ \E{\delta t^a_i \delta t^c_m} \E{\delta t^b_l \delta t^d_q} + \E{\delta t^a_i \delta t^d_q} \E{\delta t^b_l \delta t^c_m}\right\}}_{\equiv\; \mathfrak{T}}.
\end{align}
Here, we use Isserlis' theorem \citep{Isserlis_1918,Wick_1950} in order to express the four-point correlation as a sum of products of two-point correlations, i.e.,$\E{abcd}=\E{ab}\E{cd} + \E{ac}\E{bd} + \E{ad}\E{bc}$. Implicitly we introduce the shorthand notation $\E{ab} = \E{\delta t^a_i \delta t^b_l}$, where
\begin{equation}
    \E{aa}_{ij} \equiv \E{\delta t^a_i \delta t^a_j} = C^{a}_{ij} \qquad \qquad\E{ab}_{ij} \equiv \E{\delta t^a_i \delta t^b_j} = S^{ab}_{ij} = A^2_\GWB \Gamma^{ab}\Hat{S}^{ab}_{ij}.
\end{equation}
Depending on the pulsars making up each pair, the individual correlations in $\mathfrak{T}$ contribute either as the auto-covariance matrix $\mtx{C}_{a}$ or the cross-covariance matrix $\mtx{S}_{ab}$. In the decomposition of the cross-covariance matrix, $\Gamma^{ab}$ is the overlap reduction function derived from the pulsar pair correlations, which does not necessarily have to be the Hellings-Downs correlation function, depending on the angular GW power distribution. 

As pointed out above, all entries of the cross-correlation covariance matrix $\mtx{\Sigma}$ can be classified into three different cases derivation-wise (both pulsars in the pairs are the same, only one matches, all are different), that turn into five cases upon numerical implementation (the one-match cases are related to each other by symmetry of $\Gamma_{ab}$, $\Hat{S}_{ab}$ and $C_a$):
\begin{enumerate}
    \item two-match case: \hspace{0.5cm}$a=c$, $b=d$: $\quad\mathfrak{T}\longrightarrow\E{aa}_{im}\E{bb}_{lq} + \E{ab}_{iq}\E{ba}_{lm} = C^a_{im}C^b_{lq} + A_\GWB^4{\Gamma^{ab}}^2\Hat{S}^{ab}_{iq} \Hat{S}^{ab}_{lm}$
        \begin{align}
            \qquad\qquad\qquad\cov{\rho_{ab}}{\rho_{ab}} &= \sigma_{ab}^2\sigma_{ab}^2 \bigg\{ \underbrace{C^a_{im}\C^a_{ij}}_{= \delta_{mj}}\Hat{S}_{jk}^{ab}\C^b_{kl}\C_{mn}^a \Hat{S}_{np}^{ab}\underbrace{\C^b_{pq}C^b_{lq}}_{=\delta_{pl}} \nonumber \\
            & \qquad \qquad \qquad + A_\GWB^4{\Gamma^{ab}}^2\, \Hat{S}^{ab}_{iq}\C^a_{ij}\Hat{S}_{jk}^{ab}\C^b_{kl}\Hat{S}^{ab}_{lm}\C_{mn}^a \Hat{S}_{np}^{ab}\C^b_{pq} \bigg\} \nonumber\\
            &= \sigma_{ab}^2\sigma_{ab}^2
                \left\{\trc{C^{-1}_a\Hat{S}_{ab}C^{-1}_b\Hat{S}_{ba}} \right. \nonumber \\
                & \qquad \qquad \qquad  \left. + A_\GWB^4\Gamma_{ab}^2 \,\trc{C^{-1}_a\Hat{S}_{ab}C^{-1}_b\Hat{S}_{ba}C^{-1}_a\Hat{S}_{ab}C^{-1}_b\Hat{S}_{ba}} \right\},
                \label{eq:case1}
        \end{align}

    \item one-match case 
        \begin{enumerate}
            \item[\hspace{2.2cm}(a)] $a=c$, $b\neq d$: $\quad\mathfrak{T}\longrightarrow\E{aa}_{im}\E{bd}_{lq} + \E{ad}_{iq}\E{ba}_{lm} = A^2_\GWB\Gamma^{bd}C^a_{im}\Hat{S}^{bd}_{lq} + A_\GWB^4\Gamma^{ab}\Gamma^{ad}\Hat{S}^{ad}_{iq} \Hat{S}^{ba}_{lm}$
            \begin{align}
            \qquad\qquad\qquad \cov{\rho_{ab}}{\rho_{ad}} &= \sigma_{ab}^2\sigma_{ad}^2 \left\{A^2_\GWB\Gamma^{bd}\; C^a_{im}\C^a_{ij} \Hat{S}_{jk}^{ab}\C^b_{kl} \Hat{S}^{bd}_{lq} \C_{mn}^a \Hat{S}_{np}^{ad}\C^d_{pq} \right. \nonumber \\
            & \qquad \qquad \qquad \left. + A_\GWB^4\Gamma^{ab}\Gamma^{ad}\;\Hat{S}^{ad}_{iq} \C^a_{ij} \Hat{S}_{jk}^{ab}\C^b_{kl} \Hat{S}^{ba}_{lm} \C_{mn}^a \Hat{S}_{np}^{ad}\C^d_{pq}  \right\} \nonumber \\
            &=\sigma_{ab}^2\sigma_{ad}^2 \left\{A^2_\GWB\Gamma_{bd} \trc{C^{-1}_a\Hat{S}_{ab}C^{-1}_b\Hat{S}_{bd}C^{-1}_d\Hat{S}_{da}} \right. \nonumber \\
            & \qquad \qquad \qquad \left. +  A_\GWB^4\Gamma^{ab}\Gamma^{ad} \;\trc{C^{-1}_a \Hat{S}_{ab}C^{-1}_b \Hat{S}_{ba} C^{-1}_a \Hat{S}_{ad}C^{-1}_d\Hat{S}_{da}} \right\},
            \label{eq:case2}
        \end{align}
        \item[\hspace{2.2cm}(b)] $a\neq c$, $b = d$: $\quad\mathfrak{T}\longrightarrow\E{ac}_{im}\E{bb}_{lq} + \E{ab}_{iq}\E{bc}_{lm} =  A_\GWB^2\Gamma^{ac}\Hat{S}^{ac}_{im}C^b_{lq} + A_\GWB^4\Gamma^{ab}\Gamma^{bc}\Hat{S}^{ab}_{iq} \Hat{S}^{bc}_{lm}$
        \begin{align}
            \qquad\qquad\qquad\cov{\rho_{ab}}{\rho_{cb}} &= \sigma_{ab}^2\sigma_{cb}^2 \left\{ A_\GWB^2\Gamma^{ac} \Hat{S}^{ac}_{im} \C^a_{ij} \Hat{S}_{jk}^{ab}\C^b_{kl} \C_{mn}^c \Hat{S}_{np}^{cb}\C^b_{pq}C^b_{lq} \right. \nonumber \\
            & \qquad \qquad \qquad \left. + A_\GWB^4\Gamma^{ab}\Gamma^{bc} \Hat{S}^{ab}_{iq} \C^a_{ij} \Hat{S}_{jk}^{ab}\C^b_{kl} \Hat{S}^{bc}_{lm} \C_{mn}^c \Hat{S}_{np}^{cb}\C^b_{pq} \right\} \nonumber\\
            &= \sigma_{ab}^2\sigma_{cb}^2 \left\{ A_\GWB^2\Gamma^{ac} \trc{ C^{-1}_a \Hat{S}_{ab}C^{-1}_{b} \Hat{S}_{bc} C^{-1}_c \Hat{S}_{ca}} \right. \nonumber \\
            & \qquad \qquad \qquad \left. +  A_\GWB^4\Gamma^{ab}\Gamma^{bc}  \C^{-1}_a\Hat{S}_{ab} C^{-1}_b\Hat{S}_{bc}  C^{-1}_c\Hat{S}_{cb} C^{-1}_b\Hat{S}_{ba} \right\},
            \label{eq:case3}
        \end{align}

        \item[\hspace{2.2cm}(c)] $a = d$, $b\neq c$: $\quad\mathfrak{T}\longrightarrow\E{ac}_{im}\E{ba}_{lq} + \E{aa}_{iq}\E{bc}_{lm} = A_\GWB^4\Gamma^{ac}\Gamma^{ab}\Hat{S}^{ac}_{im}\Hat{S}^{ba}_{lq} + A_\GWB^2\Gamma^{bc} C^a_{iq} \Hat{S}^{bc}_{lm}$
        \begin{align}
            \qquad\qquad\qquad\cov{\rho_{ab}}{\rho_{ca}} &= \sigma_{ab}^2\sigma_{ca}^2 \left\{ A_\GWB^4\Gamma^{ac}\Gamma^{ab} \C^a_{ij} \Hat{S}_{jk}^{ab}\C^b_{kl} \Hat{S}^{ba}_{lq} \Hat{S}^{ac}_{im} \C_{mn}^c \Hat{S}_{np}^{ca}\C^a_{pq} \right. \nonumber \\
            & \qquad \qquad \qquad \left. + A_\GWB^2\Gamma^{bc}  \C^a_{ij} \Hat{S}_{jk}^{ab}\C^b_{kl} \Hat{S}^{bc}_{lm} \C_{mn}^c \Hat{S}_{np}^{ca}\C^a_{pq} C^a_{iq}\right\} \nonumber\\
            &= \sigma_{ab}^2\sigma_{ca}^2 \left\{ A_\GWB^4\Gamma^{ac}\Gamma^{ab} \trc{C^{-1}_a \Hat{S}_{ab} C^{-1}_b \Hat{S}_{ba} C^{-1}_a \Hat{S}_{ac} C^{-1}_c \Hat{S}_{ca}} \right. \nonumber \\
            & \qquad \qquad \qquad \left. +  A_\GWB^2\Gamma^{bc}  \trc{C^{-1}_a \Hat{S}_{ab}\C^{-1}_b \Hat{S}_{bc} C^{-1}_c \Hat{S}_{ca}} \right\},
            \label{eq:case4}
        \end{align}
        \end{enumerate}

    \item no-match case: \hspace{0.5cm}$a \neq b \neq c \neq d$: $\quad\mathfrak{T}\longrightarrow\E{ac}_{im}\E{bd}_{lq} + \E{ad}_{iq}\E{bc}_{lm} = A_\GWB^4\Gamma_{ac}\Gamma_{bd} \Hat{S}^{ac}_{im}\Hat{S}^{bd}_{lq} + A_\GWB^4\Gamma^{ad}\Gamma^{bc} \Hat{S}^{ad}_{iq} \Hat{S}^{bc}_{lm}$
        \begin{align}
            \qquad\qquad\qquad\qquad\cov{\rho_{ab}}{\rho_{cd}} &= \sigma_{ab}^2\sigma_{cd}^2 \left\{ A_\GWB^4\Gamma_{ac}\Gamma_{bd} \C^a_{ij} \Hat{S}_{jk}^{ab}\C^b_{kl} \Hat{S}^{ac}_{im} \C_{mn}^c \Hat{S}_{np}^{cd}\C^d_{pq}\Hat{S}^{bd}_{lq} \right. \nonumber \\
            & \qquad \qquad \qquad \left. + A_\GWB^4\Gamma^{ad}\Gamma^{bc} \C^a_{ij} \Hat{S}_{jk}^{ab}\C^b_{kl} \Hat{S}^{bc}_{lm} \C_{mn}^c \Hat{S}_{np}^{cd}\C^d_{pq} \Hat{S}^{ad}_{iq} \right\} \nonumber\\
            &= \sigma_{ab}^2\sigma_{cd}^2 \left\{ A_\GWB^4\Gamma_{ac}\Gamma_{bd} \trc{ C^{-1}_a\Hat{S}_{ab} \C^{-1}_b\Hat{S}_{bd} C^{-1}_d\Hat{S}_{dc} C^{-1}_c\Hat{S}_{ca}} \right. \nonumber \\
            & \qquad \qquad \qquad \left. + A_\GWB^4\Gamma^{ad}\Gamma^{bc} \trc{ C^{-1}_a\Hat{S}_{ab} C^{-1}_b\Hat{S}_{bc} C^{-1}_c\Hat{S}_{cd} C^{-1}_d\Hat{S}_{da} } \right\}.
            \label{eq:case5}
        \end{align}
\end{enumerate}

Comparing the terms contributing to the main diagonal entries (Equation~\eqref{eq:case1}) to the expression for the cross-correlation uncertainties $\sigma_{\alpha\beta}$ as given in Equation~\eqref{eq:delta_rho} we find that
\begin{equation}
    \cov{\rho_{ab}}{\rho_{ab}} = \sigma_{ab}^2 + A_\GWB^4\Gamma_{ab}^2 \,\trc{C^{-1}_a\Hat{S}_{ab}C^{-1}_b\Hat{S}_{ba}C^{-1}_a\Hat{S}_{ab}C^{-1}_b\Hat{S}_{ba}}.
\end{equation}
Following from that, $\varsigma$ in Equation~\eqref{eq:Sigma_effective} is given by
\begin{equation}
    \mtx{\varsigma} = \begin{cases}
         A_\GWB^4\Gamma_{ab}^2 \,\trc{C^{-1}_a\Hat{S}_{ab}C^{-1}_b\Hat{S}_{ba}C^{-1}_a\Hat{S}_{ab}C^{-1}_b\Hat{S}_{ba}} &\text{if } a=c, b=d\\
         \mathrm{Equations~\eqref{eq:case2}-\eqref{eq:case5}} & \text{otherwise.}
    \end{cases}
    \label{eq:varsigma_ppcc}
\end{equation}

\section{Demonstration with simulated data}\label{app:sec:simulation}
In order to test the behaviour of our analysis pipeline and its capability to identify and \textit{localise} individual point sources, we create simulated datasets with similar $S/N$ regime to what we observe in the actual data analysis.

Each simulated dataset has the same observation times, pulsar positions and pulsar timing model as the real dataset. 
For each MPTA pulsar we generated simulated ToAs, which include contributions from both white noise and a gravitational-wave signal. Therefore we create a \texttt{.tim} file containing idealised telescope site arrival times from the original \texttt{.par} and \texttt{.tim} files using the \texttt{tempo2} \texttt{formIdeal} plugin. Then we initialise a \texttt{libstempo} \texttt{tempo2pulsar}-object with the original \texttt{.par}-file and simulated \texttt{.tim}-file, add white noise fluctuations to the \texttt{EFAC} and \texttt{EQUAD} parameters, as well as different gravitational-wave signals to the flat residuals using the \texttt{libstempo} toolbox. 
The resulting dataset is analysed with our standard \texttt{enterprise}/\texttt{matlab} pipeline as laid out in Section~\ref{sec:methods}, assuming only white noise and a power-law gravitational-wave signal with $\log_{10}A_\GWB=-14.0$ and $\gamma_\GWB=4.31$ to be present in the data.

We investigate the presence of an isotropic background signal as well as the presence of point sources using the following different datasets:
\begin{enumerate}
    \item \textit{White noise + isotropic gravitational-wave background.}
            We inject a common red noise signal with a power-law power spectral density characterised by the values found in \cite{MPTA2024_GWB}.\footnote{\hsize=\textwidth We set the recovery $\log_{10}A_\GWB$ to different values until the $S/N$ range of the clean map matches the $S/N$ range of the clean map of the real dataset. We find that recovering the isotropic gravitational-wave background signal with a common red noise specified by $\log_{10}A_\GWB= -14.0$ produced a reasonable $S/N$ range as shown in Fig.~\ref{fig:sv_comparison_GWB}. Hence we adopted this value for the analysis across all simulations.} 
            
    \item \textit{White noise + point sources.}
        For all point source (continuous wave) injections, we adopt the same recovery amplitude and spectral index as used in the white noise + isotropic gravitational-wave background model. We position all sources at a distance of \SI{1}{\mega\parsec} and assume them to be monochromatic with $f_\gw = \SI{1e-8}{\hertz}$. The recovered $S/N$ range is adjusted via changing the chirp mass of the simulated source(s). Throughout the analysis we investigate the following scenarios:
        \begin{enumerate}
            \item single point source with $\mathcal{M}_\mathrm{c}=\SI{1e8}{\msun}$ at a position of RA~18h DEC~\SI{-45}{\degree},
            \item single point source with $\mathcal{M}_\mathrm{c}=\SI{1e8}{\msun}$ at a position of RA~6h DEC~\SI{45}{\degree},
            \item single point source with $\mathcal{M}_\mathrm{c}=\SI{1e8}{\msun}$ at a position of RA~6h DEC~\SI{-45}{\degree},
            \item two point sources $\mathcal{M}_\mathrm{c}=10^{7.9}\si{\msun}$ at positions RA~6h DEC~\SI{-45}{\degree} and RA~18h DEC~\SI{-45}{\degree}.
        \end{enumerate}
\end{enumerate}
Although these simulations are consistent with the real MPTA dataset, we emphasize that we did not include any of the various chromatic noise processes present in the observed data. Thus, we can only qualitatively compare the simulation with the actual results presented in Section~\ref{sec:results}.

\subsection{Isotropic background}
\begin{figure*}
    \centering
    \subfigure[$\SI{7}{\nano\hertz}$]{\includegraphics[width=0.32\linewidth]{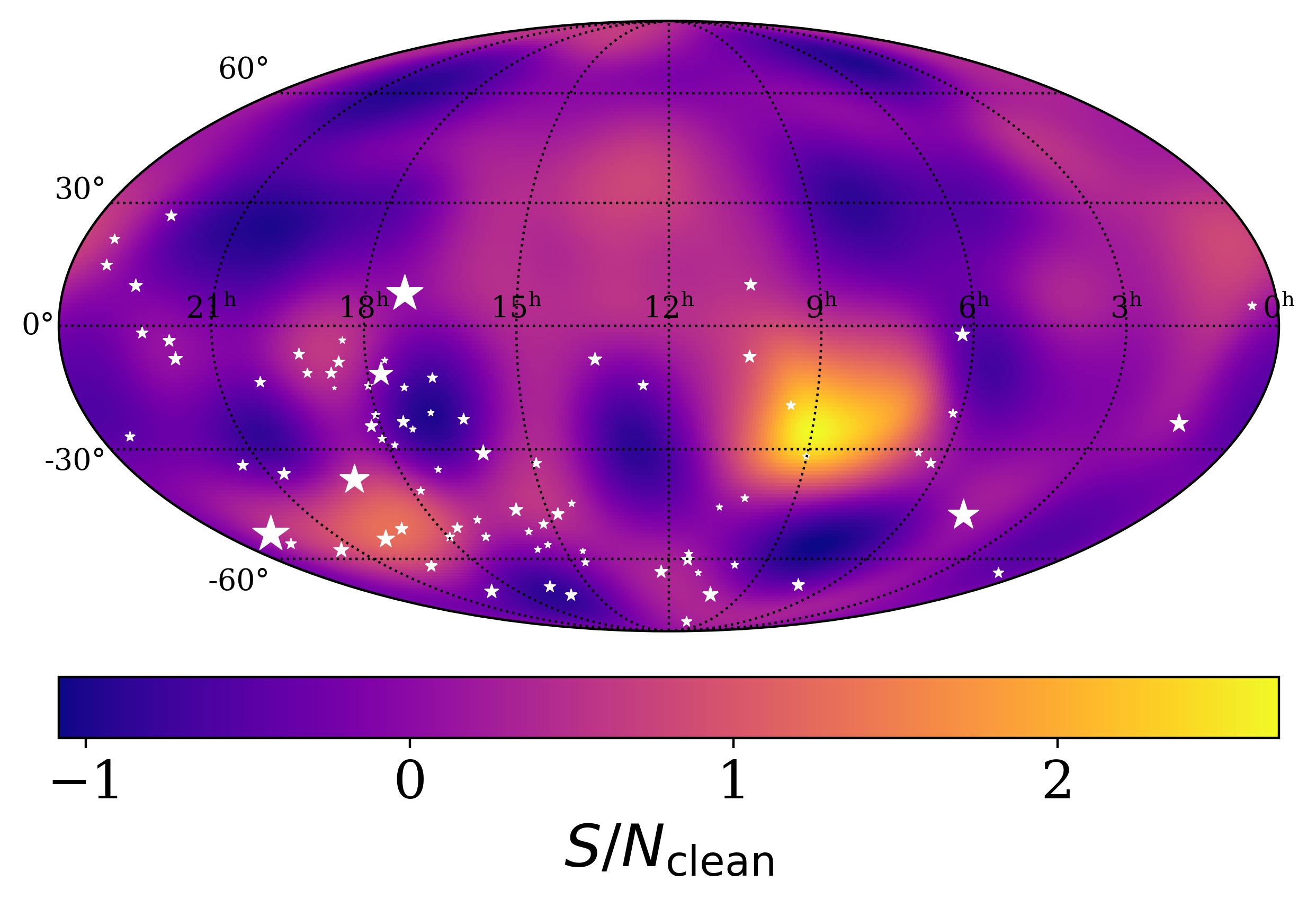}
    \label{fig:simGWB_snr1}}
    \hfill
    \subfigure[$\SI{14}{\nano\hertz}$]{\includegraphics[width=0.32\linewidth]{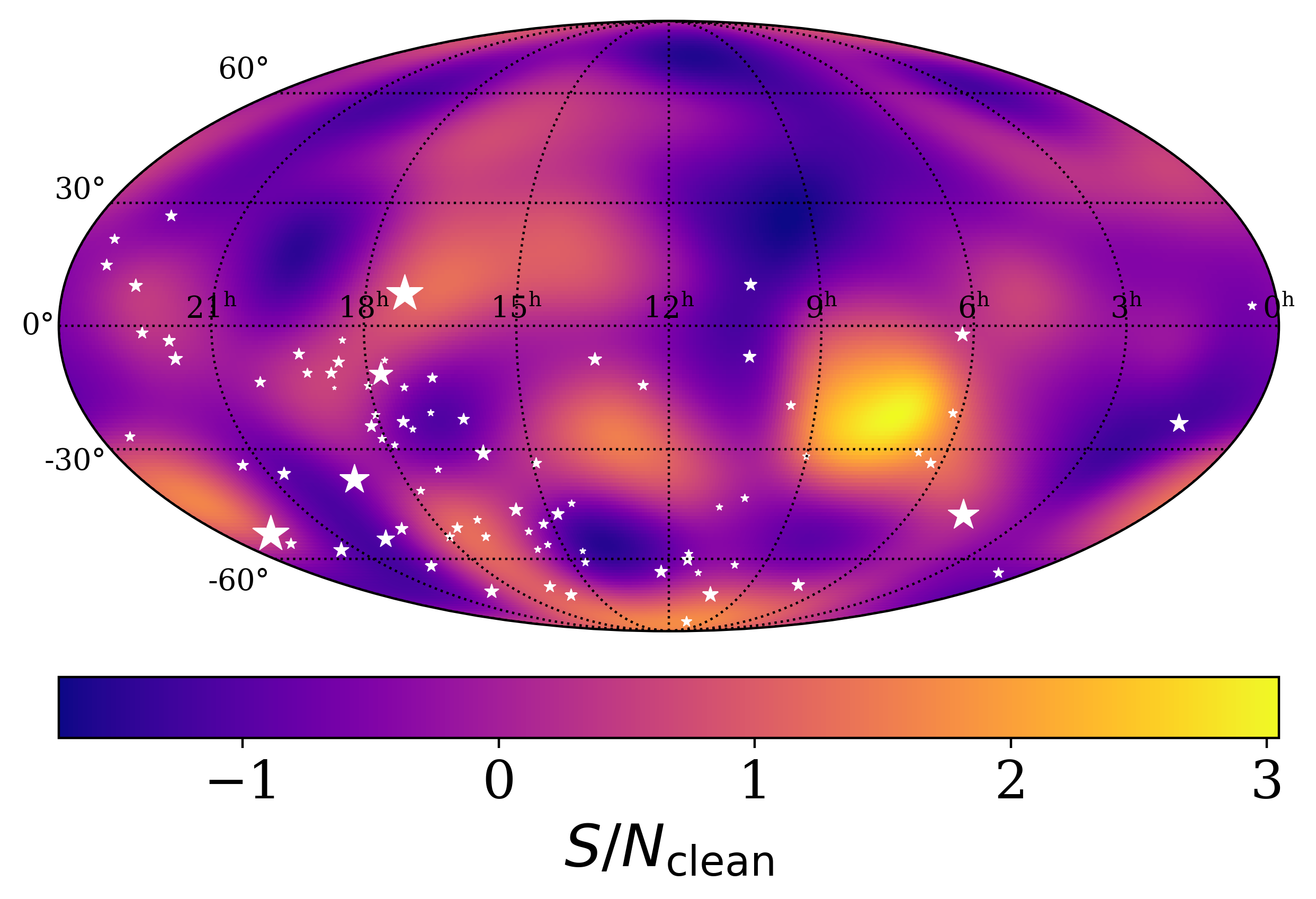}
    \label{fig:simGWB_snr2}}
    \hfill
    \subfigure[$\SI{21}{\nano\hertz}$]{\includegraphics[width=0.32\linewidth]{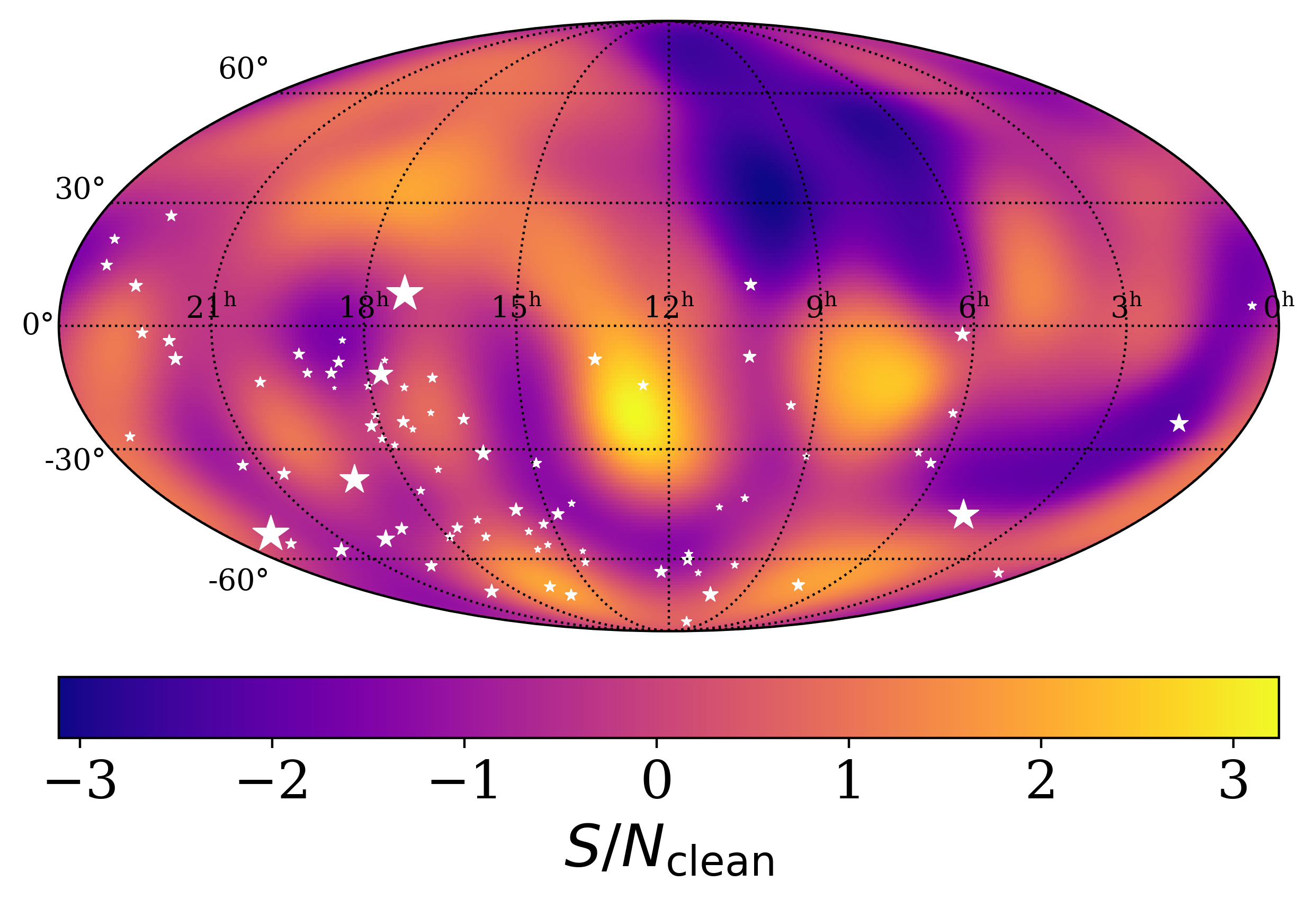}
    \label{fig:simGWB_snr3}}
    \hfill
    \caption{Regularized clean signal-to-noise ratio maps of the simulated dataset containing white noise and an isotropic gravitational wave background signal
    The white stars are the sky locations of the pulsars in the MPTA, where the size of each star is inversely proportional to the mean RMS of the best-fit timing solution.}
    \label{fig:simGWB_snr}
\end{figure*}

Fig.~\ref{fig:simGWB_snr} shows the clean map $S/N$ of the first three frequency bins for the simulated dataset containing the isotropic background signal. Similar to the results obtained with the full MPTA 4.5 year dataset, these maps exhibit an overall positive map mean with  fluctuations around that mean. 

\subsection{Point source(s)}
\begin{figure*}
    \centering
    \subfigure[single injected point source at RA~18h DEC~\SI{-45}{\degree}]{\includegraphics[width=0.48\linewidth]{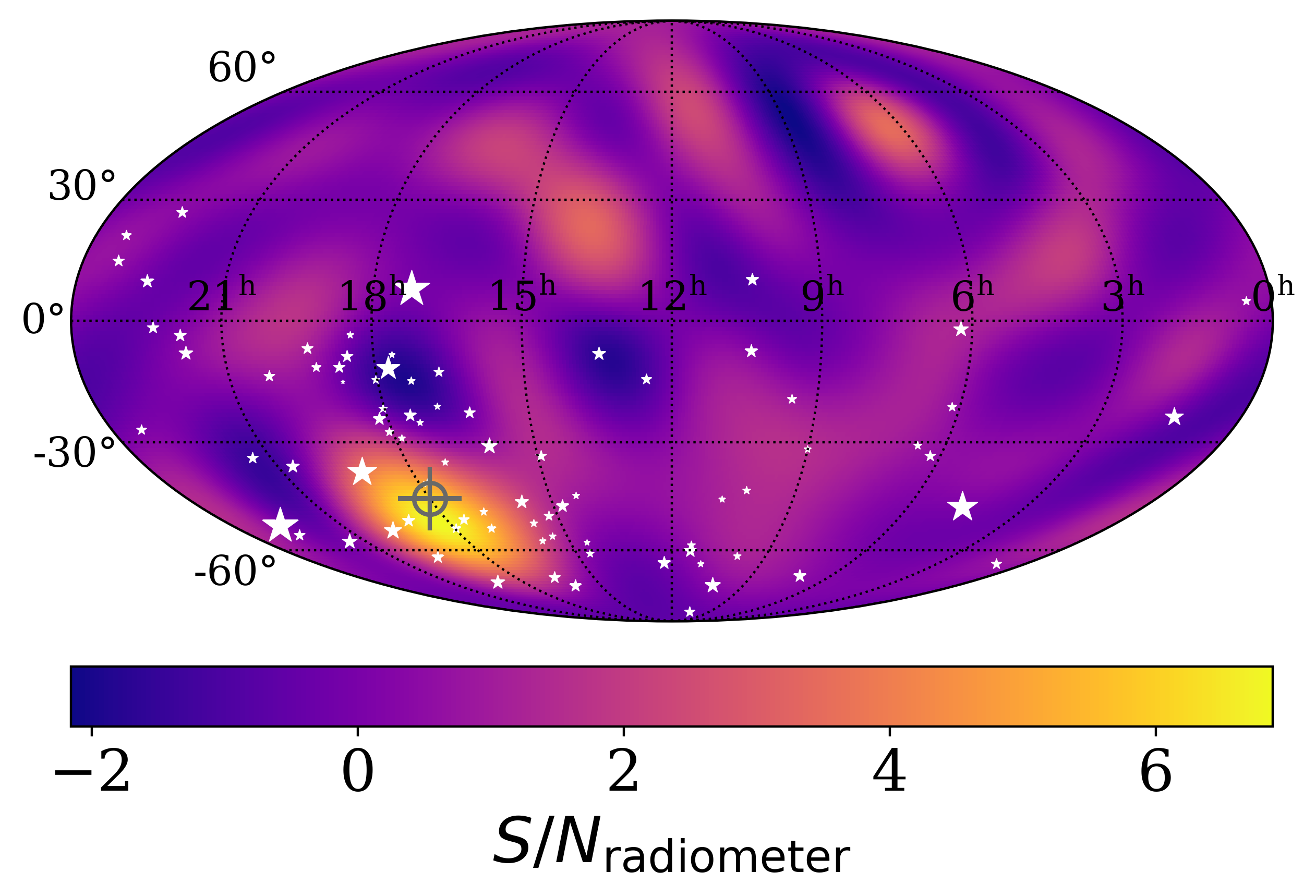}
    \label{fig:simCGW_radiometer1}}
    \hfill
    \subfigure[single injected point source at RA~06h DEC~\SI{-45}{\degree}]{\includegraphics[width=0.48\linewidth]{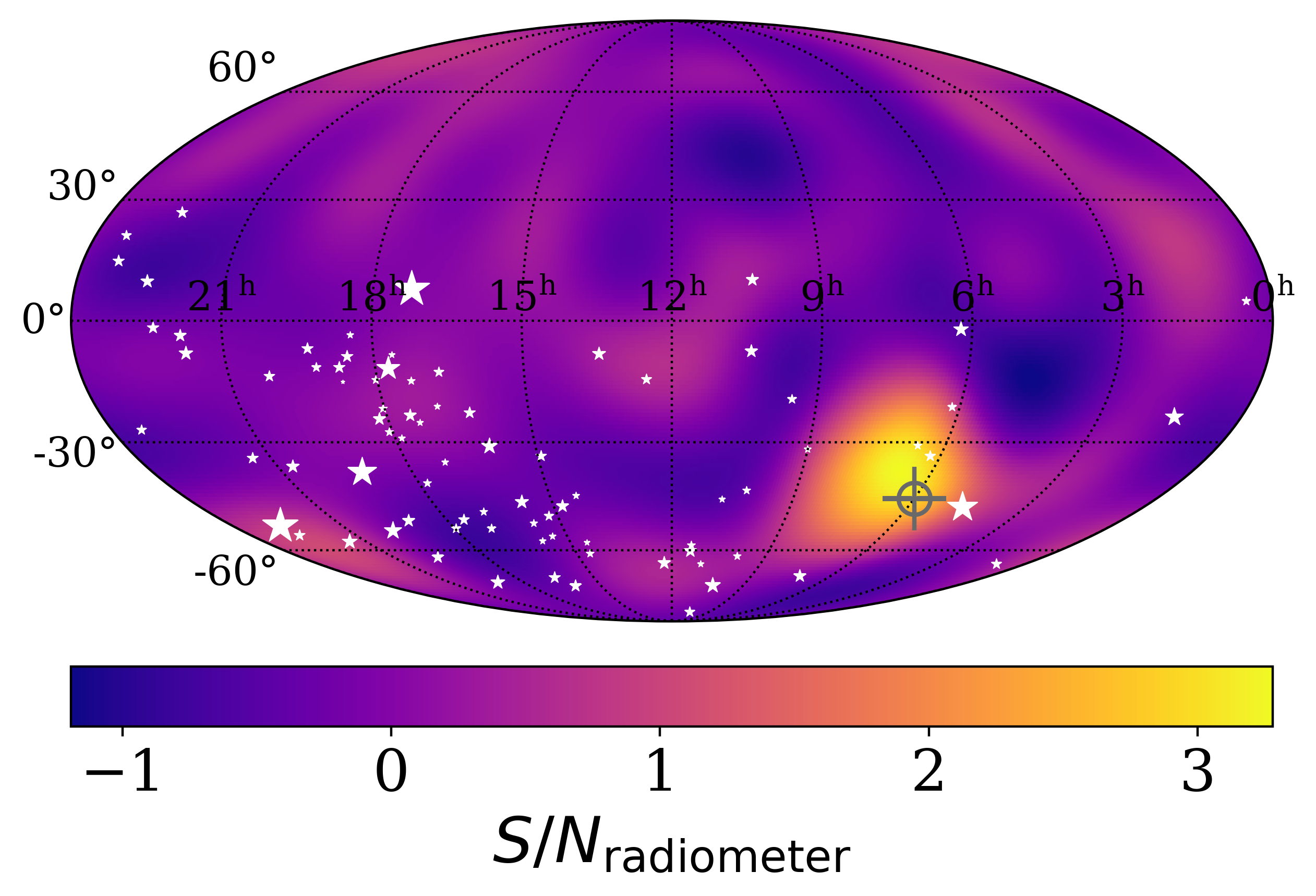}
    \label{fig:simCGW_radiometer2}}
    \hfill
    \subfigure[single injected point source at RA~06h DEC~\SI{45}{\degree}]{\includegraphics[width=0.48\linewidth]{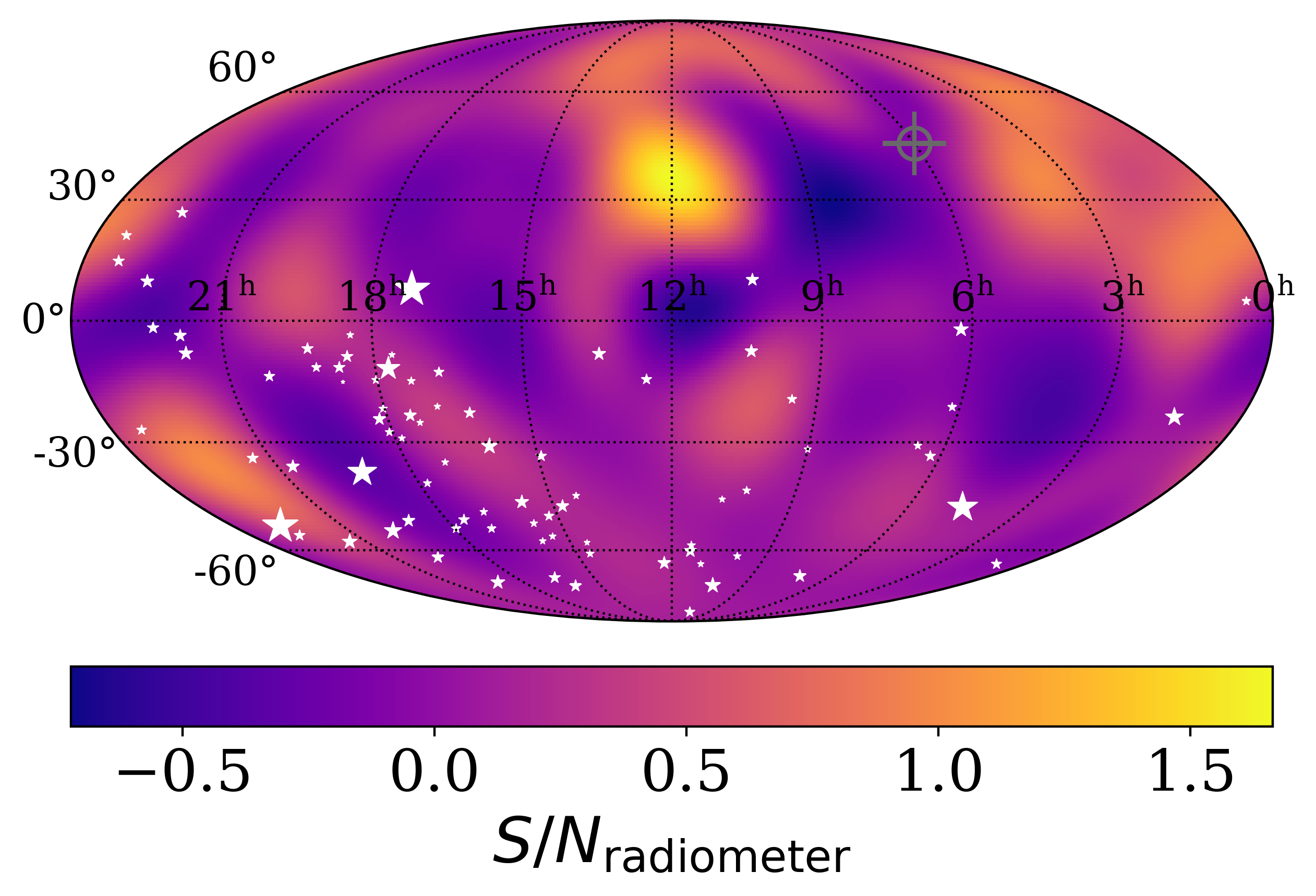}
    \label{fig:simCGW_radiometer3}}
    \hfill
    \subfigure[two injected point sources]{\includegraphics[width=0.48\linewidth]{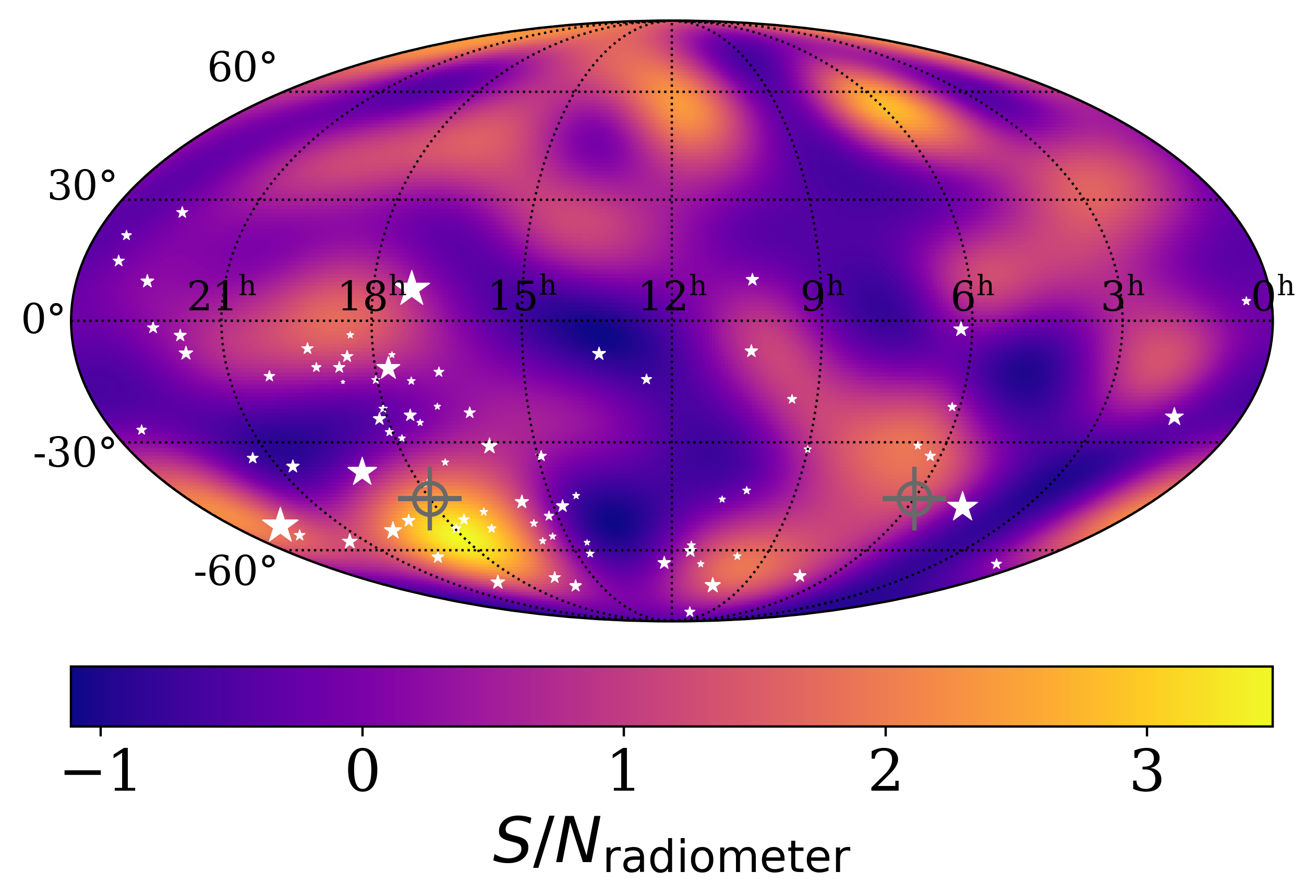}
    \label{fig:simCGW_radiometer4}}
    \caption{Radiometer maps of the first frequency bin (\SI{7}{\nano\hertz}) for the simulated data sets containing a single gravitational wave source. The injected position of the source is indicated with a crosshair.}
    \label{fig:simCGW_radiometer}
\end{figure*}

\begin{figure*}
    \centering
    \subfigure[single injected point source at RA~18h DEC~\SI{-45}{\degree}]
    {\includegraphics[width=0.48\linewidth]{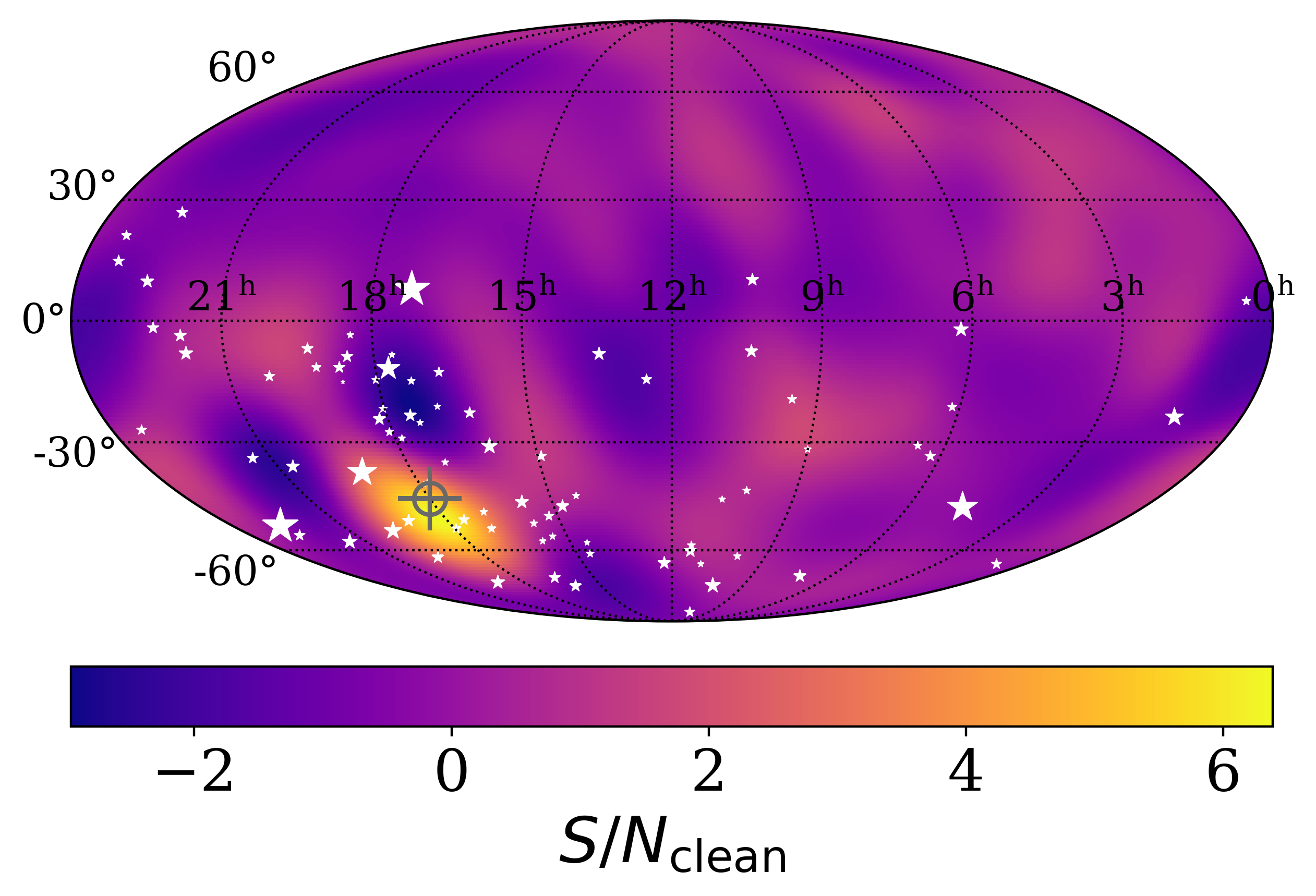}
    \label{fig:simCGW_snr1}}
    \hfill
    \subfigure[single injected point source at RA~06h DEC~\SI{-45}{\degree}]{\includegraphics[width=0.48\linewidth]{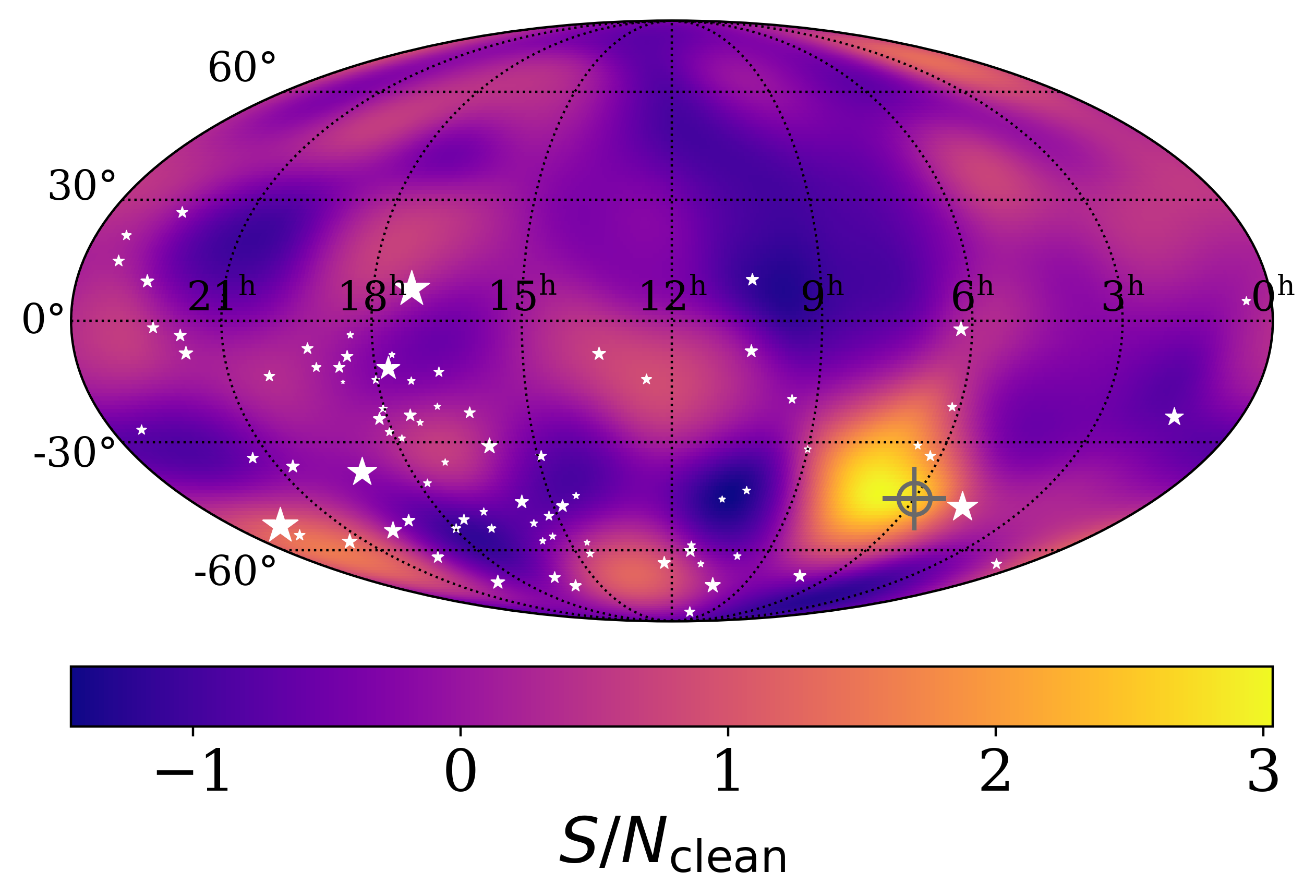}
    \label{fig:simCGW_snr2}}
    \hfill
    \subfigure[single injected point source at RA~06h DEC~\SI{45}{\degree}]{\includegraphics[width=0.48\linewidth]{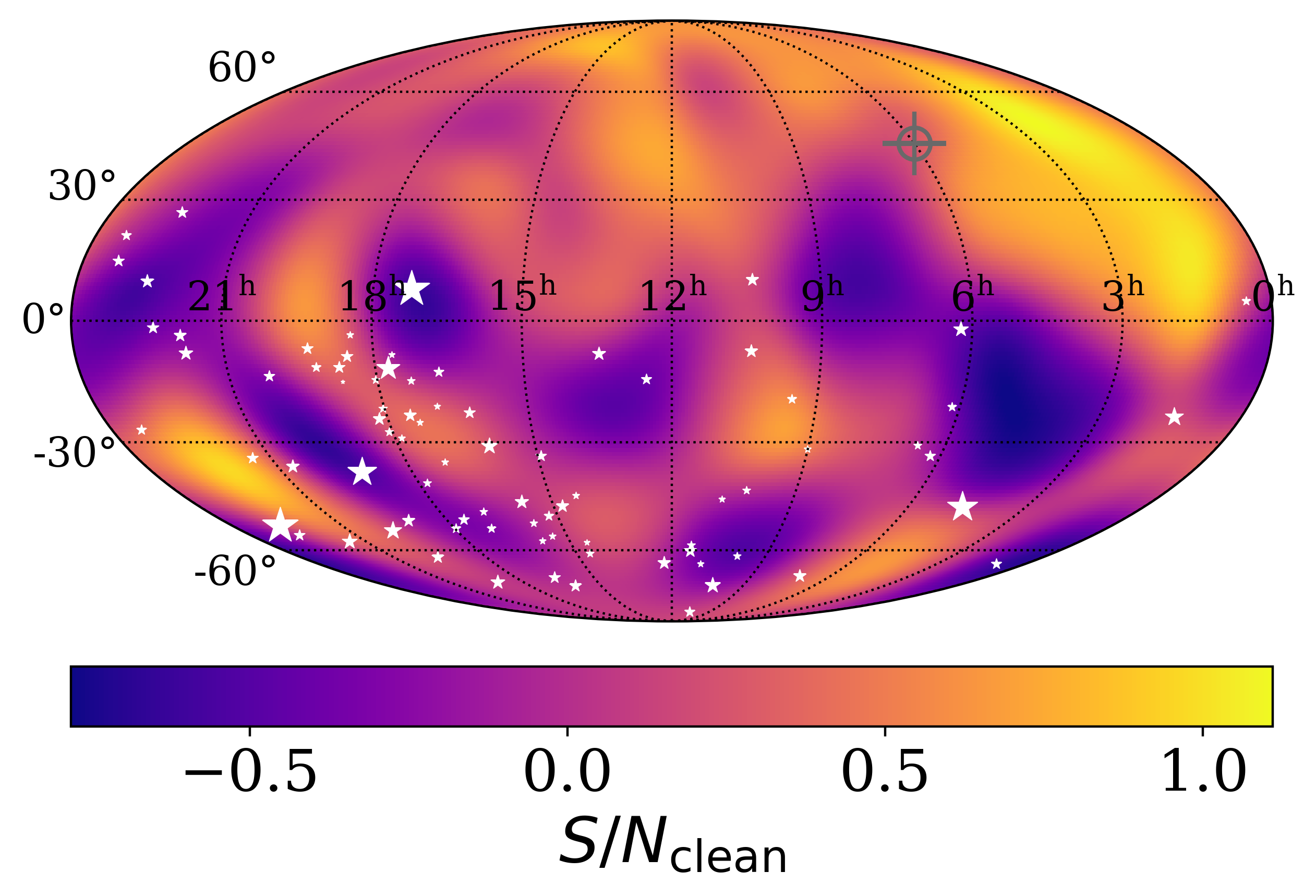}
    \label{fig:simCGW_snr3}}
    \hfill
    \subfigure[two injected point sources]{\includegraphics[width=0.48\linewidth]{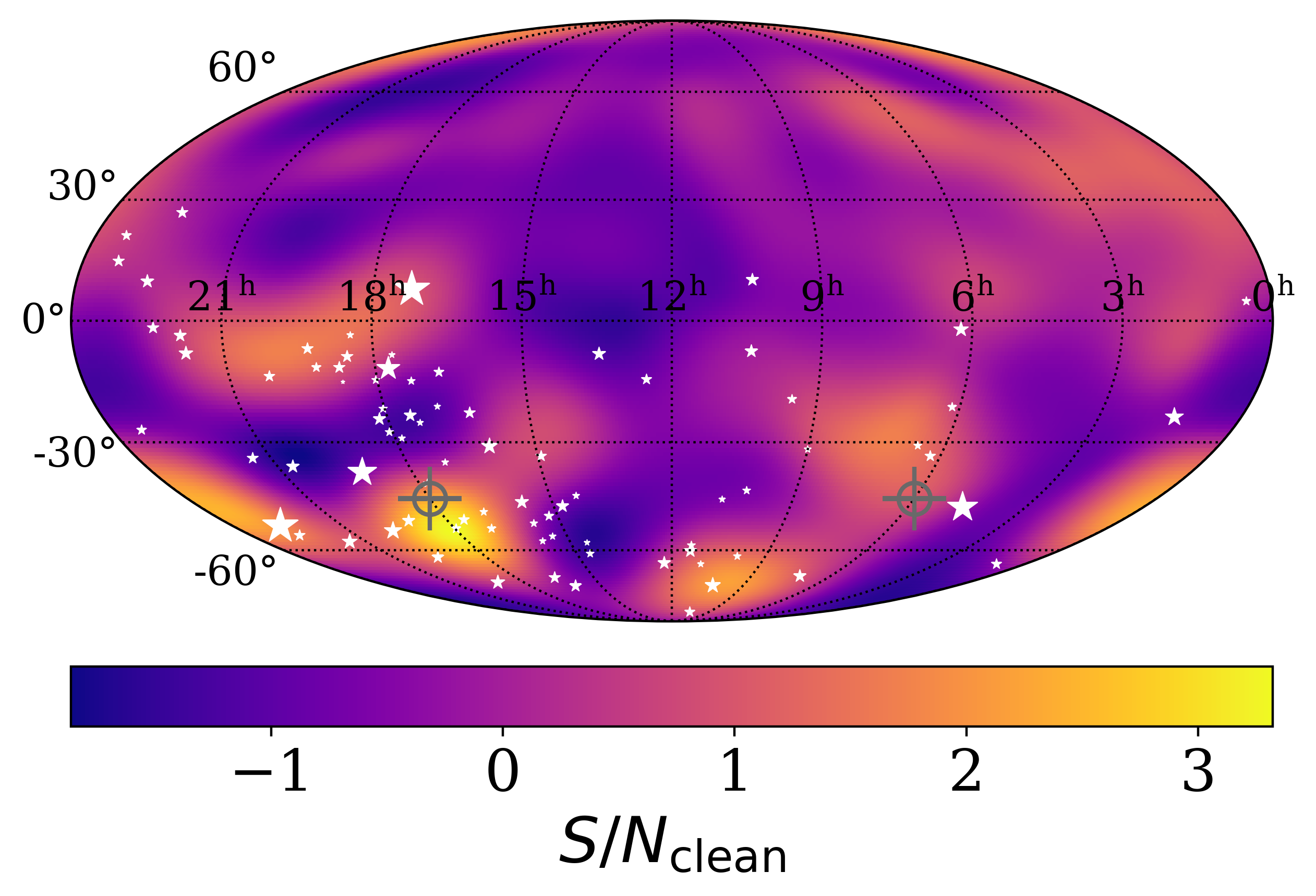}
    \label{fig:simCGW_snr4}}
    \caption{Regularized clean signal-to-noise ratio maps of the first frequency bin (\SI{7}{\nano\hertz}) for the simulated data sets containing a single gravitational-wave source. The injected position of the source is indicated with a crosshair.}
    \label{fig:simCGW_snr}
\end{figure*}

If the analysis pipeline is not only able to identify but also correctly localize the point source(s), we expect both the radiometer and clean map $S/N$ to exhibit a spatially constraint area of significantly increased $S/N$, located at a similar position as the simulated source. As the radiometer $S/N$ map (see Section~\ref{ssec:radiometer_map}) is ideal for identifying point sources, we present both the resulting radiometer and clean map $S/N$ for all three datasets mentioned above, each shown in Figs.~\ref{fig:simCGW_radiometer} and \ref{fig:simCGW_snr} respectively.

Unsurprisingly, the ``hotspot'' at the injection location is only present in the first two frequency bins, as the frequency of the simulated sources falls between the first and second frequency bin of the MPTA. Thus we present only the sky maps for the first frequency bin. In each map, the position of the injected source(s) are indicated with gray crosshairs.

As expected, sources that are injected in the Southern Hemisphere, are well localised in both the radiometer and clean $S/N$ maps, but sources injected on the Northern Hemisphere are not well resolved. That is, we are able to see a signal in the data when gravitational waves are in the northern sky, but it is difficult to differentiate an isotropic background from a point source. This is in accordance with the general understanding \citep{Taylor_2013,Ali-Haimoud_2021,Agazie_2023} that a patch of sky populated with more pulsars is in general more sensitive to the gravitational wave signal coming from that sky region.


\begin{figure*}
    \centering
    \subfigure[Mode 1]{\includegraphics[width=0.32\linewidth]{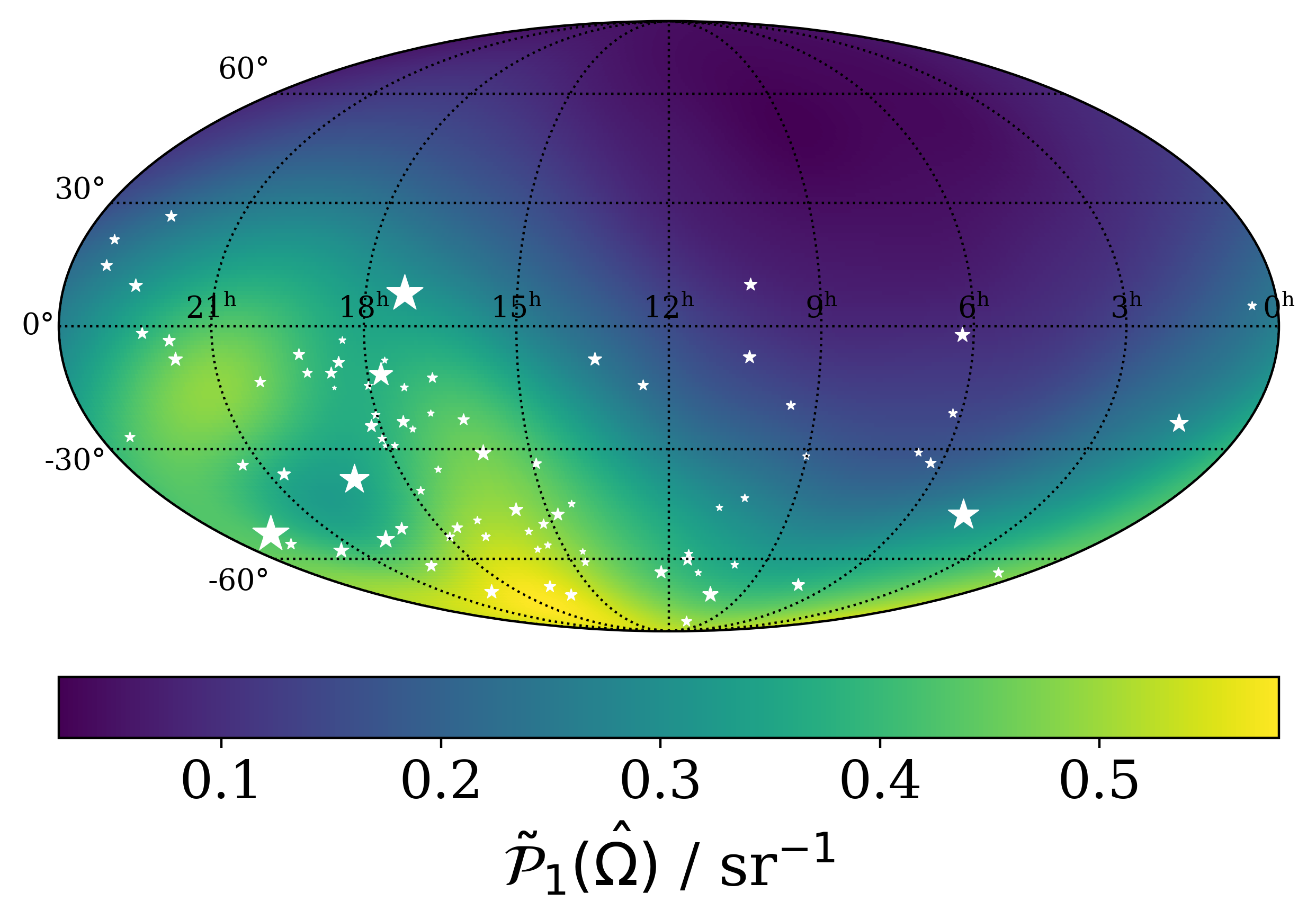}}
    \subfigure[Mode 2]{\includegraphics[width=0.32\linewidth]{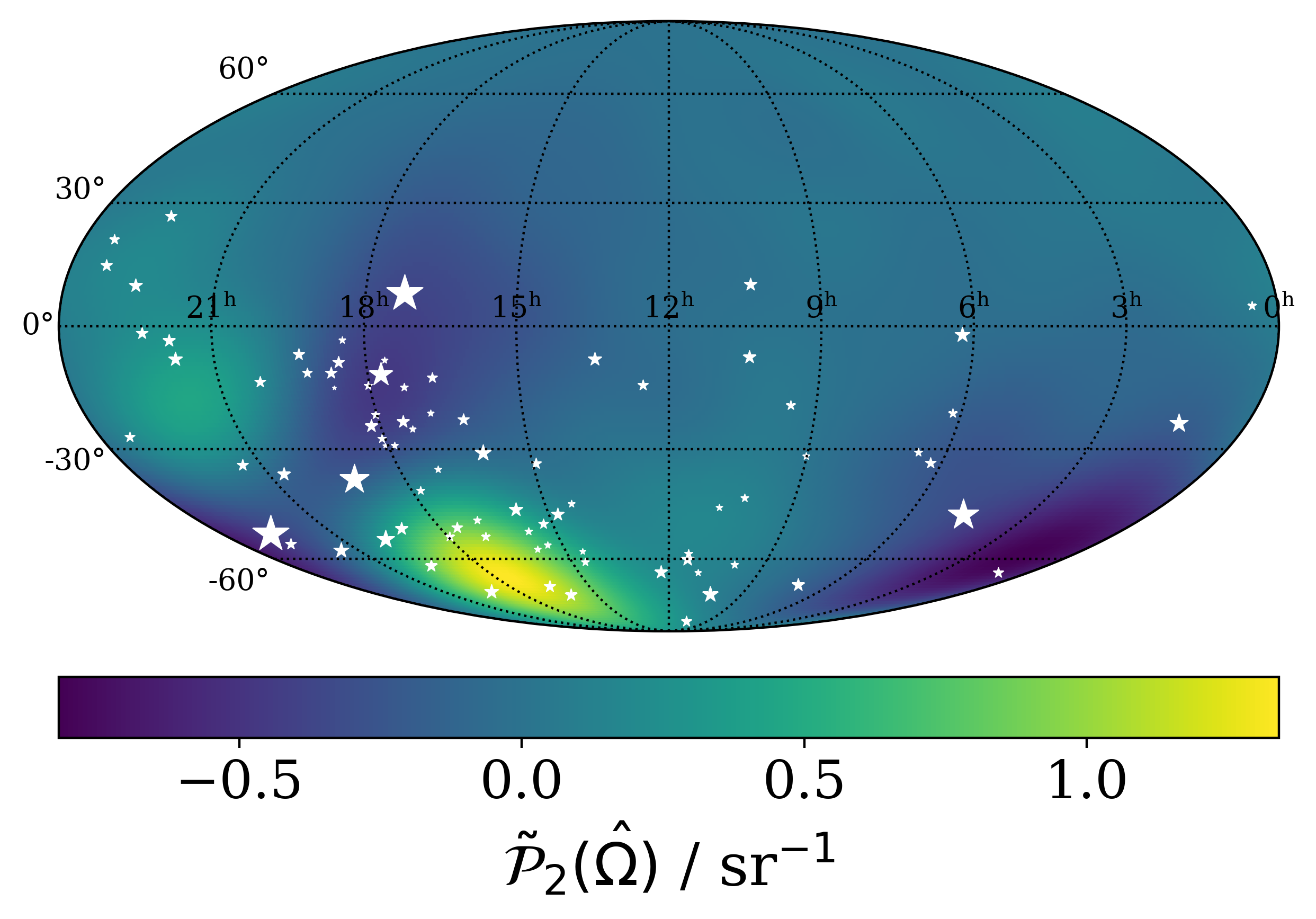}}
    \subfigure[Mode 3]{\includegraphics[width=0.32\linewidth]{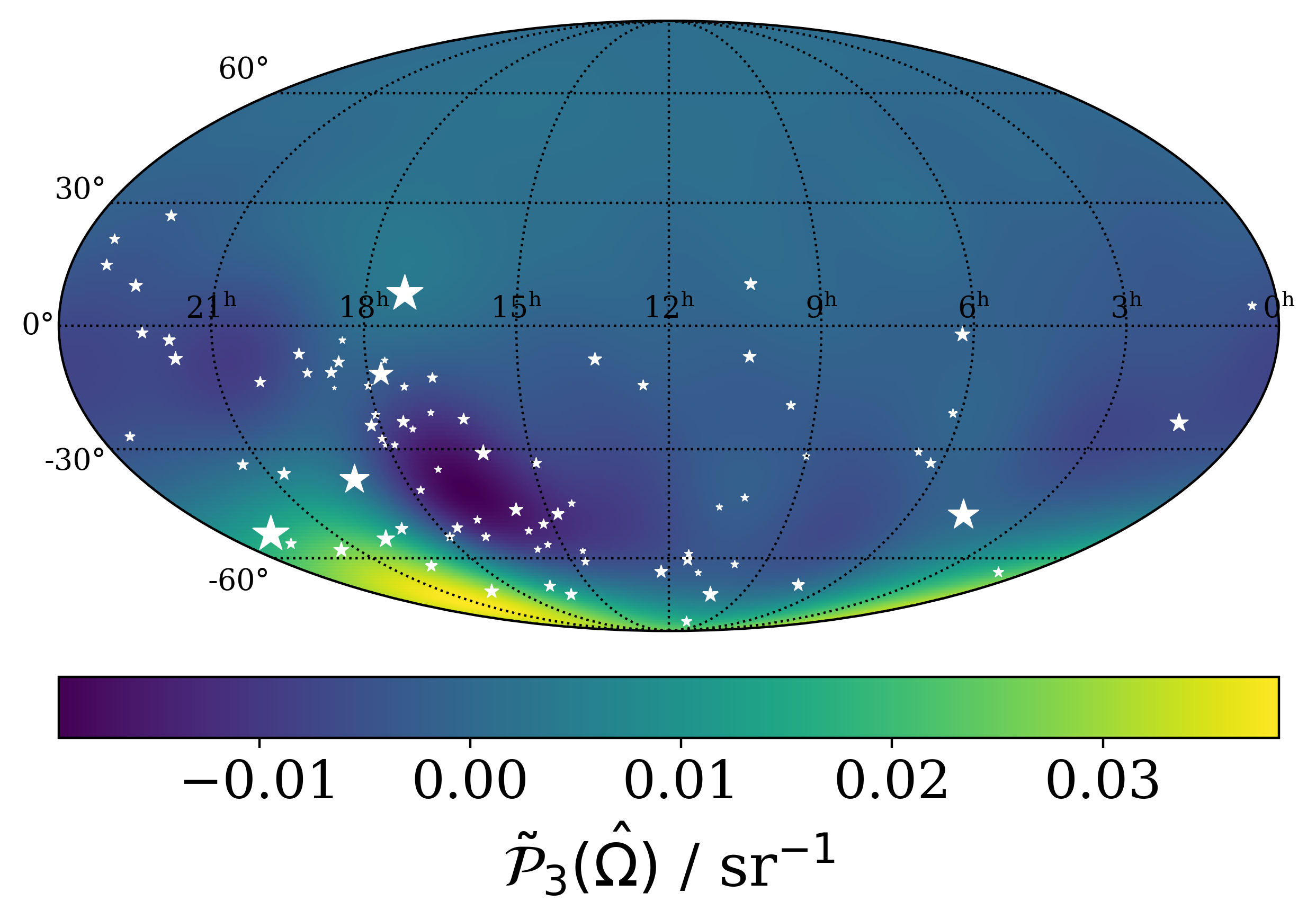}}
    \subfigure[Mode 79]{\includegraphics[width=0.32\linewidth]{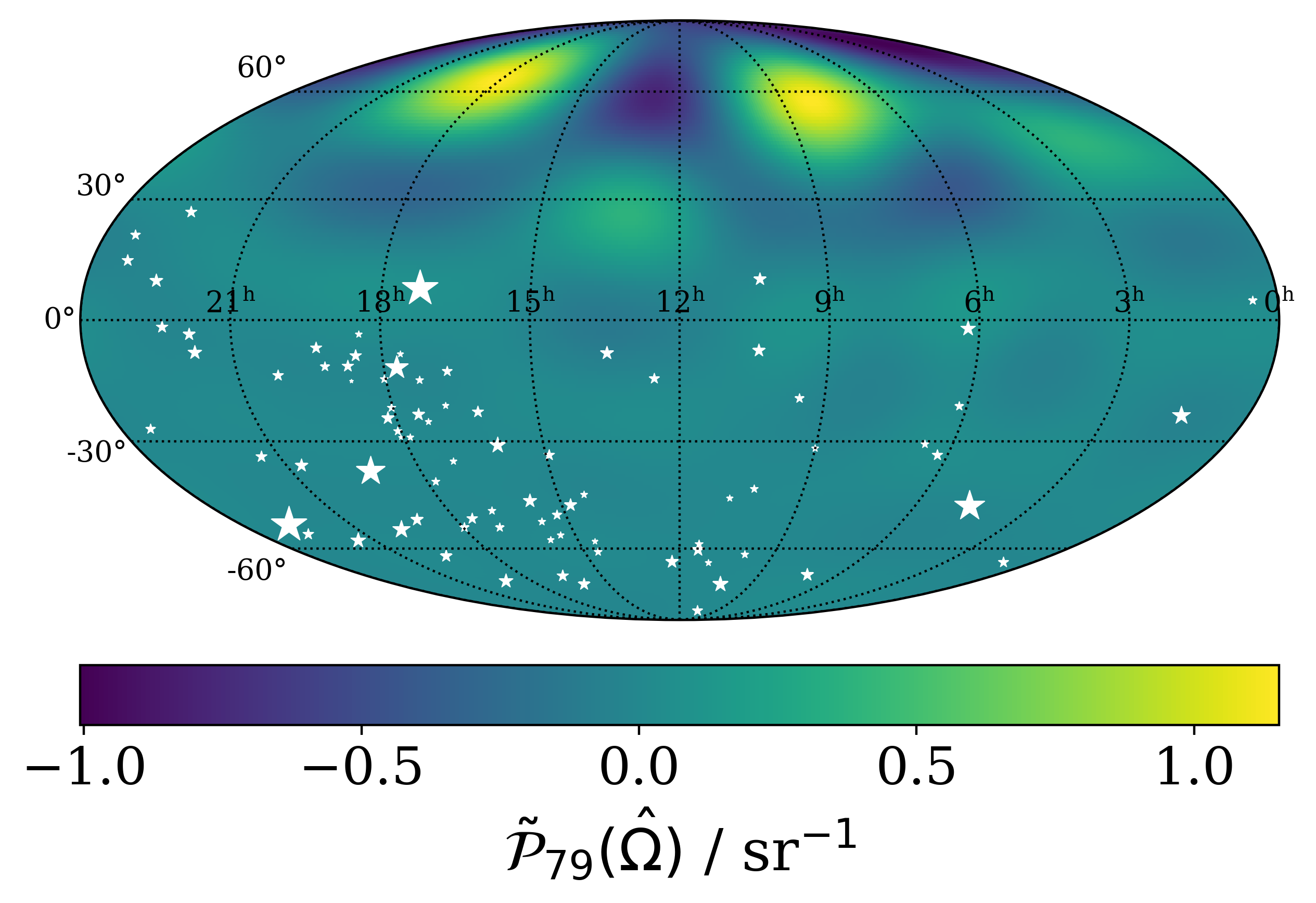}}
    \subfigure[Mode 80]{\includegraphics[width=0.32\linewidth]{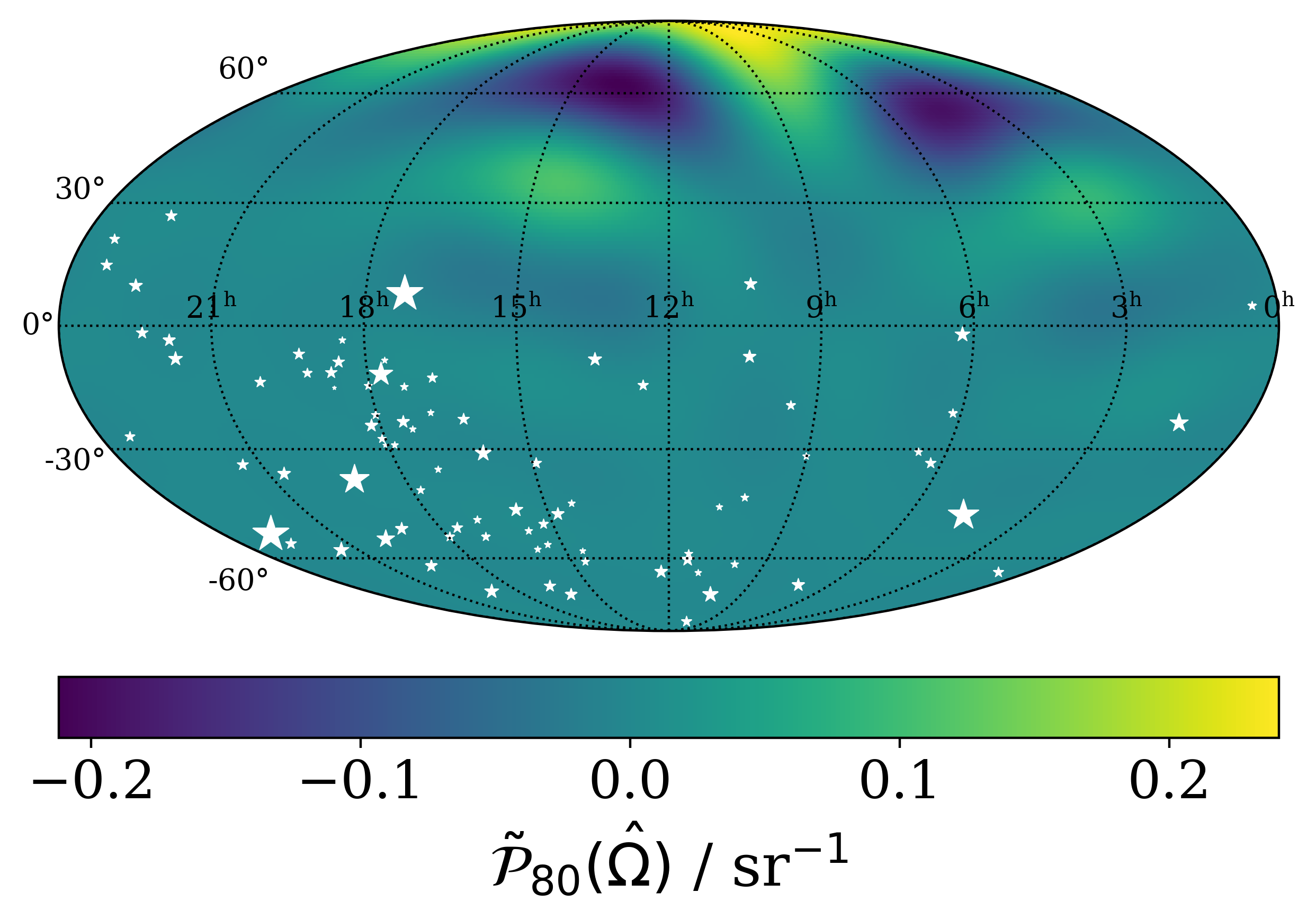}}
    \subfigure[Mode 81]{\includegraphics[width=0.32\linewidth]{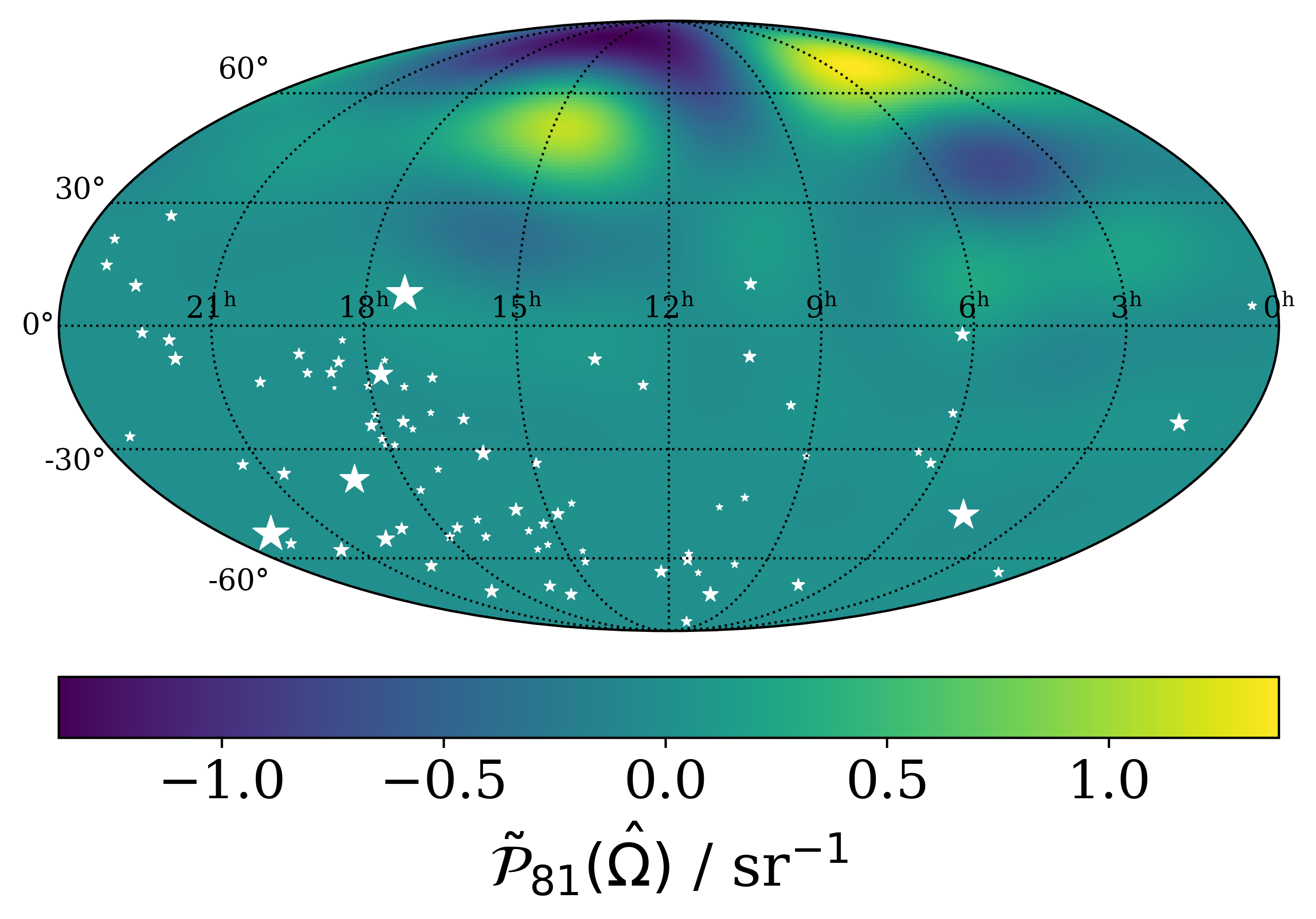}}
    \caption{Fisher matrix modes of the full MPTA 4.5 year dataset. Top row: Three most constrained modes. Bottom row: Three least constrained modes.}
    \label{fig:modes}
\end{figure*}

\section{Regularisation} \label{app:sec:regularisation}

\begin{figure*}
    \centering
    \subfigure[$i_\mathrm{max}=20$]{\includegraphics[width=0.32\textwidth]{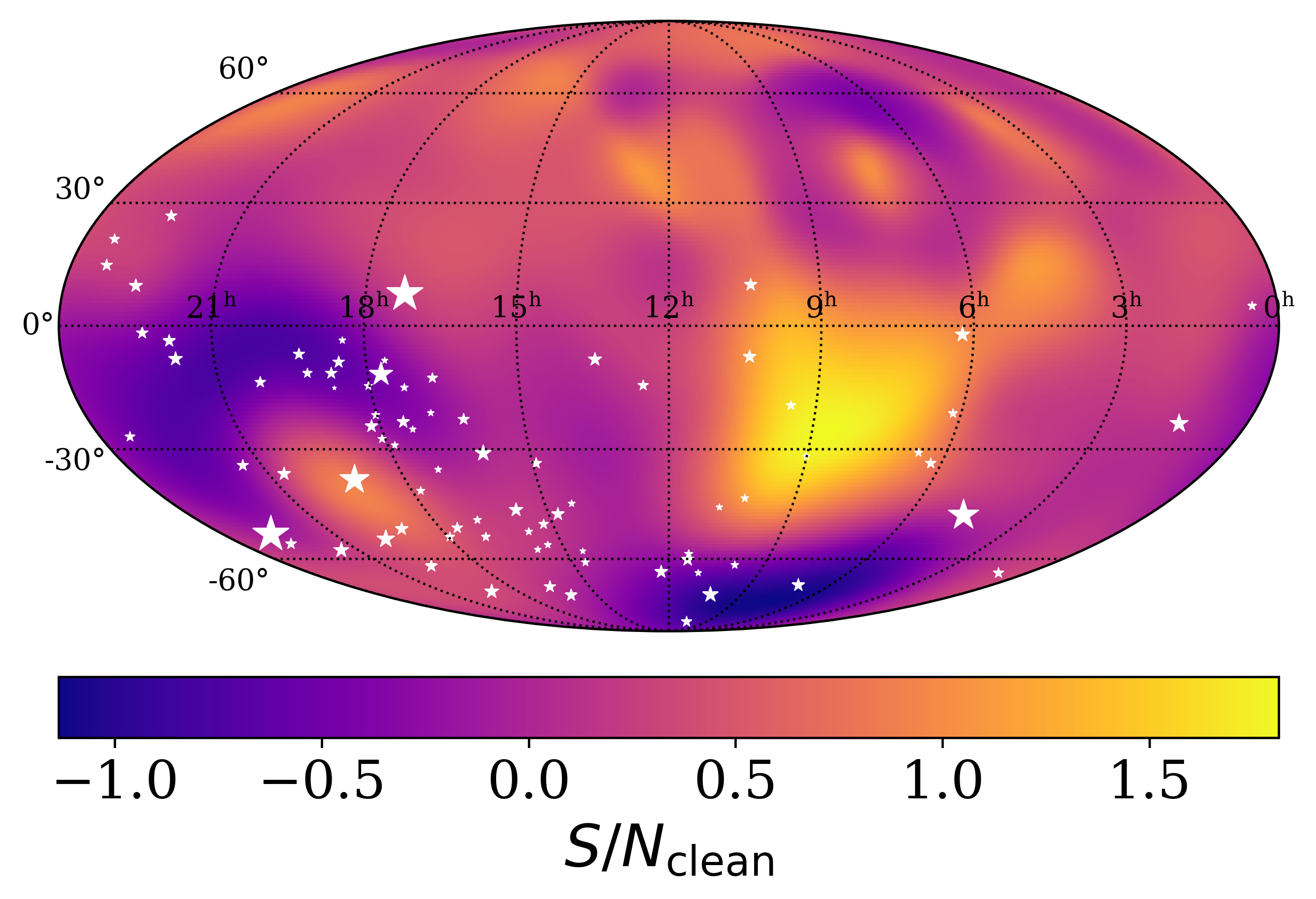}}
    \label{fig:snr_GWB_sv20}
    \hfill
    \subfigure[$i_\mathrm{max}=32$]{\includegraphics[width=0.32\textwidth]{img/simulation_GWB/snrmap_papersmall_lmax8_sv32_nside16_fbin1.png}
    \label{fig:snr_GWB_sv36}}
    \hfill
    \subfigure[$i_\mathrm{max}=54$]{\includegraphics[width=0.32\textwidth]{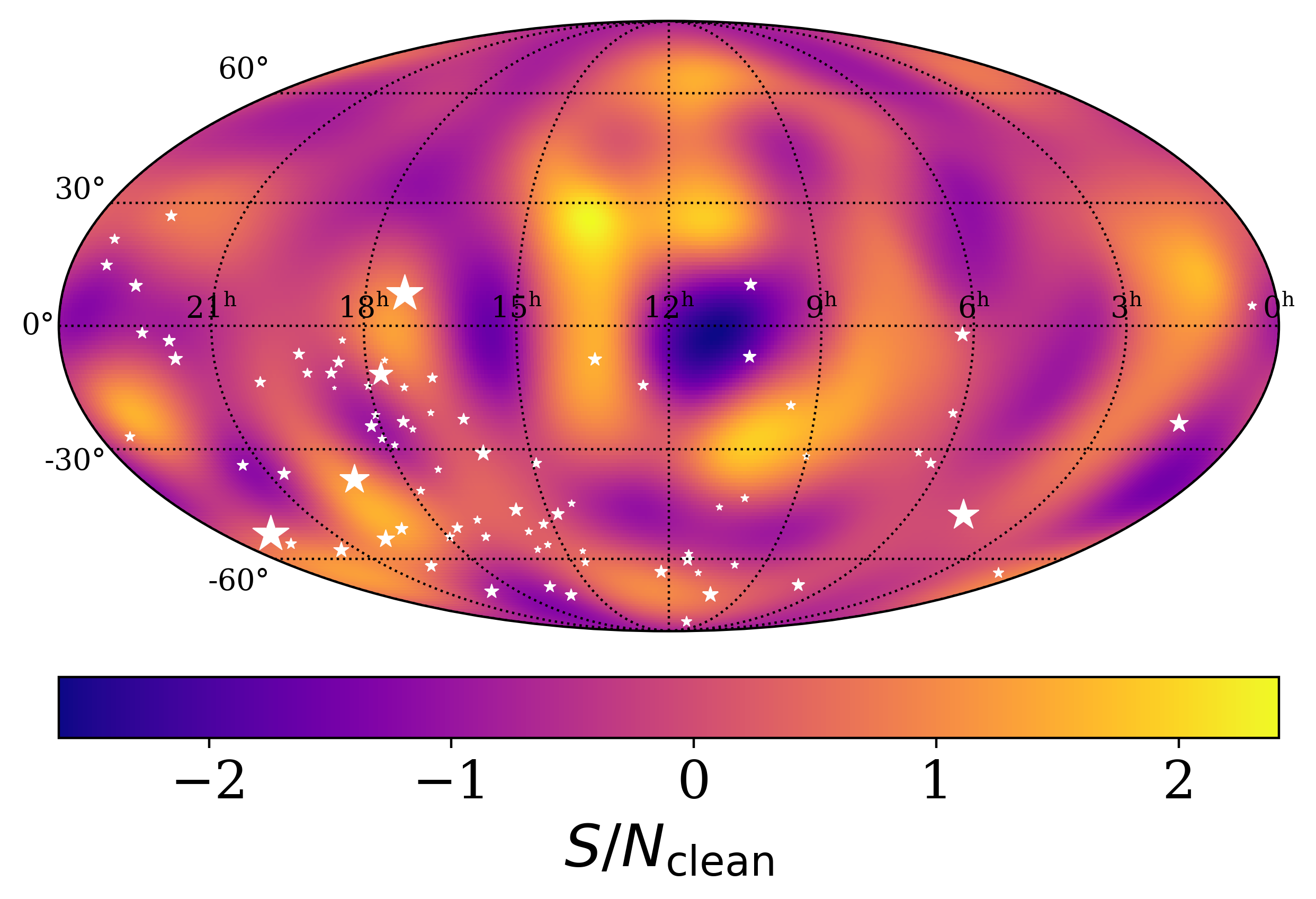}
    \label{fig:snr_GWB_sv54}}
    \hfill
    \caption{Visualisations of the clean map $S/N$ across the sky of the simulated white noise + gravitational-wave background dataset using different regularisation cut-offs.}
    \label{fig:sv_comparison_GWB}
\end{figure*}
\begin{figure*}
    \centering
    \subfigure[$i_\mathrm{max}=20$]{\includegraphics[width=0.32\textwidth]{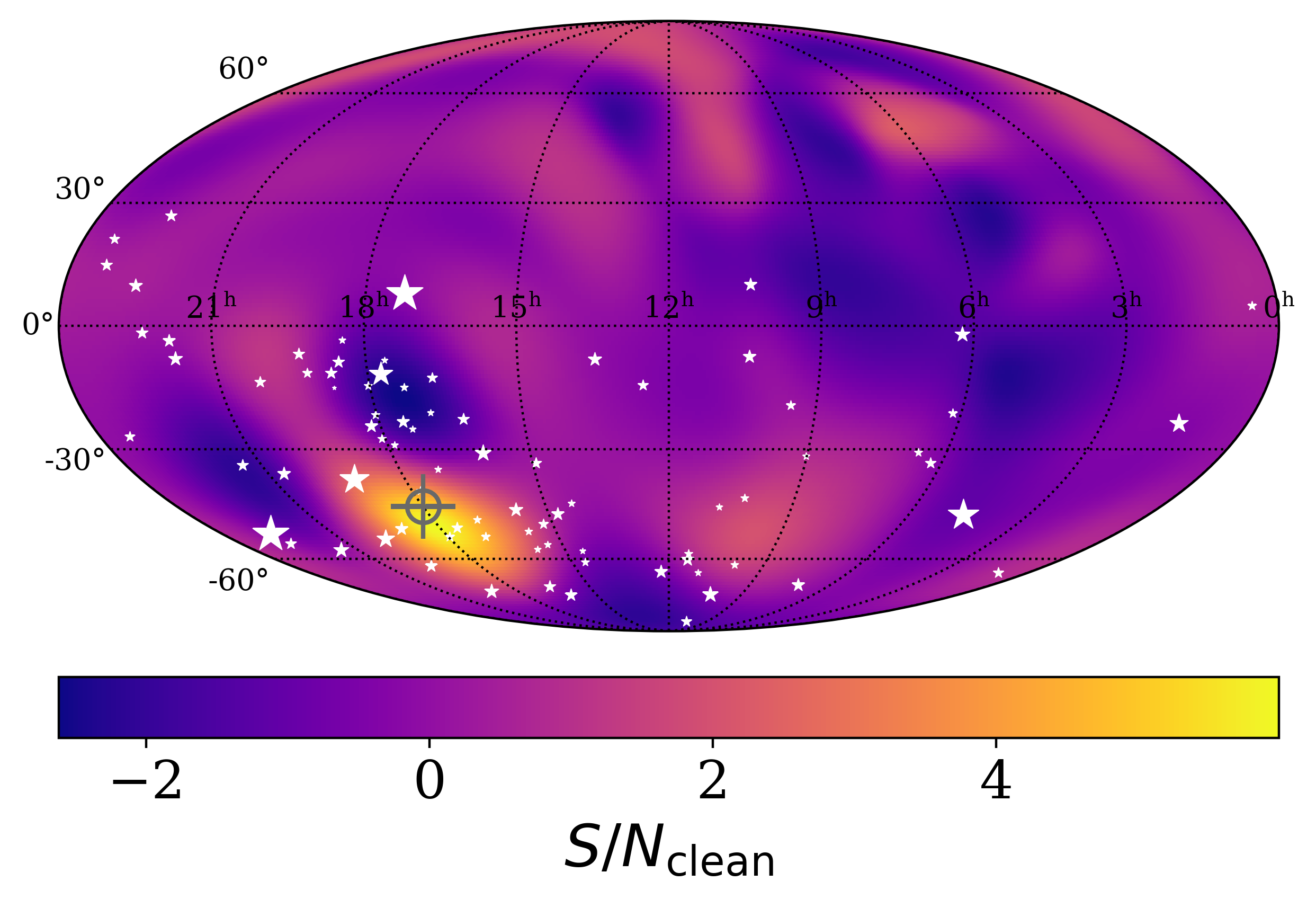}}
    \label{fig:snr_CGW_sv20}
    \hfill
    \subfigure[$i_\mathrm{max}=32$]{\includegraphics[width=0.32\textwidth]{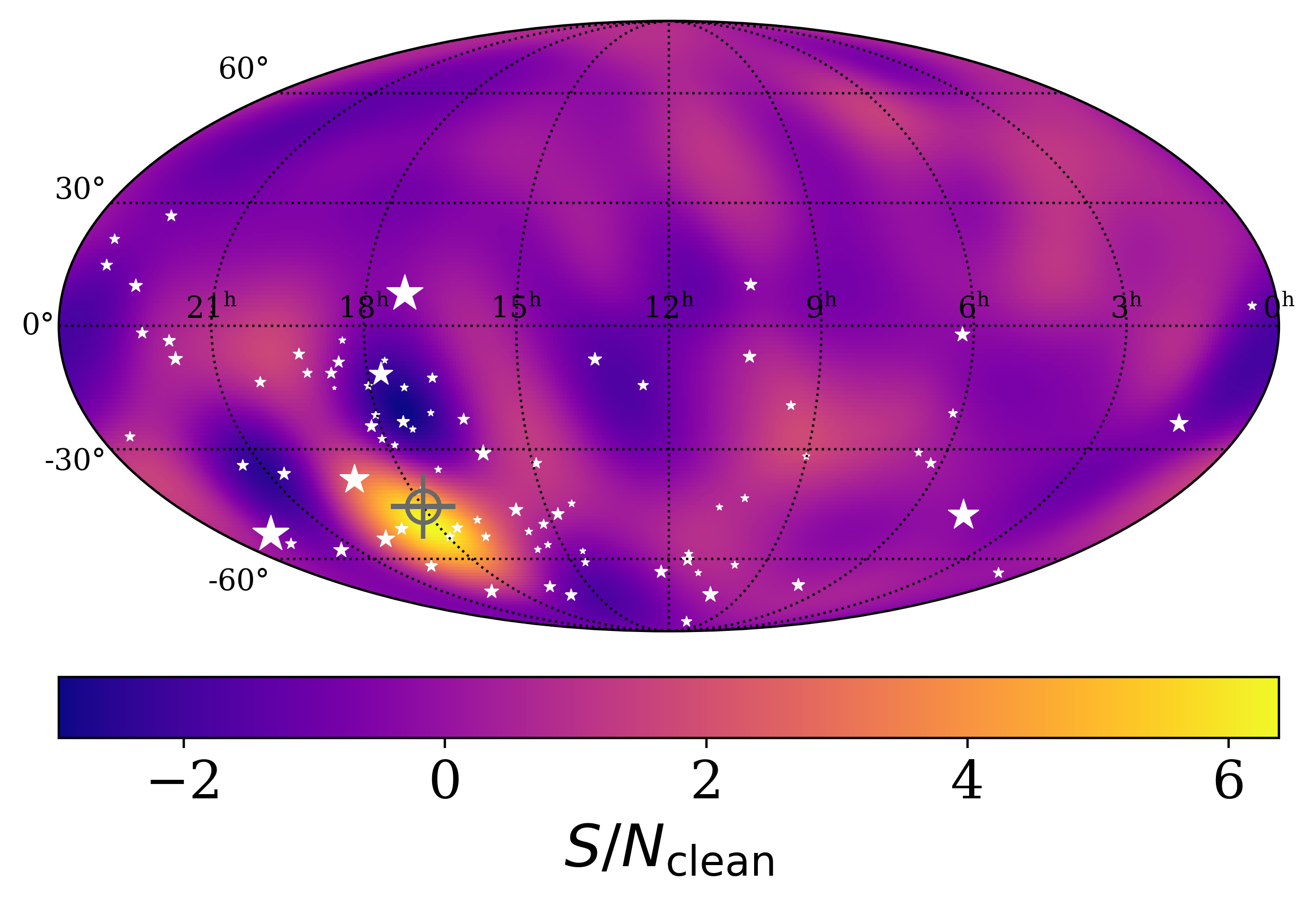}
    \label{fig:snr_CGW_sv36}}
    \hfill
    \subfigure[$i_\mathrm{max}=54$]{\includegraphics[width=0.32\textwidth]{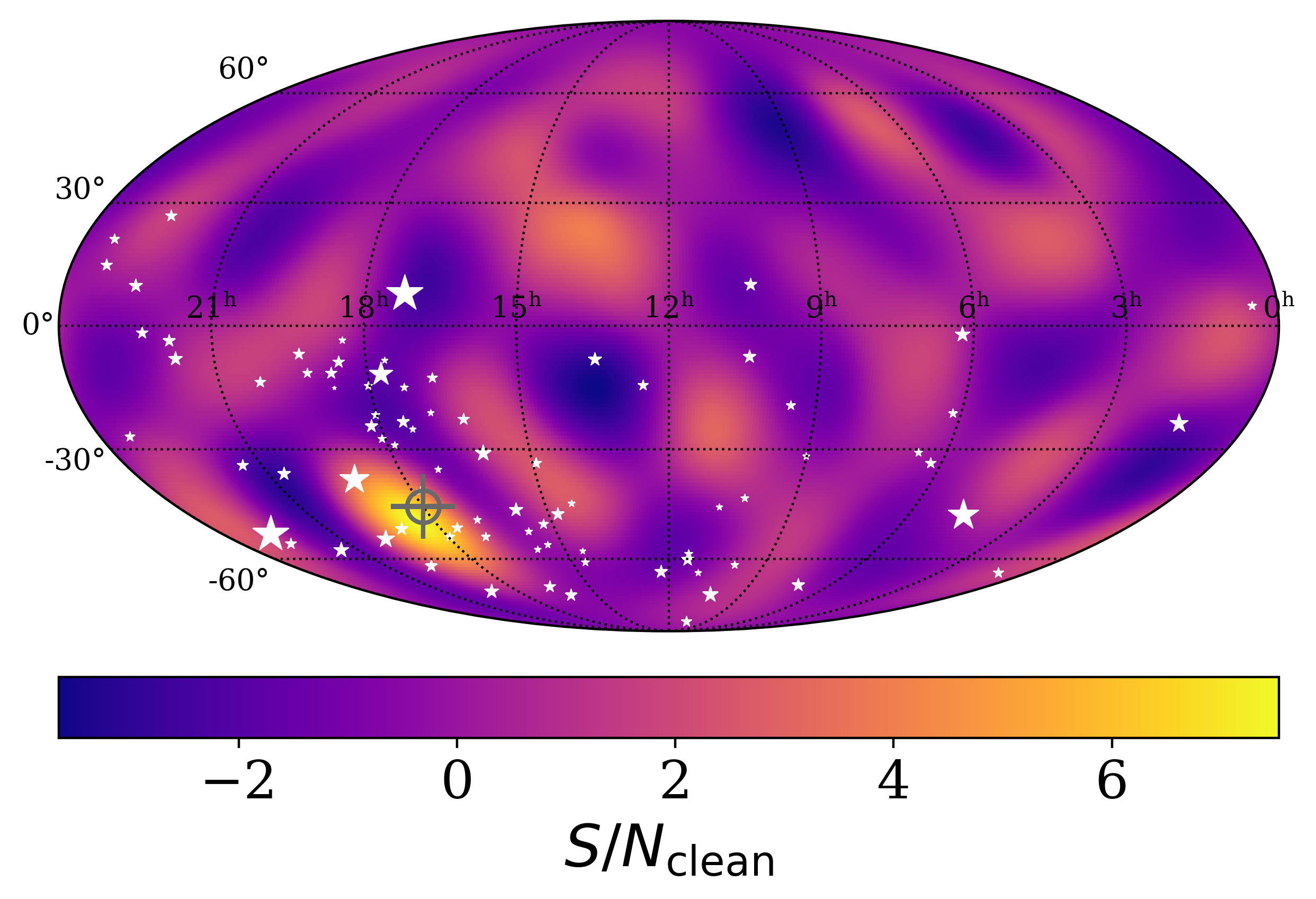}
    \label{fig:snr_CGW_sv54}}
    \hfill
    \caption{Visualisations of the clean map $S/N$ across the sky of a simulated point source (continuous wave) signal using different regularisation cut-offs.}
    \label{fig:sv_comparison_CGW}
\end{figure*}

\begin{figure}
    \centering
    \includegraphics[width=0.5\linewidth]{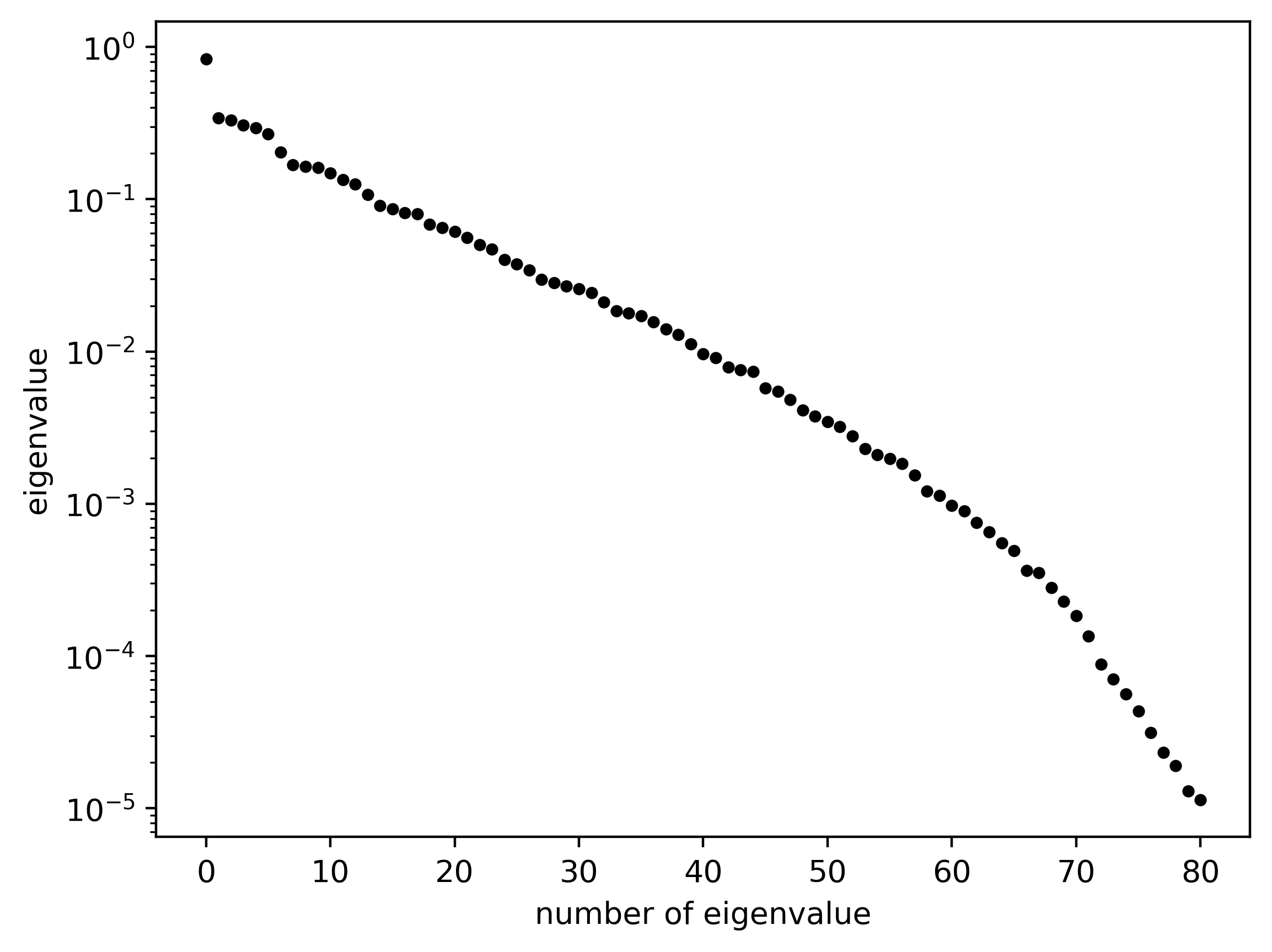}
    \caption{Singular value spectrum of the MPTA Fisher information matrix $\mtx{M}$ (cf.\  Equation~\eqref{eq:Fisher_matrix}) of the pulsar-pair correlations $\rho_{\alpha\beta}$ in the first frequency bin of the full MPTA 4.5 year data set.}
    \label{fig:singular_value_spectrum}
\end{figure}

Here we explore the impact of the regularisation cutoff on the resulting sky distributions. 
The first step is to examine the ``eigen spectrum" in Fig.~\ref{fig:singular_value_spectrum} showing the eigenvalues of the Fisher matrix in descending order.
Larger eigenvalues correspond to modes that we measure with relatively lower uncertainty. 
As noted in the main body of the manuscript, the choice of regularisation cutoff is a balancing act, improving sensitivity to some modes by throwing out other modes, thereby introducing a bias.
In this analysis, we employ a cutoff value of 32.

\begin{figure*}
    \centering
    \subfigure[$i_\mathrm{max}=20$]{\includegraphics[width=0.32\textwidth]{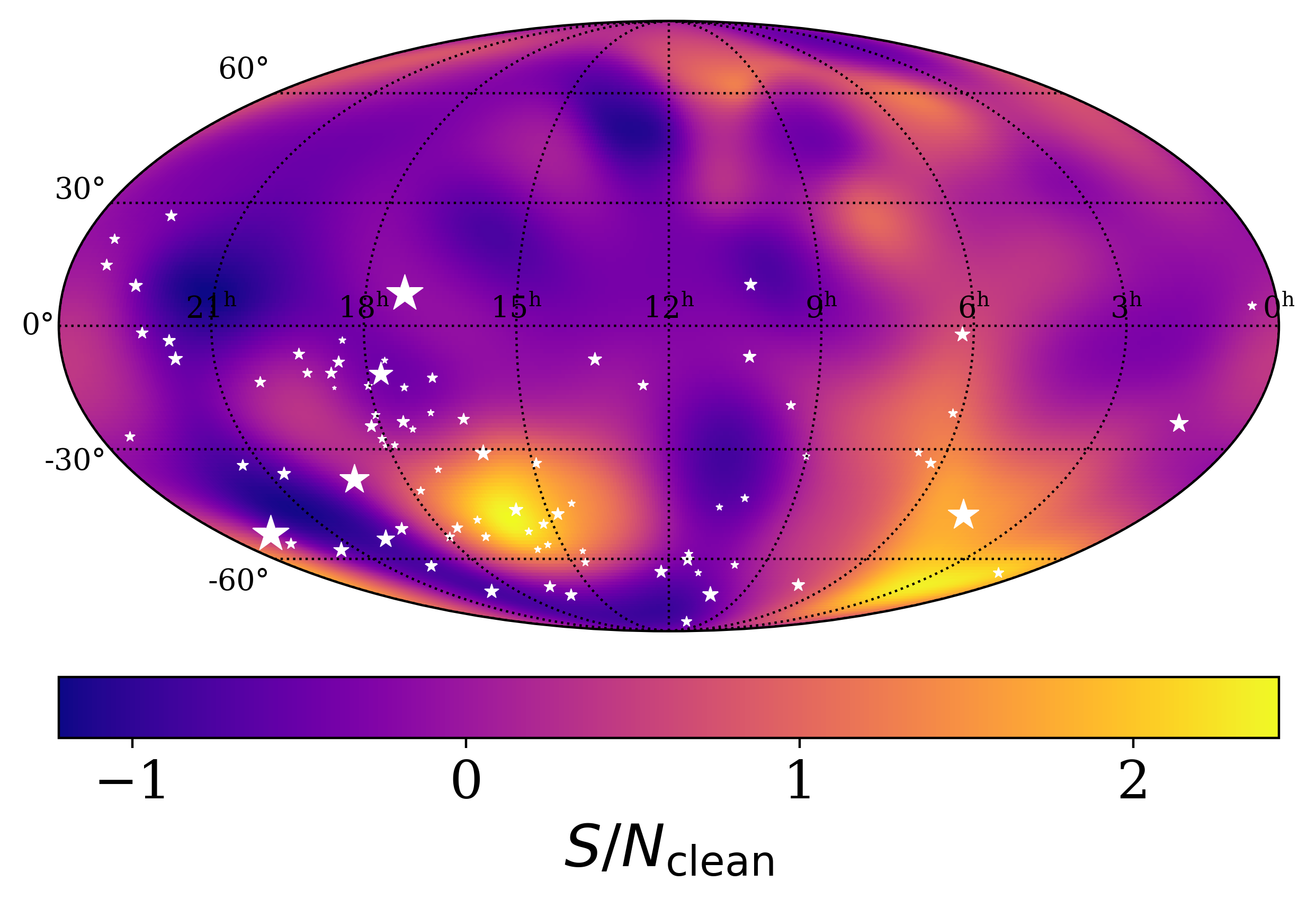}
    \label{fig:snr_full_sv20}}
    \hfill
    \subfigure[$i_\mathrm{max}=32$]{\includegraphics[width=0.32\textwidth]{img/full_dataset/snrmap_papersmall_lmax8_sv32_nside16_fbin1.png}
    \label{fig:snr_full_sv36}}
    \hfill
    \subfigure[$i_\mathrm{max}=54$]{\includegraphics[width=0.32\textwidth]{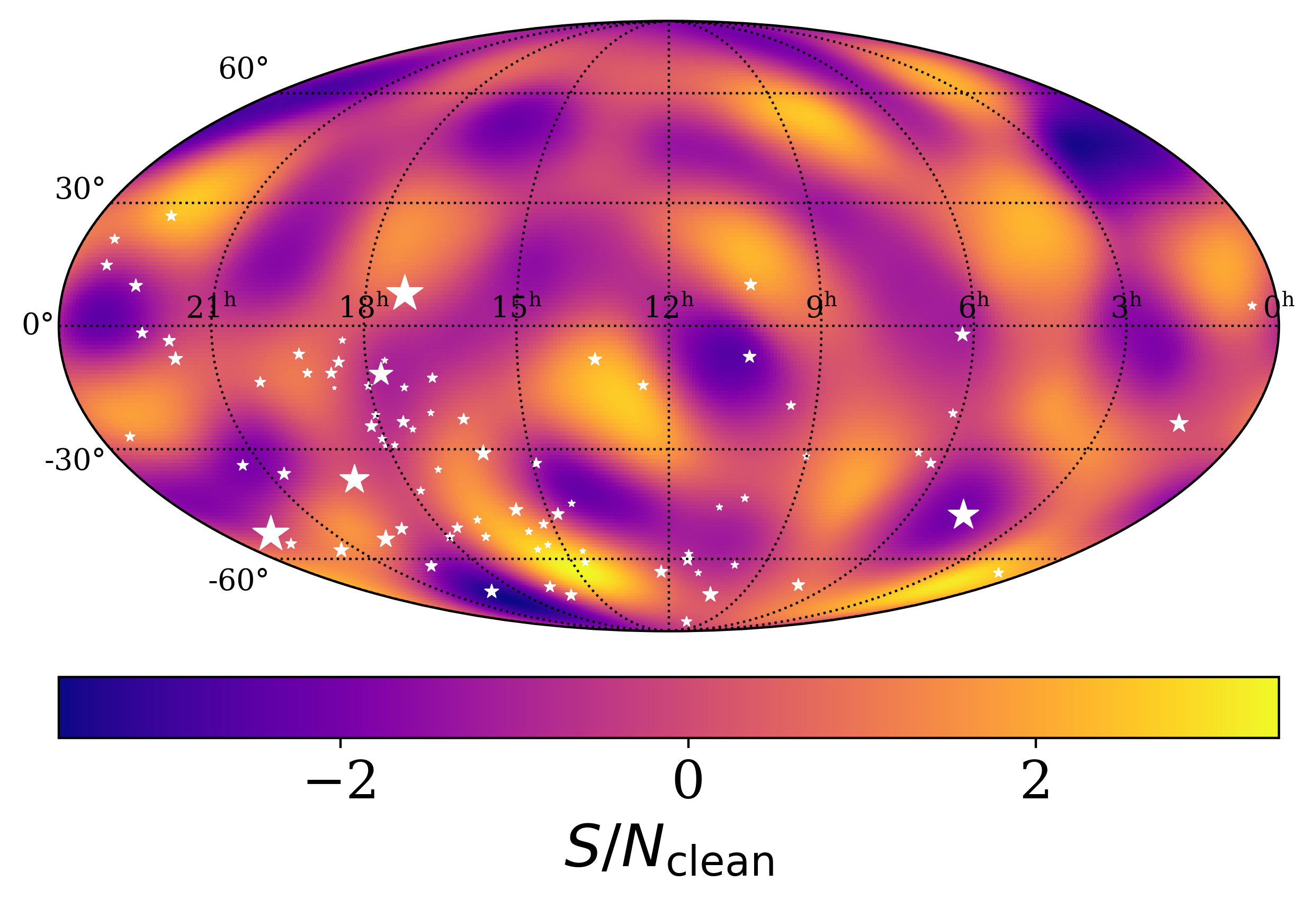}
    \label{fig:snr_full_sv54}}
    \hfill
    \caption{Realisations of the clean map $S/N$ as seen by the MPTA across the sky using different regularisation cut-offs.}
    \label{fig:sv_comparison_full}
\end{figure*}

In order to illustrate how the choice of cutoff affects our results, we create sky maps for $i_\mathrm{max}\in [1, N_\mathrm{SpH}]$ for all three frequency bins and investigate the evolution of the clean map $S/N$ distribution with varying $i_\mathrm{max}$. We show three exemplary maps for $i_\mathrm{max}=\{20, 32, 54\}$ of the first bin in Fig.~\ref{fig:sv_comparison_GWB}.
Videos showing the map evolution across all cut-offs for the three frequency bins can be found here: \url{https://doi.org/10.57891/j0vh-5g31}

Fig.~\ref{fig:sv_comparison_GWB} shows how our reconstructions of an isotropic background vary with different choices of regularisation cutoff.
Meanwhile, in Fig.~\ref{fig:sv_comparison_CGW}, we observe how our sky maps vary with regularisation cutoff for a point source (continuous wave) signal.
The more modes that we keep, the more pronounced and sharp the single source becomes. At the same time, the off-source sky exhibits increasingly large noise fluctuations.
Artifacts from the diffraction limit are visible in the right-most plot ($i_\mathrm{max}=54$).
Encouragingly, though, the location of the hotspot is stable as we vary the cutoff.
In Fig.~\ref{fig:sv_comparison_full}, we show how the actual MPTA clean maps vary with cutoff.

\section{Additional results}
\label{app:sec:additional_results}
In this appendix we show additional plots, which may be of interest to expert readers.


\subsection{Cleaned skymaps}

\subsubsection{Clean gravitational wave power skymap}
\begin{figure*}
    \centering
    \subfigure[Bin 1; $f = 1/T_\mathrm{obs} = \SI{7}{\nano\hertz}$]{\includegraphics[width=0.32\linewidth]{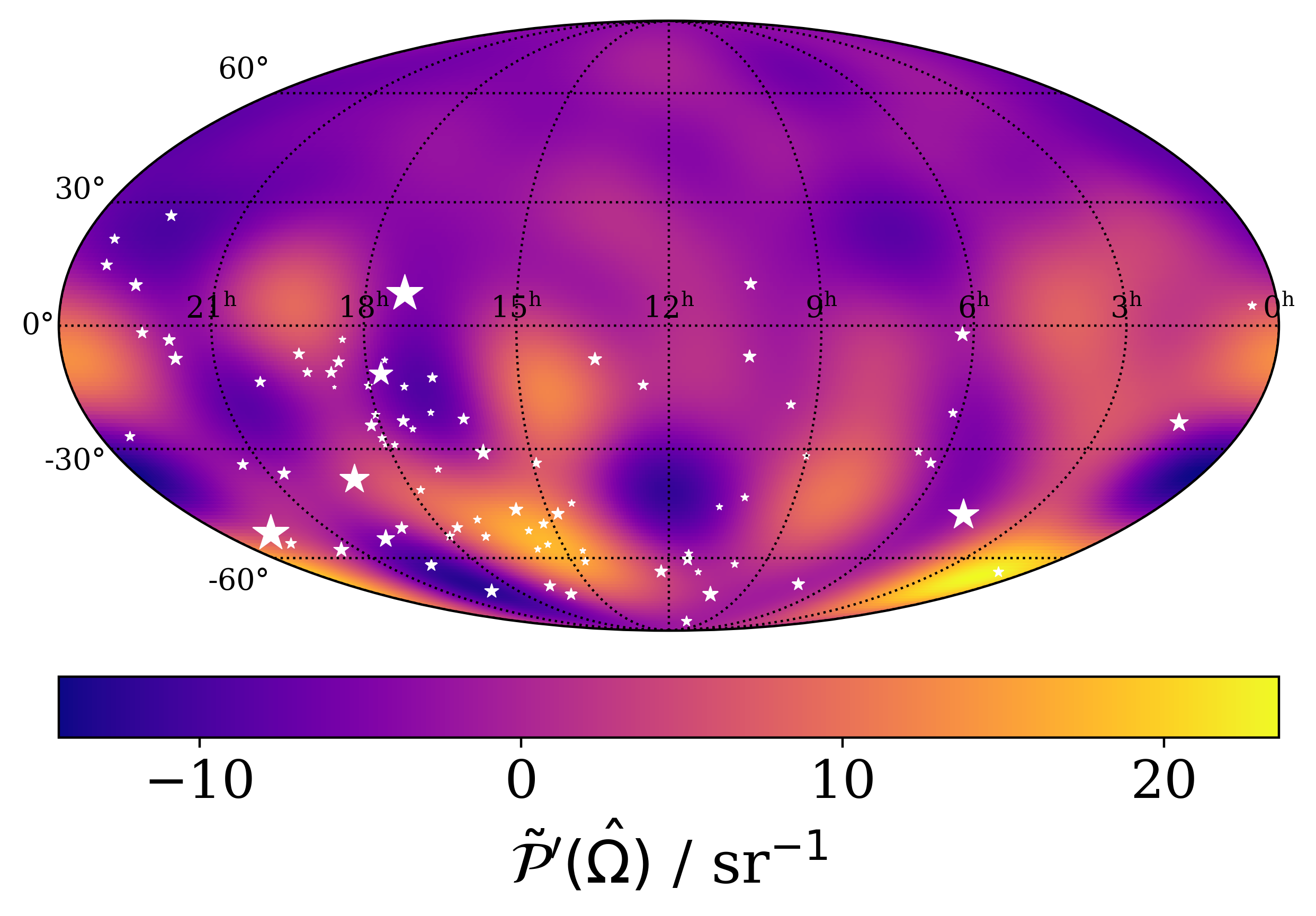}
    \label{fig:clean_data1}}
    \hfill
    \subfigure[Bin 2; $f = 1/T_\mathrm{obs} = \SI{14}{\nano\hertz}$]{\includegraphics[width=0.32\linewidth]{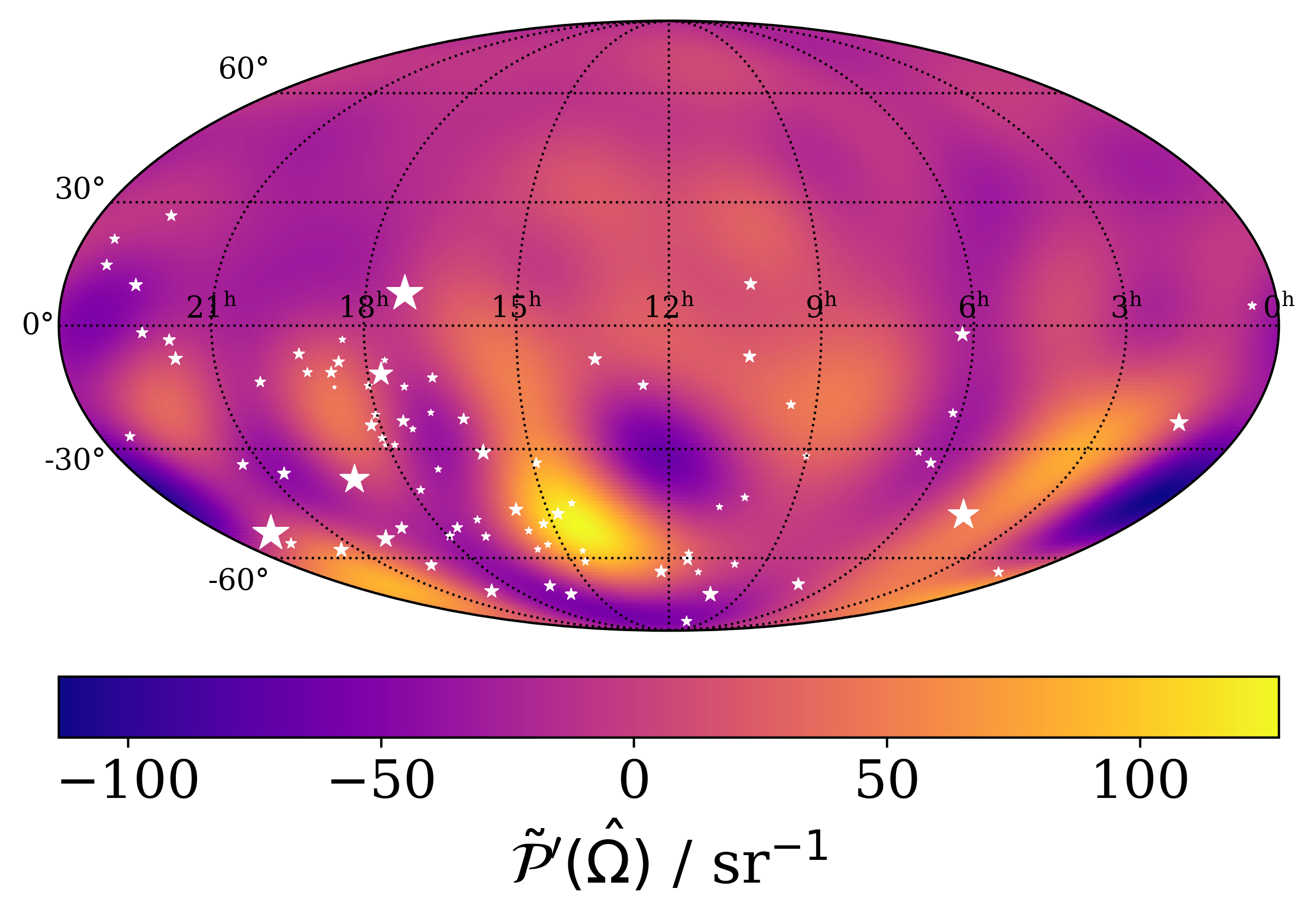}}
    \hfill
    \subfigure[Bin 3; $f = 1/T_\mathrm{obs} = \SI{21}{\nano\hertz}$]{\includegraphics[width=0.32\linewidth]{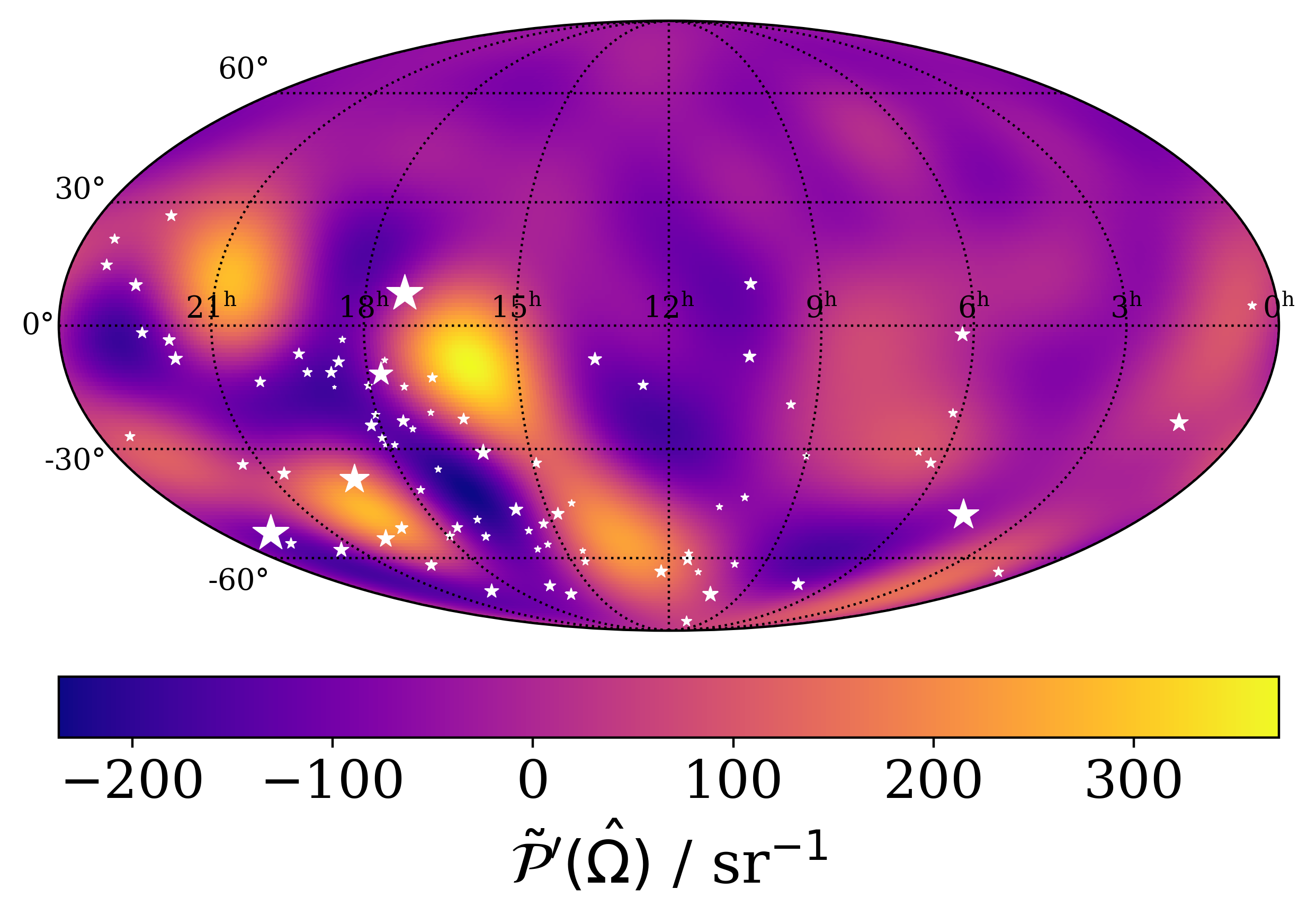}}
    \hfill
    \caption{Clean maps of the MPTA in its first three frequency bins.
        The units of the colour bar are \si{\square\correlation/\steradian}.
    }
    \label{fig:full_clean}
\end{figure*}

The clean maps associated with the S/N maps in Fig.~\ref{fig:full_clean} are shown in Fig.~\ref{fig:full_clean} across the three frequency bins. We used Equation~\ref{eq:power_sph} to derive these maps. The bright spot seen in the signal to noise ratio map in Fig.~\ref{fig:snr_data1} can also be seen in the corresponding clean map, Fig.~\ref{fig:clean_data1}.

\subsubsection{Clean sensitivity map}\label{app:ssec:sensitivity_clean}
\begin{figure*}
    \centering    
    \subfigure[Bin 1; $f = 1/T_\mathrm{obs} = \SI{7}{\nano\hertz}$]{\includegraphics[width=0.32\linewidth]{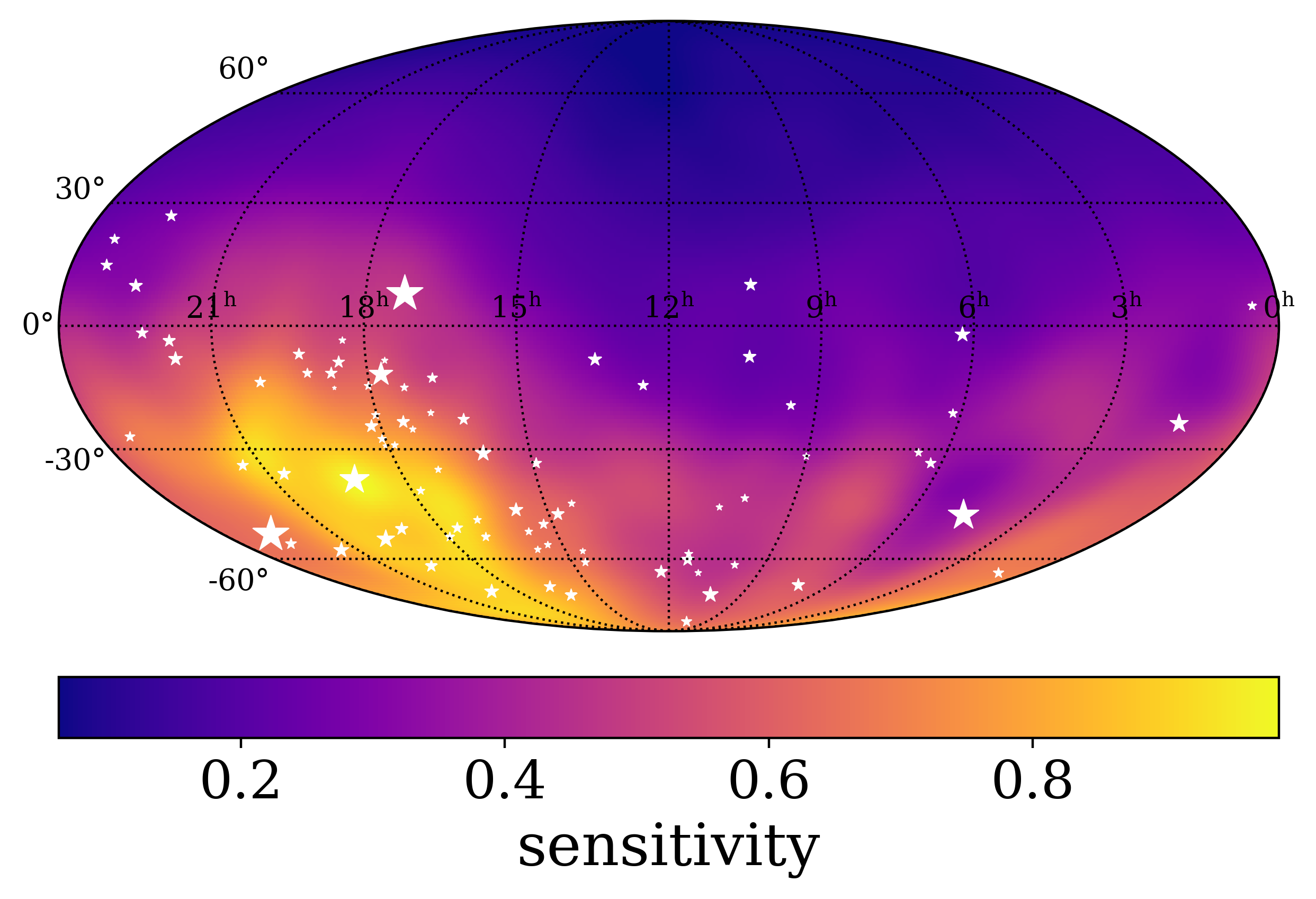}}
    \hfill
    \subfigure[Bin 2; $f = 2/T_\mathrm{obs} = \SI{14}{\nano\hertz}$]{\includegraphics[width=0.32\linewidth]{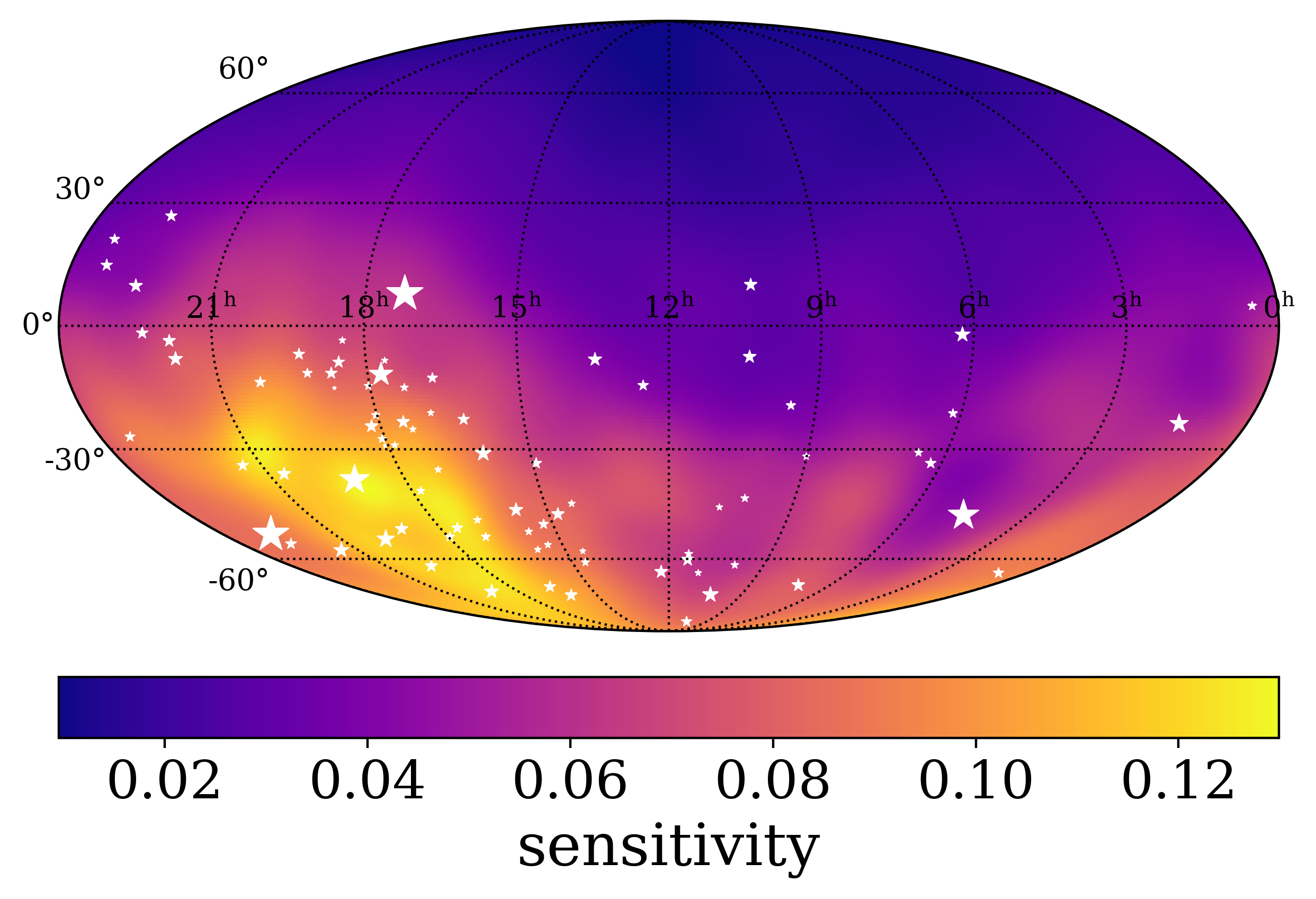}}
    \hfill
    \subfigure[Bin 3; $f = 3/T_\mathrm{obs} = \SI{21}{\nano\hertz}$]{\includegraphics[width=0.32\linewidth]{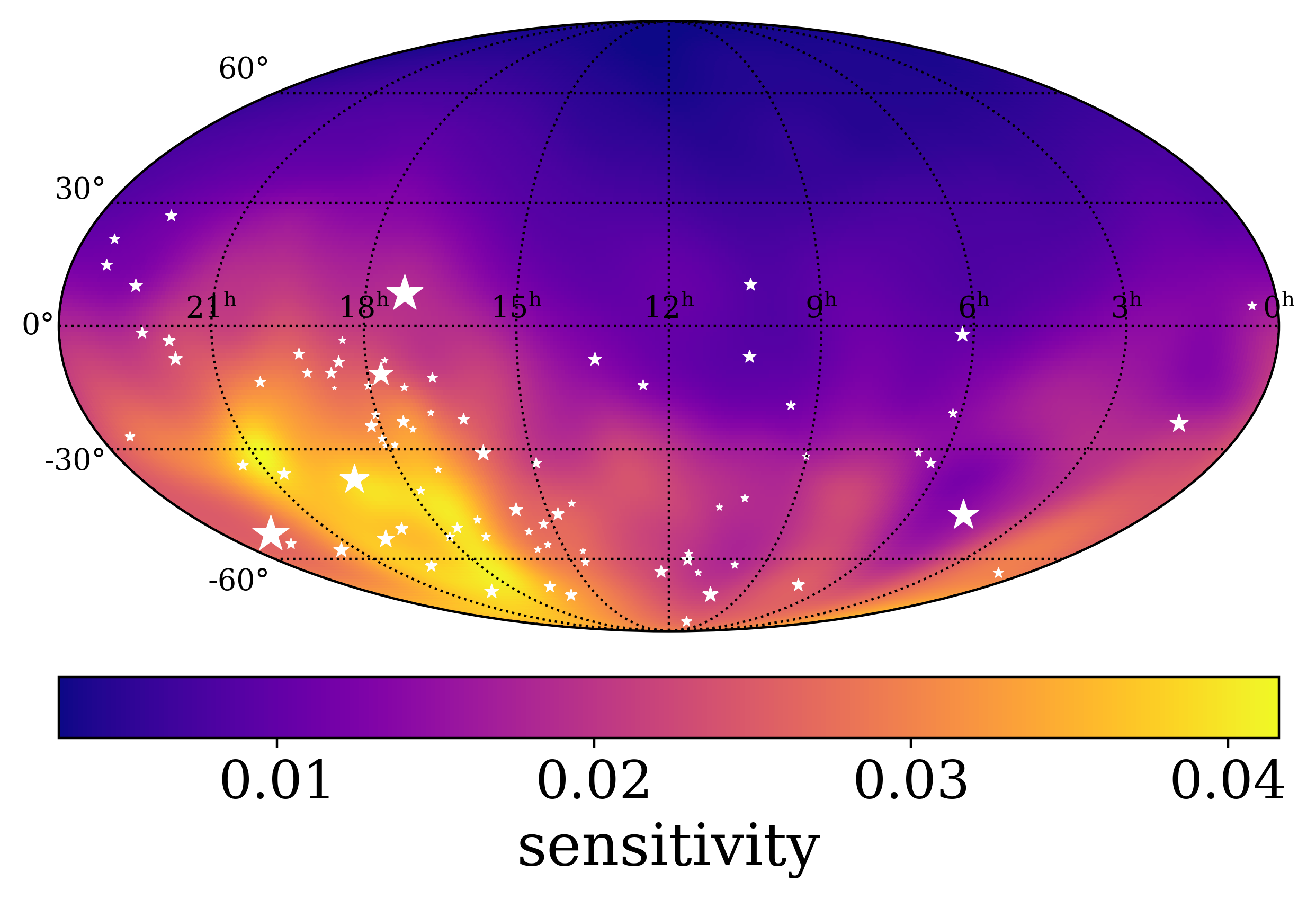}}
    \hfill
    \caption{Clean sensitivity map (defined in Eq.~\ref{eq:clean-sensitivity}) for the MPTA in its first three frequency bins.}
    \label{fig:full_sensitivity}
\end{figure*}

In order to visualise the sensitivity, $\mathcal{S}$, of the MPTA as a function of $(\text{RA}, \text{DEC})$, we define ``sensitivity maps'', which show the mean $S/N$ for a source located in some direction $\hat{n}$ with unit amplitude:
\begin{equation}
    \mathcal{P}_{\Hat{n}}(\hat\Omega) \equiv 
    \begin{cases}
        1 & \Hat{\Omega} = \Hat{n} \\
        0 & \Hat{\Omega} \neq \Hat{n}%
        \end{cases} .
\end{equation}
Formally, the sensitivity map is a plot of
\begin{equation}\label{eq:clean-sensitivity}
    \mathcal{S}_\mathrm{clean}(\hat{n}) \equiv \frac{\mathcal{P}_\text{eff}(\hat{n}) }{ \Tilde{\sigma}^{\mathcal{P}'} } = \frac{\Tilde{\mtx{M}}^{-1} \mtx{M}\; \mathcal{P}_{\Hat{n}}}{\Tilde{\sigma}^{\mathcal{P}'}}
\end{equation}
Here, $\mathcal{P}_\text{eff}(\hat{n})$ is the effective strain power after accounting for the loss of power due to regularisation. This quantity is calculated by constructing the expected dirty map $X_{\Hat{n}} = \mtx{M}\;\mathcal{P}_{\Hat{n}}$ and cleaning it with the regularised inverse of the Fisher matrix. 
Due to the pixel-wise evaluation, the sensitivity map can be written as
\begin{equation}
    \mathcal{S}(\text{RA},\text{DEC}) = \frac{\text{diag}\left(\Tilde{\mtx{M}}^{-1} \mtx{M}\right)}{\Tilde{\sigma}^{\mathcal{P}'}}.
\end{equation}
The resulting sensitivity maps are shown in Fig.~\ref{fig:full_sensitivity}. It is easier to detect gravitational-wave signals associated with high-sensitivity patches of sky. As also shown by \cite{Taylor_2013,Agazie_2023}, a PTA is expected to be most sensitive in the sky area with highest number density of pulsars, and most sensitive pulsars. This can also be seen in the maps in Fig.~\ref{fig:full_sensitivity}: They exhibit interesting structure showing that MPTA is considerably more sensitive to gravitational waves from the Southern Hemisphere than the Northern Hemisphere. As expected, the sensitivity distribution follows the number density of pulsars as well as the location of the most sensitive pulsars in the MPTA.

\subsection{Radiometer sensitivity maps}
\label{app:ssec:radiometer_sensitivity}

For completeness, we can also define the radiometer sensitivity map in analogy to the clean map sensitivity, but evaluating the radiometer $S/N$ (cf.\ Equation~\ref{eq:radiometer_snr}).  The corresponding map shows the mean $S/N$ for a source in some direction $\hat{\Omega}$ with unit amplitude:
\begin{equation}
    \eta_{\hat{n}} \equiv 
    \eta_{\hat{\Omega}}(\hat\Omega|\hat{n}) =
    \begin{cases}
        1 & \hat\Omega = \hat{n} \\
        0 & \hat\Omega \neq \hat{n}
    \end{cases}
\end{equation} 
as a function of $\hat{n}$.
Formally, the radiometer sensitivity for the pixel at position $\hat{n}$ is given as $\mathcal{S}_\mathrm{radiometer}(\hat{n}) \equiv \eta_{\hat{n}}/\sigma^\eta_{\hat{\Omega}}$. The full map can be expressed as
\begin{equation}
    \mathcal{S}_\mathrm{radiometer}(\text{RA},\text{DEC}) = \frac{\mathds{1}}{ \sigma^\eta_{\hat{\Omega}} } = \frac{\mathds{1}}{ \left(M_{\hat{\Omega}\hat{\Omega}}\right)^{-1/2} }
\end{equation}

\begin{figure*}
    \centering    
    \subfigure[Bin 1; $f = 1/T_\mathrm{obs} = \SI{7}{\nano\hertz}$]{\includegraphics[width=0.32\linewidth]{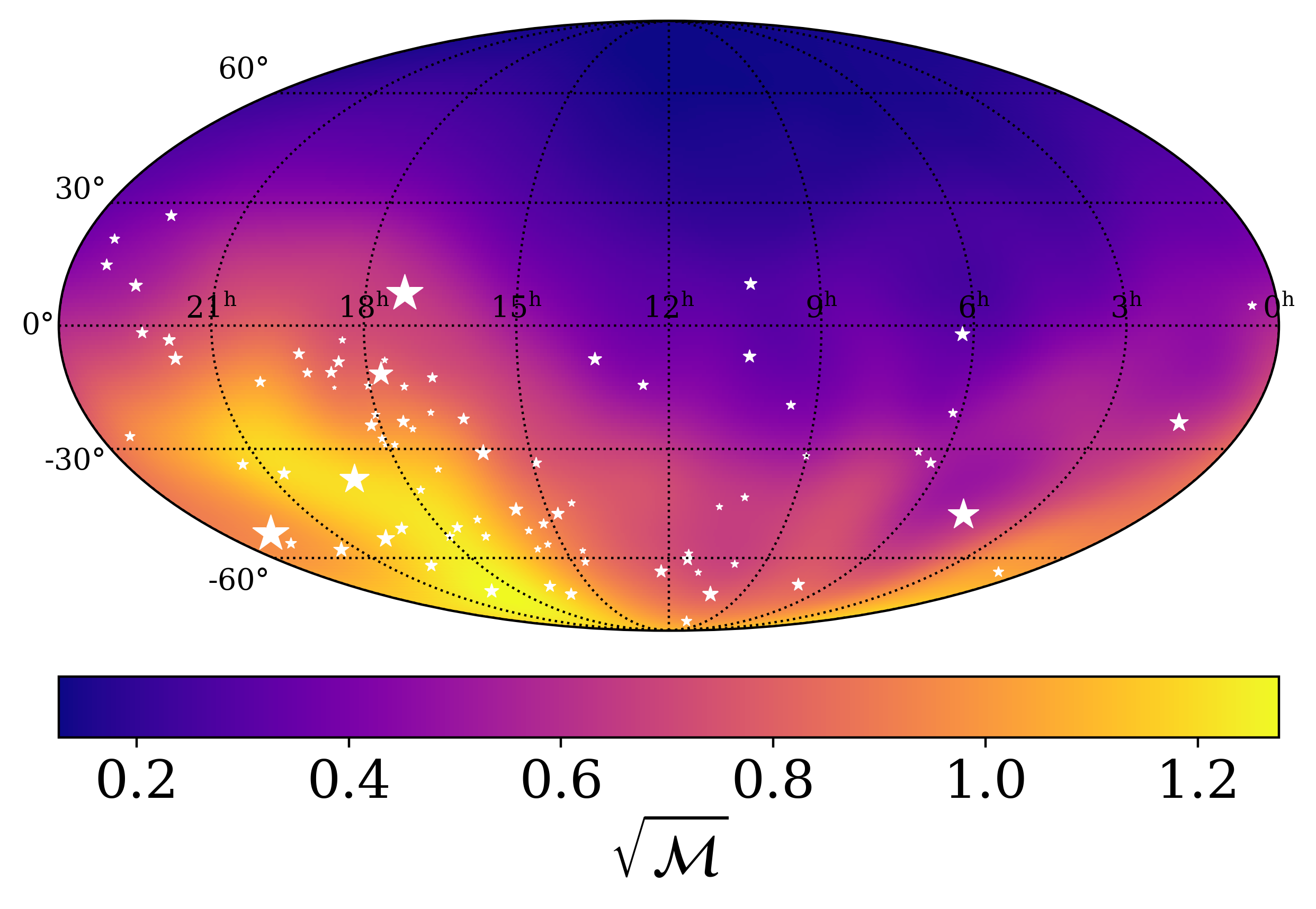}}
    \hfill
    \subfigure[Bin 2; $f = 2/T_\mathrm{obs} = \SI{14}{\nano\hertz}$]{\includegraphics[width=0.32\linewidth]{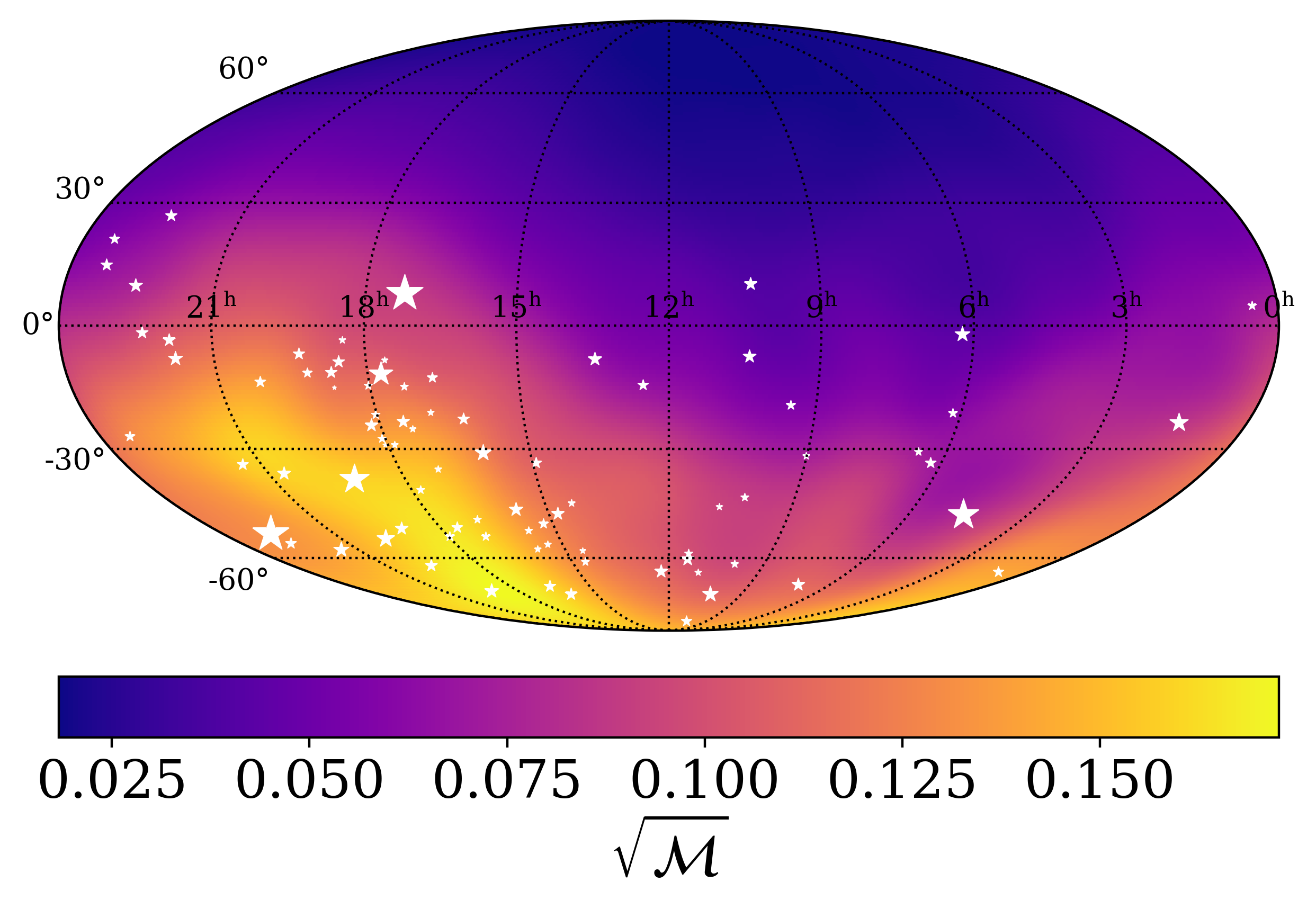}}
    \hfill
    \subfigure[Bin 3; $f = 3/T_\mathrm{obs} = \SI{21}{\nano\hertz}$]{\includegraphics[width=0.32\linewidth]{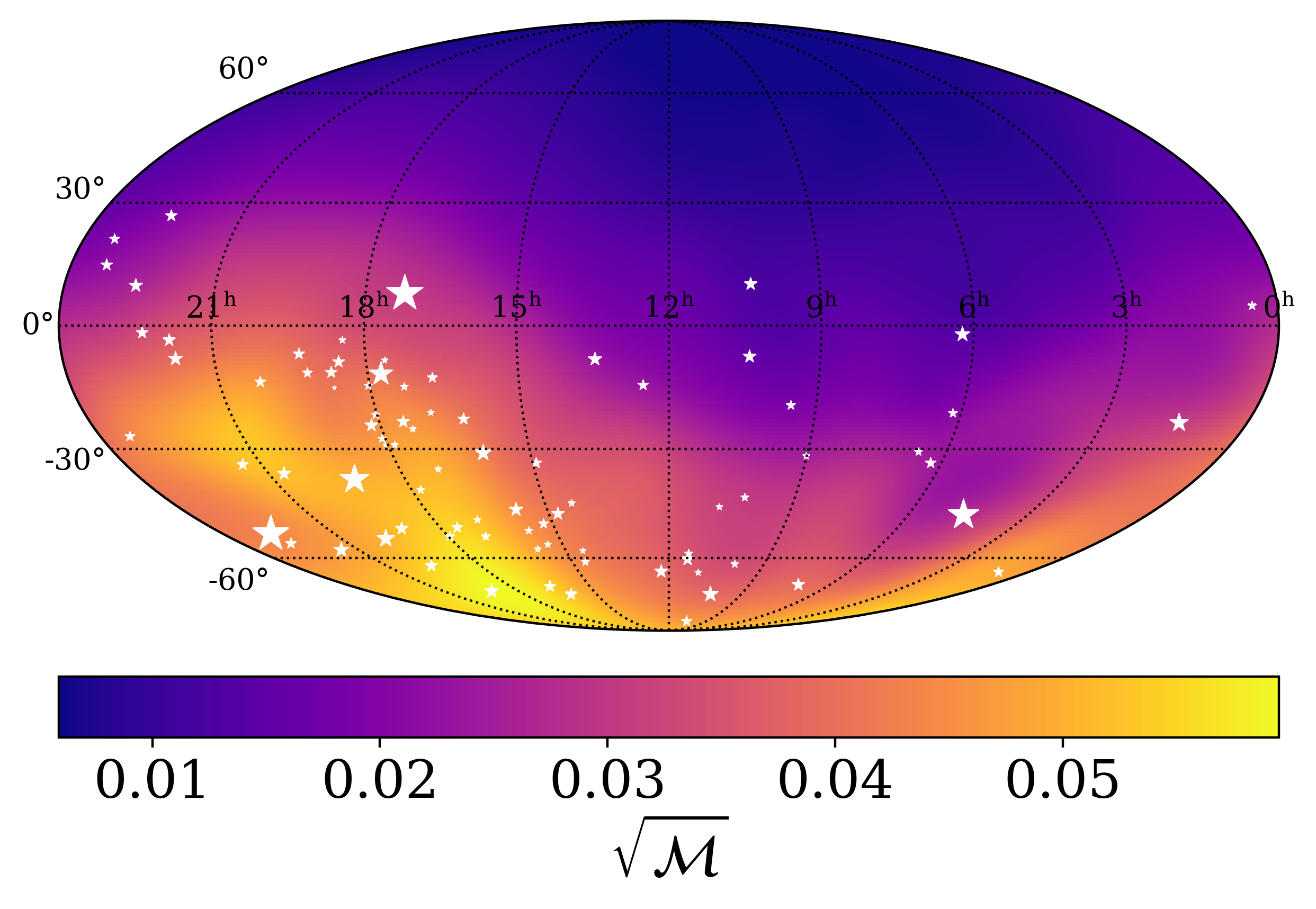}}
    \hfill
    \caption{Radiometer sensitivity map (defined in Eq.~\ref{eq:radiometer-sensitivity}) for the MPTA in its first three frequency bins.}
    \label{fig:radiometer_sensitivity}
\end{figure*}

These maps are shown in Fig.~\ref{fig:radiometer_sensitivity}. 

\section{Dropout analysis} \label{app:ssec:dropout}

\begin{figure*}
    \centering
    \subfigure[dropout J0437$-$4715]{\includegraphics[width=0.32\textwidth]{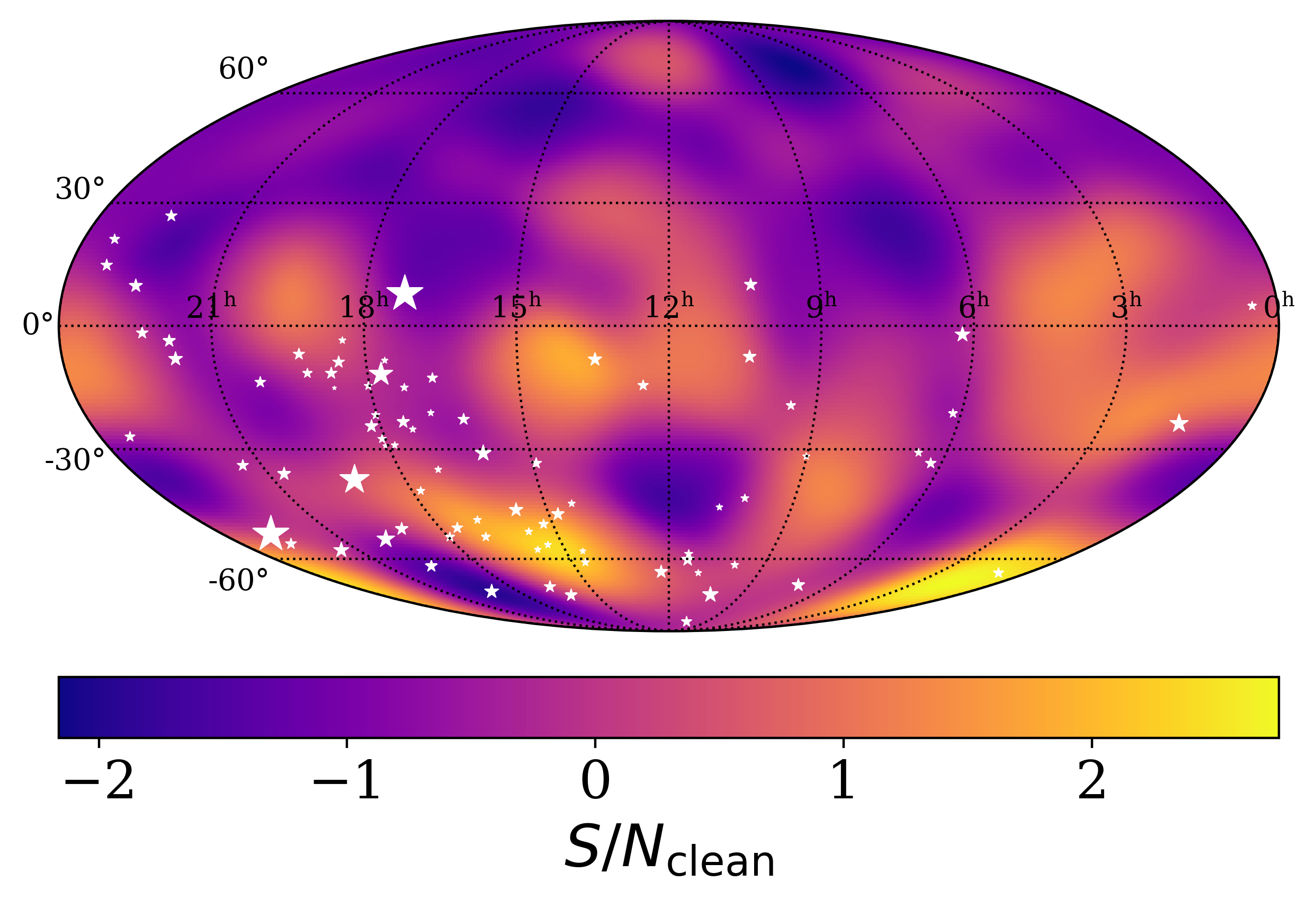}}
    \subfigure[dropout J2241$-$5236]{\includegraphics[width=0.32\textwidth]{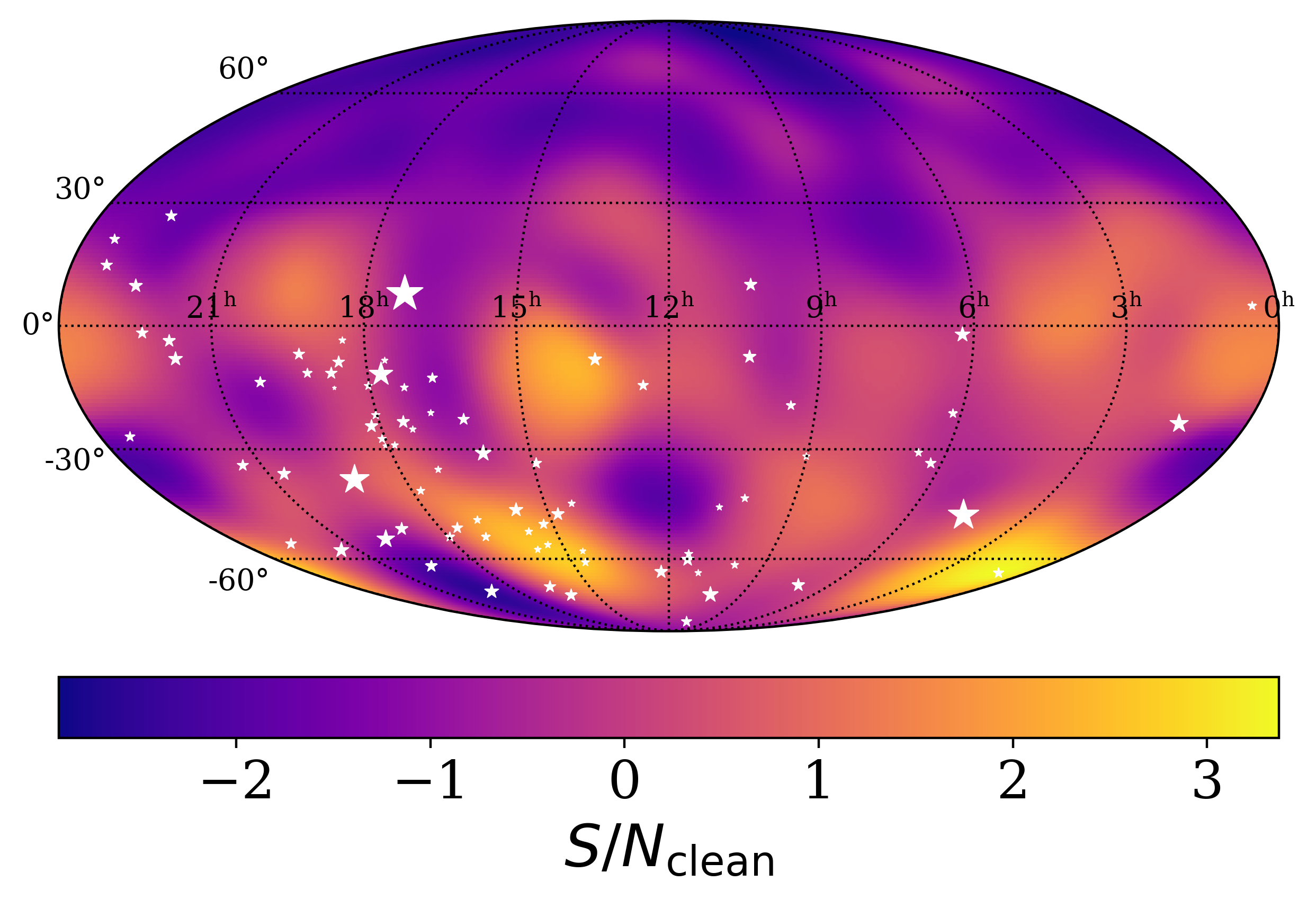}}
    \subfigure[dropout J1909$-$3744]{\includegraphics[width=0.32\textwidth]{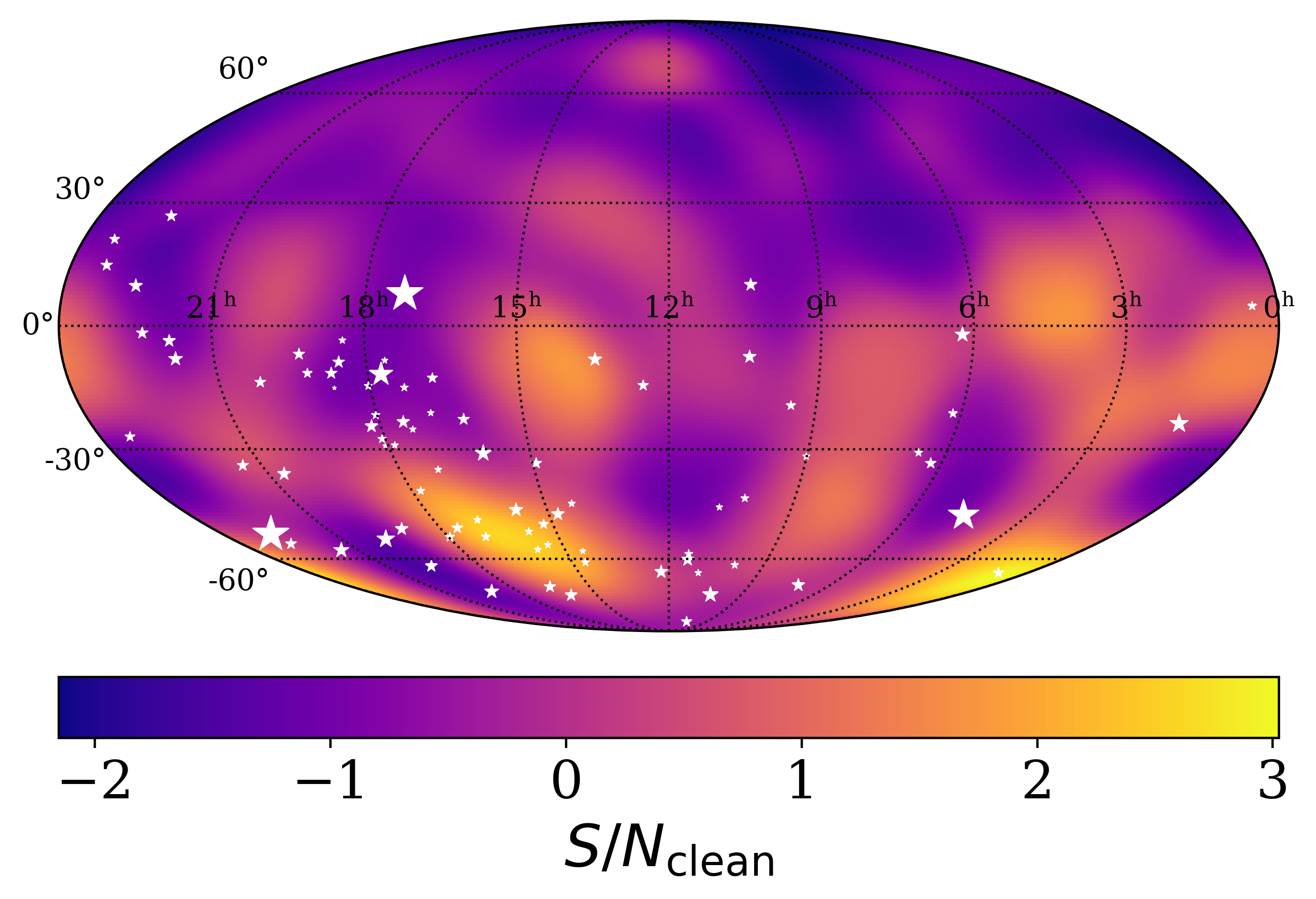}}
    \subfigure[dropout J2129$-$5721]{\includegraphics[width=0.32\textwidth]{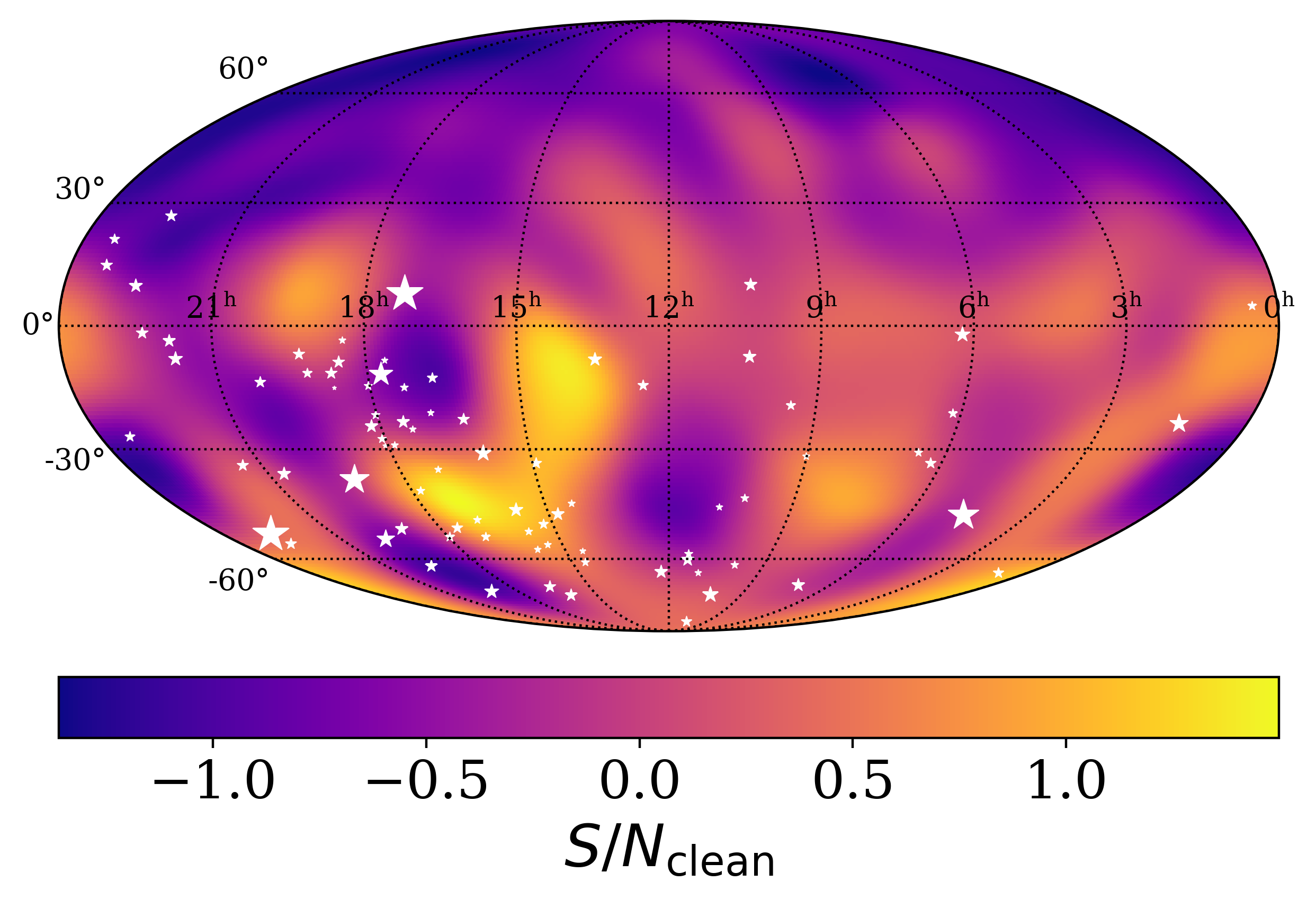}}
    \subfigure[full dataset]{\includegraphics[width=0.32\textwidth]{img/full_dataset/snrmap_papersmall_lmax8_sv32_nside16_fbin1.png}}
    \caption{Skymaps of the clean map $S/N$ in the first frequency bin resulting from the dropout analysis described in Section~\ref{app:ssec:dropout}.}
    \label{fig:dropout}
\end{figure*}

In this section we investigate how our sky maps depend on different pulsars by performing dropout analyses.
We identify the pulsars with the potentially largest influence on the sky maps based on the mean ToA uncertainty, overall timing residual mean square and the volatility of their noise models.  
The three pulsars with the lowest mean ToA uncertainty are (in ascending order) J0437$-$4715, J2241$-$5236 and J1909$-$3744. The respective clean map $S/N$ sky maps of the first frequency bin are shown in the top row of Fig.~\ref{fig:dropout}. As expected, the $S/N$ range changes slightly upon removing each pulsar. Nevertheless, the position and size of the patches with higher and lower $S/N$ remain similar compared to the sky map of the full dataset. Thus we conclude that our result is robust against the influence of these most precisely measured pulsars.

Additionally we test the influence of J2129$-$5721 on the sky map, due to the peculiarity of its noise model. As pointed out by \cite{MPTA2024_GWB}, it shows an extremely steep-spectrum achromatic red noise. On the one hand, the same noise is not present in pulsars that are at low angular separation to it (i.e.\ J1909$-$3744 and J2241$-$5236). On the other hand, 16 years of PPTA observations did not allow us to constrain any achromatic red noise in this pulsar. Thus, \cite{MPTA2024_GWB} presents the most sensitive solution excluding the achromatic red noise process for this pulsar. Since J2129$-$5721 is located at a key angular separation to other, more sensitive pulsars in the MPTA, we also test our results against removing it from the data set. The resulting sky map in the first frequency bin is shown in the bottom row of Fig.~\ref{fig:dropout}. Again we find that the $S/N$ range is decreased compared to the full data set. But in this case, the position of the $S/N$ patches change significantly. Most striking, the hotspot located at RA~1h DEC~\SI{-70}{\degree} almost completely vanishes, while the areas around RA~14h DEC~\SI{-10}{\degree} and RA~17h DEC~\SI{-45}{\degree} become more prominent as a ``triangle'' of increased $S/N$ areas in the Western Hemisphere.

\end{document}